\documentclass[12pt]{article}
\usepackage{amsmath,epsfig}

\newcommand{\rd}{\,\mathrm{d}}
\newcommand{\bra}[1]{\left\langle #1 \right|}
\newcommand{\ket}[1]{\left| #1 \right\rangle}
\newcommand{\gev}{~{\rm GeV}}

\newcommand{\rt}{\mathrm{T}}
\newcommand{\rl}{\mathrm{L}}
\newcommand{\rtl}{\mathrm{TL}}
\newcommand{\rli}{\mathrm{L1}}
\newcommand{\rlii}{\mathrm{L2}}
\newcommand{\rliii}{\mathrm{L12}}
\newcommand{\rcs}{\mathrm{CS}}
\newcommand{\rgj}{\mathrm{GJ}}
\newcommand{\rdh}{\mathrm{DH}}
\newcommand{\rsh}{\mathrm{SH}}
\DeclareMathOperator{\tr}{Tr}
\allowdisplaybreaks[4]

\newcommand{\beq}[1]{\begin{equation}\label{#1}}
\newcommand{\eeq}{\end{equation}}
\newcommand{\beqar}[1]{\begin{eqnarray}\label{#1}}
\newcommand{\eeqar}{\end{eqnarray}}
\newcommand{\nn}{\nonumber}
\newcommand{\dd}{{\rm d}} 

\newcommand{\ep}{\varepsilon}

\newcommand{\si}{\sigma}

\begin{document}

\hfill TPR-00-05
\vspace{3cm}
\begin{center}
{\Large \bf Nuclear Enhanced Higher-Twist Effects in the Drell-Yan Process}
\vspace{1cm}
\end{center}
\centerline{R.~J.~Fries, A.~Sch{\"a}fer, E. Stein}
\vspace{0.6 cm}
\centerline{\em  Institut f\"ur Theoretische Physik , 
Universit\"at Regensburg, 
D-93040 Regensburg, Germany}
\vspace{0.4 cm}
\centerline{and}
\vspace{0.6 cm}
\centerline{B.~M\"uller}
\vspace{0.4 cm}
\centerline{\em  Department of Physics, Duke University, 
Box 90305, Durham, North Carolina 27708-0305}
\vspace{1cm}
\bigskip


\bigskip
\noindent{\bf Abstract:}
We calculate the Drell-Yan cross section, resolving the full 
kinematics of the lepton pair, at high transverse momentum for hadron nucleus 
collisions. We use the general framework of Luo, Qiu and Sterman to calculate 
double scattering contributions that are of twist-4 and demonstrate their 
nuclear enhancement. By comparing single and double scattering at RHIC 
energies we find that double scattering 
gives contributions of comparable size. 
We also show that the angular dependence of the Drell-Yan pair discriminates
between the various double scattering contributions.
\bigskip
\eject

\section{Introduction}

In recent times the interest in understanding hard processes involving nuclei,
like electron nucleus collisions ($e+A$) and hadron nucleus collisions 
($h+A$), has rapidly increased. The relativistic heavy ion collider (RHIC) 
will provide a new tool for experimental studies of $h+A$ and $A+A$ physics. 
This experimental advance provides a challenge to theorists for developing
a description of such reactions strictly in terms of the underlying
fundamental field theory, i.e.\ quantum chromodynamics (QCD). 

The usual way to proceed in QCD is to use factorization theorems
\cite{coll89} that permit a rigorous treatment
of hard processes. In this approach, 
incalculable soft contributions are factorized
into universal, process independent distribution functions.
Measurement of the distribution functions in one process then
allows the prediction of cross sections for other processes.
An example is the Drell-Yan (DY) process where two hadrons 
scatter to produce a lepton pair of large invariant mass $Q^2$.
If $Q^2$ is large enough the factorization theorems state, that the inclusive
Drell Yan cross section can be expressed as the product of a calculable 
hard part and universal twist-2 parton distribution functions that can be 
measured, for instance in deeply inelastic lepton nucleon scattering.
Such universal distribution functions have been successfully
extracted from existing data.
The natural question arises how this 
picture is changed for hadron-nucleus collisions.
Differences between the twist-2 distribution of a single nucleon and those of a nucleus of mass number $A>1$ were  first discovered by the 
European Muon Collaboration. 
This phenomenon is known as the EMC effect at large values of the Bjorken 
scaling variable $x$ and as shadowing at lower values of $x$ 
\cite{Arneodo:1994wf}.

However, strong interaction dynamics cannot solely be described in terms of 
twist-2 structure functions. 
These functions are simple one-particle correlators and
have the interpretation of a probability to find a parton in the 
hadron with certain momentum fraction $x$. These distribution
functions are measured when the hard probe scatters off a single parton in the
nucleon. 
Of course, the probe may scatter off more than one parton in the target.
Such processes are related to multi-parton correlators and describe corrections
to the single-parton scattering process. In general they are suppressed by 
powers of the large scale, and are called contributions of 
higher twist.

If we consider processes involving large nuclei it is quite obvious that 
multiple scattering will be enhanced and even dominate in the limit 
$A \rightarrow \infty$. 
The classical example for enhancement of multiple scattering in 
$h+A$ processes is the Cronin 
effect, i.e.\ the anomalous $A$-dependence of the Drell-Yan cross section
at large values of the transverse momentum
of the lepton pair \cite{Boymond:1974ns}.
Instead of a scaling of the cross section with the volume $\sim A$
as expected in the twist-2 case, one observes an $A$-dependence parametrized by
$\sigma^{h+A} = A^\alpha \sigma^{h+p}$ with $\alpha \approx 4/3$ 
\cite{Alde:1990im}.
This is a clear signal for multiple scattering and shows that higher-twist
multiple scattering is grossly enhanced at large momentum transfer.

The literature contains a wide variety of approaches dealing with multiple 
scattering, see e.g.~\cite{bod89}, but a rigorous formulation in the language 
of perturbative QCD was attempted only recently.
In a series of papers, Luo, Qiu and Sterman (LQS) and recently also Guo, 
showed how to extend the factorization theorem beyond leading twist to get a 
handle on double scattering in terms of perturbative QCD \cite{Luo94,Luo94b}. 
Their twist-4 matrix elements are correlators of two partons in the nucleus 
and can describe scattering off partons from different nucleons. 
This approach predicts a stronger dependence on the size of the nucleus
than the linear scaling with $A$ of single scattering. 
LQS argue that ---  due to the geometrical structure of the matrix elements
--- some of these twist-4 nuclear matrix elements pick up an additional factor
$A^{1/3}$, compensating for some of the inherent suppression of higher 
twist corrections. This nuclear enhancement is able to describe the observed 
scaling with $A^{4/3}$ in certain kinematical regions.

In this paper we discuss nuclear enhanced double scattering contributions to 
DY pair production off nuclear targets 
\begin{equation}
  \label{process}
  h(P_2) + A (P_1 A) \longrightarrow l^+ l^- + X
\end{equation}
in the framework of the LQS approach. 
It was already shown some time ago by Guo \cite{Guo98,Guo98b} that 
higher-twist effects in the process (\ref{process}) are enhanced at large 
transverse momentum. Extending the calculations of Guo \cite{Guo98},
we have calculated the full kinematical dependence of the cross section
$\dd \sigma/\dd Q^2 \,\dd q_\perp^2 \,\dd y \,\dd\Omega$
including the angular distribution of the lepton pair and evaluated it, 
in particular, for $p+A$ scattering at RHIC. 
Our most important results were already published as a letter \cite{fmss99}. 
In this contribution we present the complete analytical results together with 
a brief review of the LQS approach and some technical details. We also give 
some numerical results not contained in \cite{fmss99} and an extended 
discussion. 
We will argue that nuclear enhanced double scattering corrections cannot be 
neglected at RHIC. 

\begin{figure}[t]
  \label{scattering}
  \epsfig{file=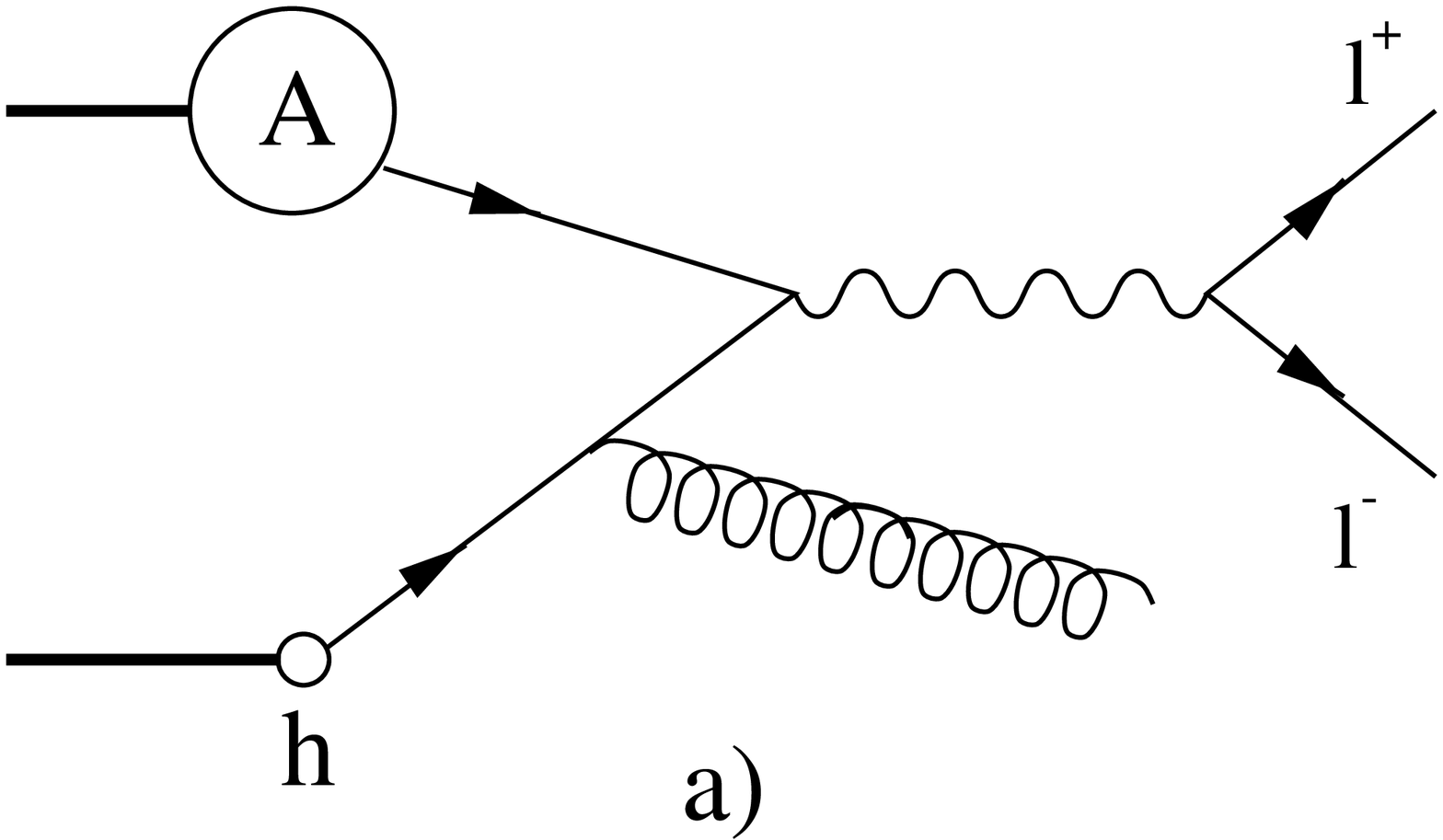,width=6cm}\hspace{2cm}
  \epsfig{file=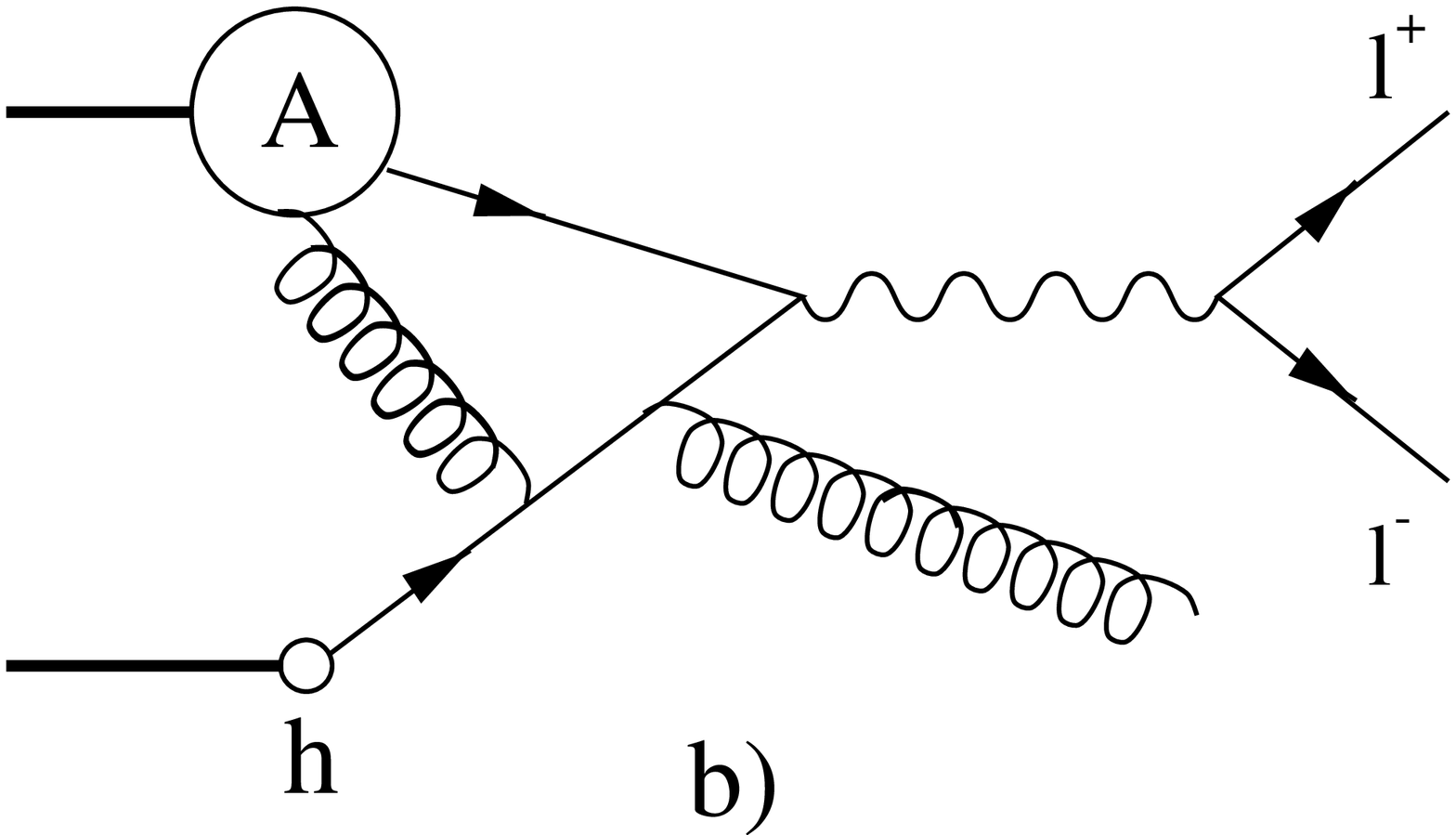,width=6cm}
  \caption{Single- (a) and double-scattering (b) schematically for a 
    hadron $h$ colliding with a nucleus $A$ producing a $\mu^+\mu^-$ pair. 
    One additional unobserved parton is radiated in the leading contribution 
    to high transverse momentum $q_\perp$.}
\end{figure}

In Fig.~\ref{scattering} we show typical Feynman diagrams contributing to the
DY process. 
Single scattering means that, e.g., a quark from the single hadron 
and an antiquark from the nucleus annihilate into a
virtual photon in a hard process. To obtain large transverse momentum
an additional unobserved parton, here a gluon, has always to be radiated. 
Double scattering in our example implies that the quark can pick up an 
additional gluon from the nucleus. 
The generalized twist-4 factorization theorem of LQS states
that two distinct scattering reactions occur for double scattering.
One is referred to as double-hard (DH) scattering. Here the parton from the 
single hadron undergoes two subsequent independent hard scattering reactions
 in the target. 
The other one is called soft-hard (SH) scattering. Here the parton from the 
single hadron interacts with the collective soft gluon field in the nucleus 
(it can e.g.\ be thought of as feeling the color-magnetic Lorentz force) before
producing the virtual photon via the final hard scattering. At large 
transverse momentum interference of these two mechanisms can be neglected.

The double-hard reaction resembles the classical double scattering 
picture, while soft-hard scattering may have an analogy in plasma physics,
where a hard scattering event can be accompanied by initial- or final-state
interaction with the mean field.
The first process might be accounted for by event generators in simulations 
of heavy-ion reactions whereas the implementation of soft-hard scattering has 
yet to be formulated in a fully consistent manner (see \cite{bmp99} for a 
conceptual discussion).

It would be helpful to find observables with vanishing or at least very small 
contributions from single scattering. 
The measurement of such observables would permit the identification of 
twist-4 effects at larger values of $Q^2$ where perturbation theory is more
reliable and thus provide a clearer evidence for the occurrence of double 
scattering.
We find these observables in the angular distribution of DY pairs, namely the 
helicity amplitude $W_\Delta$ and the Lam-Tung relation 
$W_{\Delta\Delta}=W_\rl/2$ \cite{lt80}, both introduced in section 2.
The Lam-Tung relation is analogous to the Callan-Gross relation in 
deep-inelastic electron nucleon scattering (DIS).
Furthermore it would be helpful to find observables where double-hard and 
soft-hard contributions can be disentangled. These can also be found in the 
angular distribution of DY pairs. While SH scattering leads to non-trivial 
angular patterns, DH scattering does not. 
Any anomalous $A$-dependent angular distribution therefore would be a 
clear sign of non trivial twist-4 correlations.

Although we will focus on $h+A$ collisions, our work has been primarily 
motivated by the fact that higher-twist effects or multiple scattering should
be most prominent in nucleus-nucleus ($A+A$) collisions at high energy, 
as they will be routinely studied at RHIC. 
Two microscopic, QCD-inspired descriptions of
nucleus-nucleus collisions at RHIC energies and beyond have been proposed in
recent years. In one, the parton cascade picture \cite{ekl89,kgbm92,kg95},
the collision is modeled as a sequence of incoherent binary scattering events
among partons (``minijets''), which can be cast in a probabilistic formulation
as a multiple scattering process. In the other one, the random mean-field
model \cite{mv94} the collision is described as a nonlinear interaction
among coherent color fields carried by the colliding nuclei. We hope that the
study of the interplay between soft-hard and multiple-hard processes in a
simpler framework, such as $h+A$ interactions, will help to elucidate the
competition between these processes in $A+A$ collisions.

The determination of multiparton correlation functions in nuclei is also an 
important step towards the application of QCD to nuclear physics in general. 
The generic enhancement of higher-twist effects in reactions involving nuclei 
facilitates their extraction and give the chance for a deeper theoretical
understanding. They encode the essential aspects of the difference in the 
quark-gluon structure between isolated nucleons and nuclei. From the practical
point of view a better understanding of such medium effects is necessary to
be able to interpret the so called ``hard probes'' of high-energy heavy-ion
collisions in a convincing manner.

The unintegrated Drell-Yan cross section is actually the classic example
for a typical two-scale process. Both $Q^2$ and $q_\perp$, the mass and 
transverse momentum of the photon, are detected. The perturbative
treatment then gives rise to logs of the type $\log^2(Q^2/q_\perp^2)$
which can become large if the scales differ substantially. In such a case
they have to be resummed. Resumming large logs in the
presence of additional soft gluons contained in the twist-4 correlators
is a delicate task that could  not be solved yet. We therefore
follow a less ambitious procedure and require that
both scales are of the same order, thus reducing the
original two-scale problem to an effective one-scale problem. In addition the
LQS formalism in its present form requires $q_\perp^2$ to be not much smaller 
than $Q^2$ in order to neglect interference terms that are not calculated up 
to now and which would introduce additional twist-4 matrix elements. 
Recently Guo, Qiu and Zhang proposed a method to describe the region
of low transverse momentum and how to match it to the perturbative calculation
at high $q_\perp$ \cite{gqz99}.

Finally we would like to comment on the experimental situation.
There exist data for angular coefficients on $\pi+A$ scattering, e.g. from the 
NA10 \cite{na10} and E615 \cite{e615} experiments which are not yet completely
understood in terms of conventional QCD-treatment.
These experiments measure pair masses $Q \approx 4\ldots8.5 \gev$ and 
transverse momenta $q_\perp< 3\gev$ which is certainly at the limit of 
where we could still trust our calculation. We will focus on future RHIC 
perspectives where we could hopefully reach higher 
transverse momenta.

The paper is organized as follows:
In section 2 we give the general definition of the 
Drell-Yan kinematics, we introduce our observables --- helicity amplitudes and 
angular coefficients --- and explain possible choices of the photon rest frame.
In section 3 we remind the reader of the known twist-2 results
for the helicity amplitudes. Section 4 is devoted to the calculation of the 
double-scattering contribution. We explain some essential features of the 
factorization procedure of LQS.
In particular we discuss the distinction between double-hard and soft-hard 
scattering and give an introduction of how to calculate them practically.
Finally section 5 contains our numerical results along with a discussion 
and we close with  a summary of our final conclusions. In the appendices we 
have collected some technical details referred to in the text.

\section{General Definitions}

\subsection{Kinematics}

Let a nucleus with momentum $P_1^\mu$ per nucleon collide with a hadron
with momentum $P_2^\mu$. We study lepton pair decay of virtual photons with 
momentum $q^\mu$ and mass $Q^2=q^2$. In the hadron 
center-of-mass (c.m.) frame we take   
\begin{equation}
  \label{momenta}
  P_1^\mu =\frac{1}{2}(\sqrt{S},0,0,\sqrt{S}) ,\quad P_2^\mu =\frac{1}{2}
  (\sqrt{S},0,0,-\sqrt{S})
\end{equation}
where $\sqrt{S}$ is the total center-of-mass energy.
The differential cross section for Drell-Yan pair production is then given by
\begin{equation}
  \label{cross}
  d \sigma = \frac{\alpha^2_{em}}{2 S Q^4} L_{\mu\nu} W^{\mu\nu} 
  \frac{\dd^4 q}{(2 \pi)^4} \dd\Omega.
\end{equation}
The solid angle $\Omega$ is parametrized by the polar and azimuthal angles 
$\theta$ and $\phi$ of one decay lepton in the rest frame of the virtual 
photon. We have the freedom to choose the axes in the photon rest frame 
conveniently. We will discuss this point after eq.~(\ref{as}).

The hadronic tensor follows the standard definition and is given by
\begin{equation}
  \label{wmunu}
  W_{\mu\nu} = \int \dd^4 x \, e^{i q\cdot x} 
  \bra{P_1 P_2} j_{\mu}(x) j_\nu(0)\ket{P_1 P_2} \; .
\end{equation}
The leptonic tensor is
\begin{equation}
  \label{lmunu}
  L^{\mu\nu} = 2 \left(l^{\mu}_1 l^{\nu}_2 + l^{\mu}_2 l^{\nu}_1 \right)-
             g^{\mu\nu} Q^2 \; ,
\end{equation}
where $l_1$ and $l_2$ are the momenta of the leptons.

We rewrite the $q$-dependence of the cross section in terms of rapidity $y$, 
transverse momentum $q_\perp$ and mass $Q$ of the virtual photon in the hadron 
c.m.~frame. Introducing the hadronic Mandelstam invariants 
\begin{eqnarray}
  \label{mandelh}
  S &=& (P_1 + P_2)^2 \nn \\
  T &=& (P_1 -q)^2 \\
  U &=& (P_2 -q)^2 \nn
\end{eqnarray}
the rapidity $y$ of the photon is
\begin{equation}
\label{rap}
  y =  \frac{1}{2} \ln\left(\frac{q^0 + q^3}{q^0 - q^3}\right)
  = \frac{1}{2} \ln\left(\frac{Q^2-U}{Q^2 - T}\right)\; .
\end{equation}
Likewise we can express the transverse momentum of the lepton pair
in the hadron c.m. system by
\begin{equation}
\label{transm}
  q_\perp^2 = \frac{(Q^2-U)(Q^2 - T)}{S} - Q^2 \; .
\end{equation}
and we can rewrite the differential cross section as:
\begin{equation}
\label{cross1}
  \frac{\dd \sigma}{\dd Q^2 \,\dd q_\perp^2 \,\dd y \,\dd\Omega} = 
  \frac{\alpha^2_{em} \pi}{4 S Q^4} L_{\mu\nu} W^{\mu\nu} 
  \frac{1}{(2 \pi)^4} 
\end{equation}

\subsection{Helicity amplitudes and their frame dependence}

To discuss the angular distribution of the lepton pair it is
convenient to introduce helicity amplitudes $W_{\sigma,\sigma'}$ defined
as 
\begin{equation}
\label{helicity}
  W_{\si,\si'} = \epsilon_{\mu}^{(\sigma)} \epsilon_{\nu}^{(\sigma')*}
  W^{\mu\nu}
\end{equation}
where $\epsilon_{\mu}^{(\si)}$, $\sigma \in\{ \pm 1, 0\}$ is a set of 
polarization vectors of the virtual photon with respect to the axes defined 
by the spatial unit vectors {\bf X}, {\bf Y}, {\bf Z} in its rest frame:
\begin{equation}
\label{vectors}
  \ep_{\mu}^{(\pm)} = \frac{1}{\sqrt{2}}(\mp X - i Y)_\mu , \qquad
  \ep_{\mu}^{(0)} = Z_\mu
\end{equation}
Following \cite{lt78} it is advantageous to introduce a complete set of 
structure functions 
\begin{equation}
  \label{helicity2}
  \begin{split}
  W_\rt &= W_{1,1} \\
  W_\rl &= W_{0,0} \\
  W_\Delta &= \frac{1}{\sqrt{2}}\left(W_{1,0} + W_{0,1}\right) \\
  W_{\Delta\Delta} &= W_{1,-1}
  \end{split}
\end{equation}
which are referred to as the transverse, longitudinal, spin flip, and 
double spin flip helicity  amplitudes.
Due to parity conservation other combinations are related to these by 
$W_{\si,\si'} = (-)^{\si+\si'} W_{-\si,-\si'}$.
In terms of these helicity structure functions the cross section 
can be written as
\begin{multline}
  \label{cross2}
  \frac{\dd \sigma}{\dd Q^2 \,\dd q_\perp^2 \,\dd y \,\dd\Omega} =
  \frac{\alpha^2_{em}}{64 \pi^3 S Q^2}  \\
  \left( W_\rtl\left(1+\cos^2\theta\right) +
     W_\rl\left(\frac{1}{2}-\frac{3}{2}\cos^2\theta\right) +
     W_\Delta\left(\sin 2\theta\cos\phi \right) +
     W_{\Delta\Delta}\left(\sin^2\theta\cos 2\phi \right)\right).
\end{multline}

\begin{figure}[t]
  \centerline{\epsfig{file=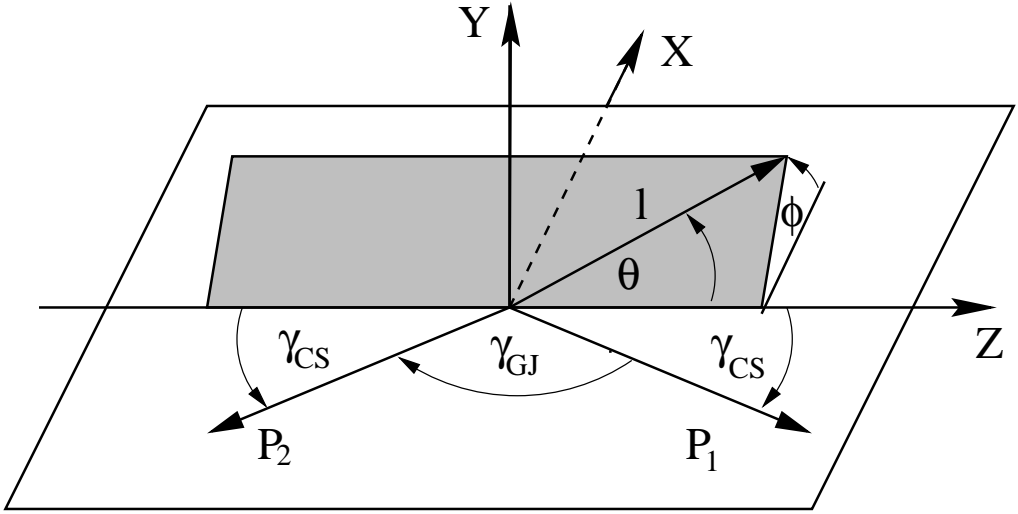,width=13.5cm}}
  \caption{Definition of the angles $\theta$ and $\phi$ in the rest frame
    of the lepton-pair. The frame shown here is the Collins-Soper frame.}
  \label{framec}
\end{figure}

The angular dependence can be integrated out which reduces the cross section 
to 
\beq{total}
\frac{\dd \sigma}{\dd Q^2 \,\dd q_\perp^2 \,\dd y } =
\frac{\alpha^2_{em}}{64\pi^3 S Q^2} \left(\frac{16}{3}\pi \right) W_\rtl
\end{equation}
with $W_\rtl = W_\rt + \frac{1}{2} W_\rl = -\frac{1}{2} W_{\mu}^{\mu}$.
The ratio of the angular dependent and angular integrated cross sections is 
usually parametrized in two standard forms \cite{na10,cs77}: 
\begin{equation}
  \label{angcoeff}
  \begin{split}
  \frac{16 \pi}{3} 
  \left(\frac{
  \frac{\dd \sigma}{\dd Q^2 \,\dd q_\perp^2 \,\dd y \,\dd\Omega}}
  {\frac{\dd \sigma}{\dd Q^2 \,\dd q_\perp^2 \,\dd y}} \right) 
   & = (1 + \cos^2\theta) + A_0 \left(\frac{1}{2} - \frac{3}{2} \cos^2\theta
  \right) + A_1 \sin 2\theta \, \cos\phi +
  \frac{A_2}{2} \sin^2\theta \, \cos 2\phi = \\
   & = \frac{4}{\lambda+3} \left( 1+ \lambda \cos^2\theta + \mu \sin 2\theta \,
  \cos \phi + \frac{\nu}{2} \sin^2 \theta \, \cos 2\phi \right)
  \end{split}
\end{equation}
in terms of two sets of angular coefficients
\begin{align}
  \label{as}
  A_0 &= \frac{W_\rl}{W_\rtl} & A_1 &= \frac{W_\Delta}{W_\rtl} & 
  A_2 &= \frac{2\,W_{\Delta\Delta}}{W_\rtl}, \\
  \lambda &= \frac{2-3A_0}{2+A_0} &  \mu &= \frac{2 A_1}{2+A_0} &
  \nu &= \frac{2 A_2}{2+A_0}.
\end{align}
At this point it is necessary to comment on the proper choice of coordinate 
axes in the photon rest system with respect to which the polar and azimuthal 
angles $\theta$ and $\phi$ are defined. Clearly the helicity amplitudes and 
angular coefficients depend on the chosen frame.
Two prominent examples are
the Collins-Soper frame (CS) and the Gottfried-Jackson frame (GJ) which 
are discussed in detail in \cite{lt78}.
When we boost from the hadron c.m.\ frame to a lepton c.m.\ frame,
the collinearity of the hadron 
momenta ${\mathbf P}_1$ and ${\mathbf P}_2$ is lost and they
span a plane which we identify with the $\mathbf X$-$\mathbf Z$-plane. 
We are still free to fix $\mathbf Z$ within this plane. 
In the Collins-Soper frame we choose $\mathbf Z$ to bisect the angle between 
${\mathbf P}_1$ and $-{\mathbf P}_2$, see Fig.~\ref{framec}. The angle between
$\mathbf Z$ and ${\mathbf P}_1$ and between $\mathbf Z$ and $-{\mathbf P}_2$ is
called $\gamma_\rcs$  In the Gottfried-Jackson frame $\mathbf Z$ lies in the 
direction of ${\mathbf P}_1$. We have introduced the the angle $\gamma_\rgj$
between ${\mathbf P}_1$ and ${\mathbf P}_2$ for short notations in that case.
Obviously we have $2 \gamma_\rcs + \gamma_\rgj = \pi$. See Appendix 7.1 for 
more details. Also note that for $q_\perp= 0$ both frames are equal and 
$\phi$ is no longer defined. 

In order to obtain the results in both frames in a straightforward way we use 
the projectors given by Mirkes \cite{mir92}
\begin{align}
  \label{mirkesproj}
  P^{\mu\nu}_{\rtl} &= -g^{\mu\nu} &
  P^{\mu\nu}_{\rli} &= \frac{P_1^\mu P_1^\nu}{E_1^2}  \nn \\ 
  P^{\mu\nu}_{\rliii} &= \frac{P_1^\mu P_2^\nu +P_2^\mu P_1^\nu}{E_1 E_2} &
  P^{\mu\nu}_{\rlii}  &= \frac{P_2^\mu P_2^\nu}{E_2^2} 
\end{align}
where $E_1$ and $E_2$ are the energies of the hadrons in the lepton c.m.\ 
frame. In terms of hadronic Mandelstam variables and the photon momentum 
these are
\begin{equation}
  E_1 = \frac{Q^2-T}{2Q}=\frac{e^{-y}}{2Q}\sqrt{S(Q^2+q_\perp^2)} ,\qquad
  E_2 = \frac{Q^2-U}{2Q}=\frac{e^{y}}{2Q}\sqrt{S(Q^2+q_\perp^2)}.
\end{equation}
Contracting the hadronic tensor with these projectors gives amplitudes 
$T_\alpha = P^{\mu\nu}_\alpha W_{\mu\nu}$, $\alpha \in \{ \rtl,\rli,\rlii,
\rliii \}$ which are just linear combinations of the helicity amplitudes in 
the different frames. These can be found by linear transformations
\begin{equation}
  \label{mirkes2}
  \left( \begin{array}{c} W_{\rtl}^{(\beta)} \\ W_\rl^{(\beta)} \\ 
           W_\Delta^{(\beta)} \\ W_{\Delta\Delta}^{(\beta)} 
         \end{array} 
  \right) = M_{(\beta)}  
  \left( \begin{array}{c} T_{\rtl} \\ T_\rl1 \\ T_\rl2 \\ T_\rl12
         \end{array} 
  \right),
\end{equation}
with $\beta \in \{ \rcs, \rgj \}$. The transformation matrices 
$M_{(\beta)}$ follow from simple geometry and are given in Appendix 7.1 for 
the CS and GJ frames.

\section{Leading twist results}

To make this paper self contained we give the leading order
results for the angular coefficients.
Due to twist-2 factorization the hadronic tensor is given by a convolution
of a perturbatively calculable partonic tensor and two twist-2 distribution 
functions:
\begin{equation}
  \label{tw2fac}
  W_{\mu\nu} = g^2 e_q^2 \sum_{a,b}\int \frac{\dd \xi_1}{\xi_1} 
  \int \frac{\dd \xi_2}{\xi_2}
  H_{\mu\nu}^{a+b}(\xi_1,\xi_2) f_{a/A}(\xi_1) f_{b/h}(\xi_2) 
  2\pi \delta(s+t+u-Q^2).
\end{equation}
The sum runs over partons $a$ and $b$ in the nucleus $A$ and the single hadron
$h$ respectively and $\xi_1$, $\xi_2$ are their momentum fractions. We denote 
the strong coupling by $g$ and $e_q$ is the fractional electric  charge of the
quark flavour coupling to the virtual photon.
The $f(\xi)$ correspond to twist-2 parton distribution functions
defined in the usual way as
\begin{equation}
  \label{pardis}
  \begin{split}
  f_{q/h} (\xi) &= \frac{1}{P^+}
  \int \frac{\dd y^-}{2\pi} P^+ e^{i \xi P^+ y^-} \,
  \frac{1}{2}\bra{P} \bar{q}(0) \gamma^+ {\cal P}q(y^-) \ket{P}
   \\
  f_{g/h}(\xi) &= \frac{1}{\xi P^+} \int \frac {\dd y^-}{2\pi} 
  e^{i \xi P^+ y^-} \bra{P} F_a^{+\nu}(0) {\cal P} 
  F_{a \, \nu }^{\hspace{1.5 ex} +} (y^-) \ket{P} 
  \end{split}
\end{equation}
for quarks and gluons, respectively in a hadron moving with 
momentum $P^+$ along the light cone \cite{coll89}. $q$ and $F$ denote quark 
fields and gluon field strength tensors and
${\cal P}$ is the path-ordered exponential 
\begin{equation}
  \label{parallel}
  {\cal P} = P \exp \left\{ ig \int_{y^-}^0 \dd x^- A^+(x^-)\right\}
\end{equation}
which makes the definition gauge invariant.
In the following we will not write this path-ordered exponential
explicitly, but it is assumed to be inserted between all factors in a 
product of operators referring to different space-time points. In the 
$A^+ = 0$ gauge this term will disappear but we will always use covariant 
gauge. Note that we also suppressed the dependence
of the distribution function on the factorization scale $\mu^2$.

The partonic tensor $H_{\mu \nu}^{a+b}$ represents the hard part of 
the process. 
We have introduced partonic Mandelstam variables by
\begin{eqnarray}
  \label{mandpar}
  s &=& (\xi_1 P_1 + \xi_2 P_2)^2 = \xi_1\xi_2 S \nn \\
  t &=& (\xi_1 P_1 -q)^2 = \xi_1 (T-Q^2) + Q^2 \\ 
  u &=& (\xi_2 P_2 -q)^2 = \xi_2 (U-Q^2) + Q^2. \nn
\end{eqnarray}
The $\delta$-function in eq.~(\ref{tw2fac}) stems from the on-shell condition 
for the unobserved emitted parton with momentum $l$. It can be used for trivial
integration of $\xi_2$ via
\begin{equation}
  \label{delta}
  \delta(l^2)=\delta(s+t+u-Q^2) = \frac{1}{\xi_1 S + U -Q^2} 
  \delta\left(\xi_2 + \frac{Q^2 + \xi_1(T-Q^2)}{\xi_1 S + U - Q^2}\right).
\end{equation}

\begin{figure}[t]
  \label{twist2fig}
  \epsfig{file=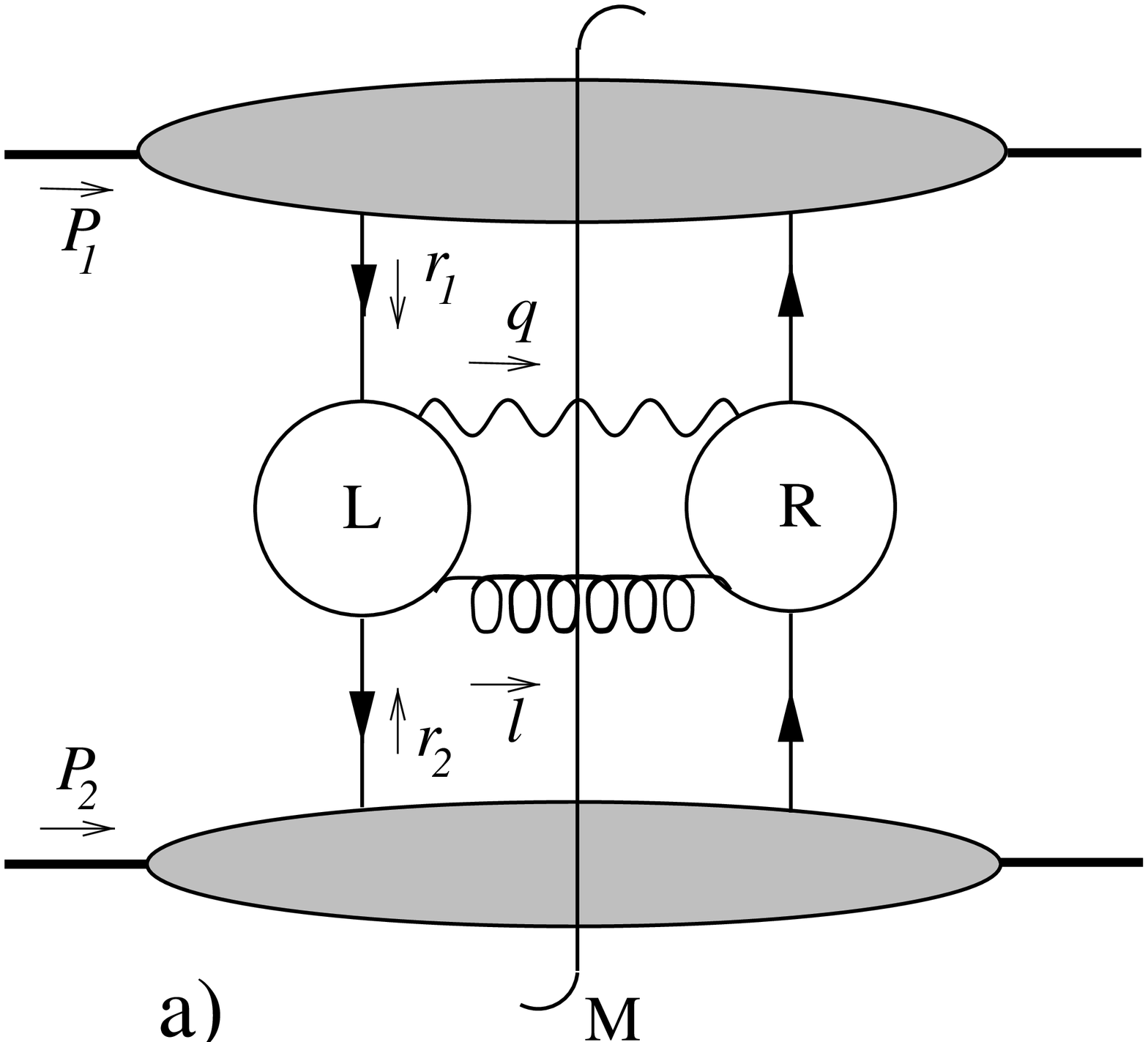,width=5.5cm}
  \epsfig{file=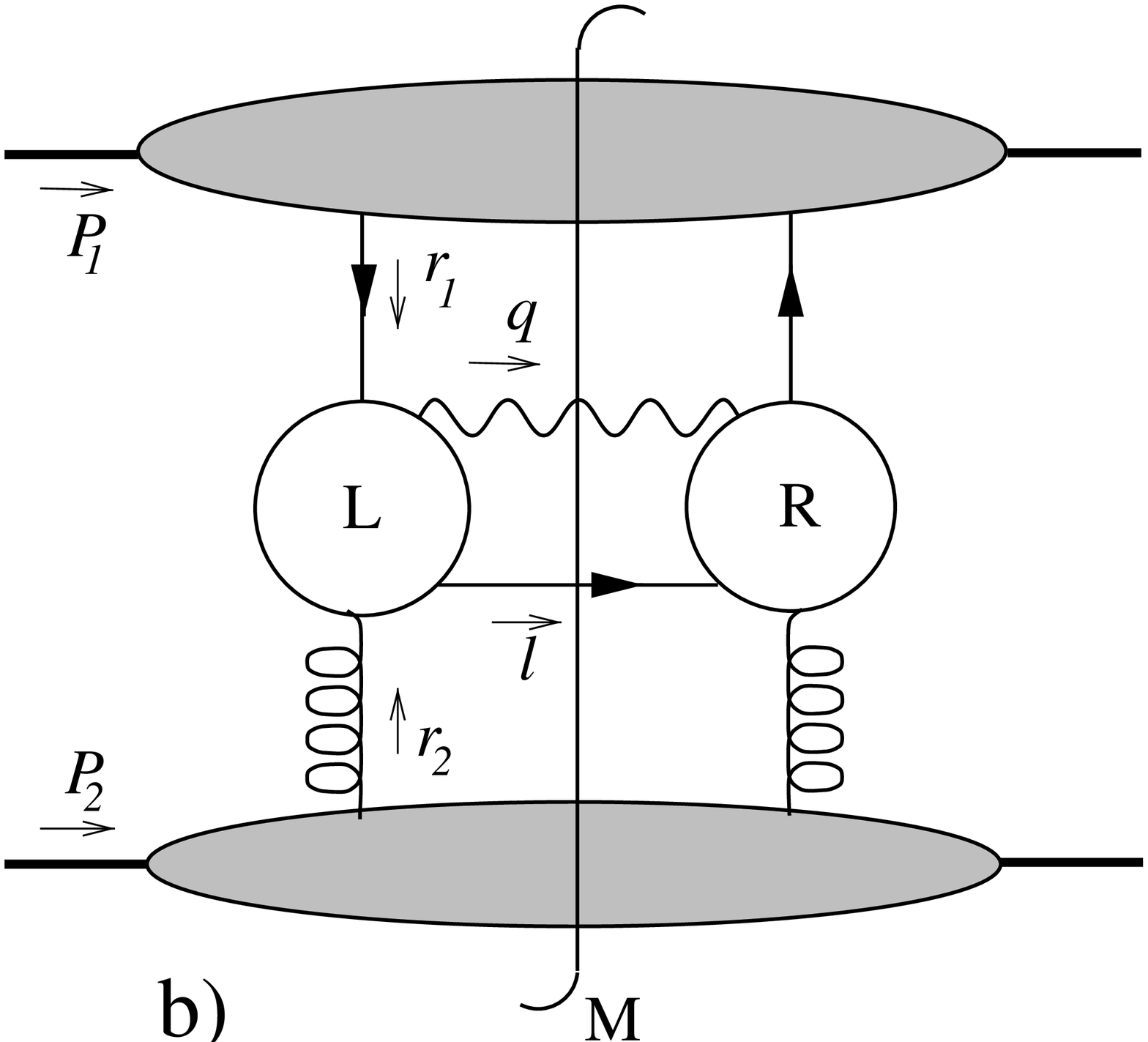,width=5.5cm}
  \epsfig{file=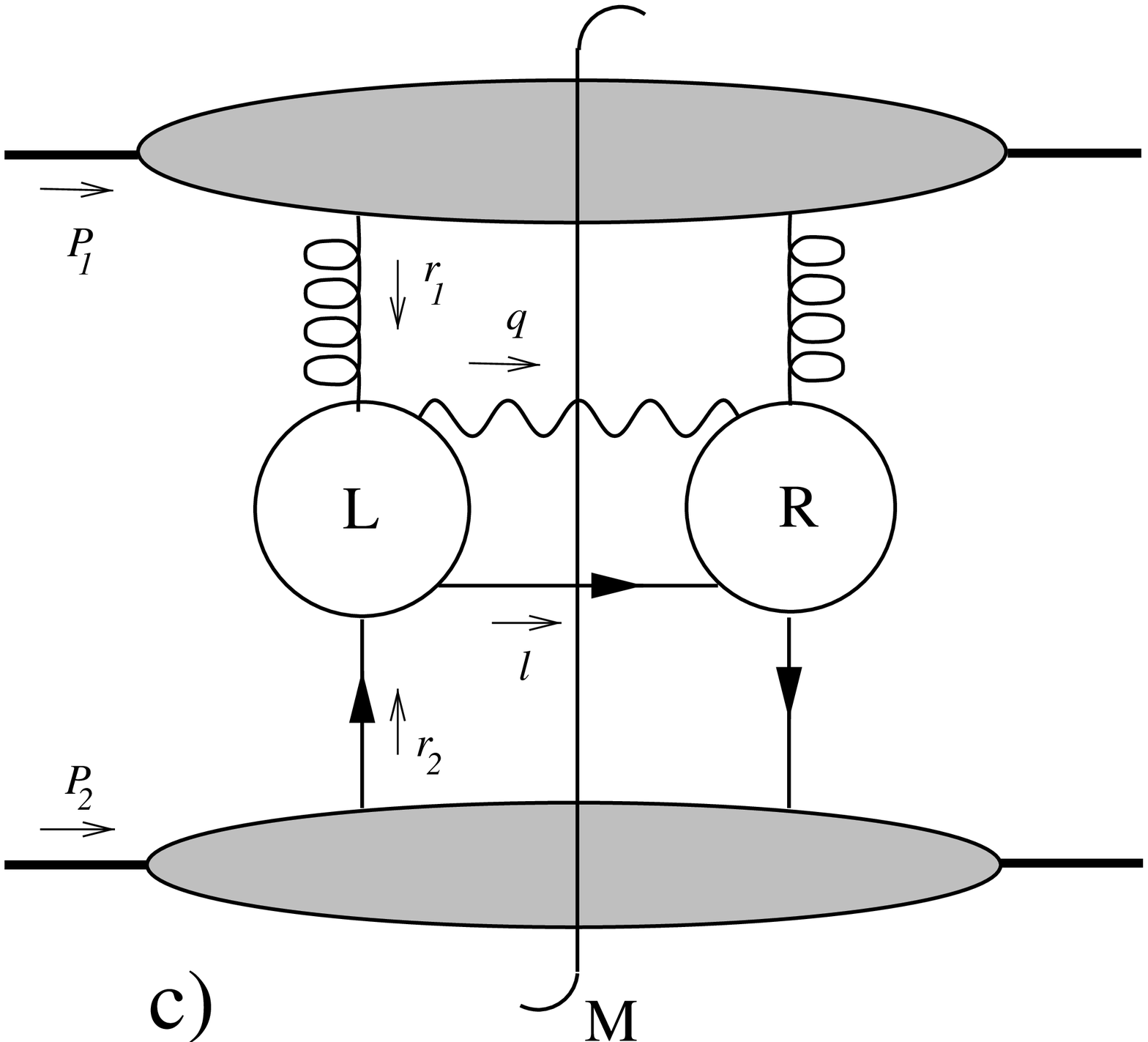,width=5.5cm}
  \caption{Twist-2 contributions to order $\alpha_s$: annihilation (a), 
    and Compton processes (b and c). The blobs L and R stand for 
    all possible tree-level diagrams.}
\end{figure}

We can decompose the partonic tensor in a way completely analogous to the 
decomposition of the hadronic tensor $W_{\mu\nu}$ in terms of helicity 
amplitudes. Therefore we define the four projections $H_\rtl^{a+b}, 
H_\rl^{a+b}, H_{\Delta}^{a+b}, H_{\Delta\Delta}^{a+b}$ which are obtained in 
the same way as in eqs.~(\ref{helicity},\ref{helicity2}).
This allows us to write hadronic tensor and helicity amplitudes as
\begin{equation}
  \label{t2result}
  W_\alpha = 8\pi^2 \alpha_s e_q^2 \sum_{a,b}\int\limits_B^1 
  \frac{\dd \xi_1}{\xi_1} \frac{1}{\xi_1 (Q^2-T) -Q^2}
  f_{a/A}(\xi_1) f_{b/h}(\bar{\xi}_2) H_\alpha^{a+b}(\xi_1, \bar{\xi}_2).
\end{equation}
with  $B = \frac{-U}{S+T-Q^2}$, $\bar{\xi}_2 = -\frac{Q^2 + \xi_1(T-Q^2)}{
\xi_1 S + U - Q^2}$, $\alpha_s=g^2/4\pi$ and $\alpha \in \{ \mu\nu, TL, L, 
\Delta, \Delta\Delta \}$.
There are three kinds of partonic processes contributing to the leading-twist 
process at large transverse momentum: quark-antiquark annihilation 
($\bar q+ q$) and Compton processes with a gluon from the single hadron 
($q+g$) or the nucleus ($g+q$). They are shown in Fig.~\ref{twist2fig}.

The cross section for the leading twist DY pair production 
at large transverse momentum is well-known \cite{mir92,Cleymans:1979ip}. 
For completeness we give the results in the Collins-Soper frame in Appendix 
7.2. The relation $H_{\Delta\Delta} = \frac{1}{2} H_\rl$ is precisely the 
so-called Lam-Tung relation \cite{lt80}. It can be expressed in terms of 
angular coefficients as $A_0=A_2$ or $2\nu-(1-\lambda)=0$. This relation holds
for all partonic subprocesses and is not specific for the Collins-Soper frame 
but holds in any photon rest frame as long as the $\mathbf Z$-axis lies in the 
reaction plane.
Corrections arise at next-to-leading order (NLO) \cite{mir92}.
Note that $H_{\Delta}^{a+b}$ picks up an additional sign compared with 
$H_{\Delta}^{b+a}$ in contrast to the other helicity amplitudes. 
This is due to the inversion of 
the orientation of the angle $\gamma_\rcs$ if we interchange the roles of 
$P_1$ and $P_2$. For symmetric collisions, e.\,g.\ $p+p$, this leads to the 
leading-twist result $W_\Delta=0$.

\section{Twist-4 Factorisation} 

\subsection{Soft-hard and double-hard scattering}

\begin{figure}[t] 
  \epsfig{file=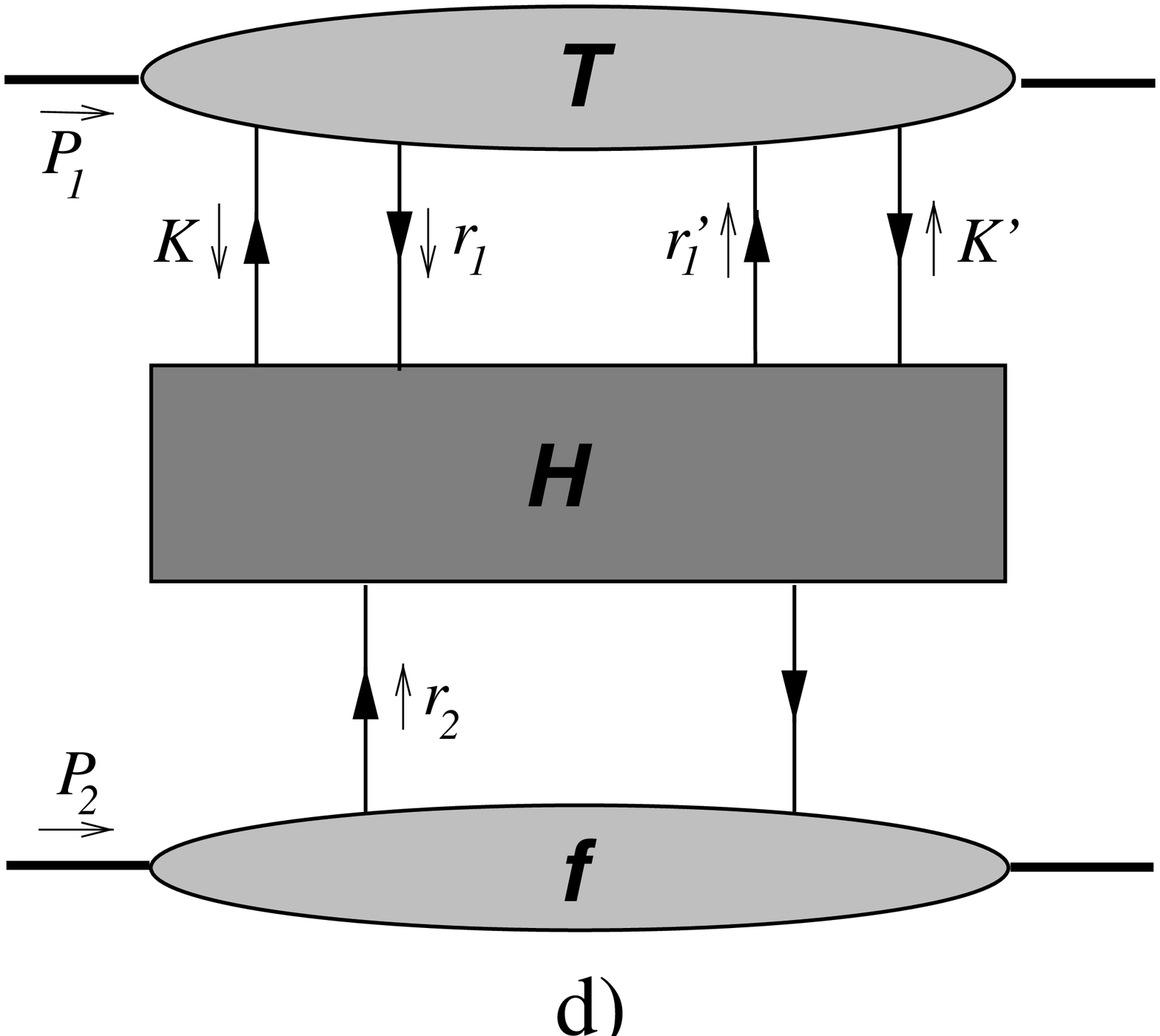,width=5cm}  
  \epsfig{file=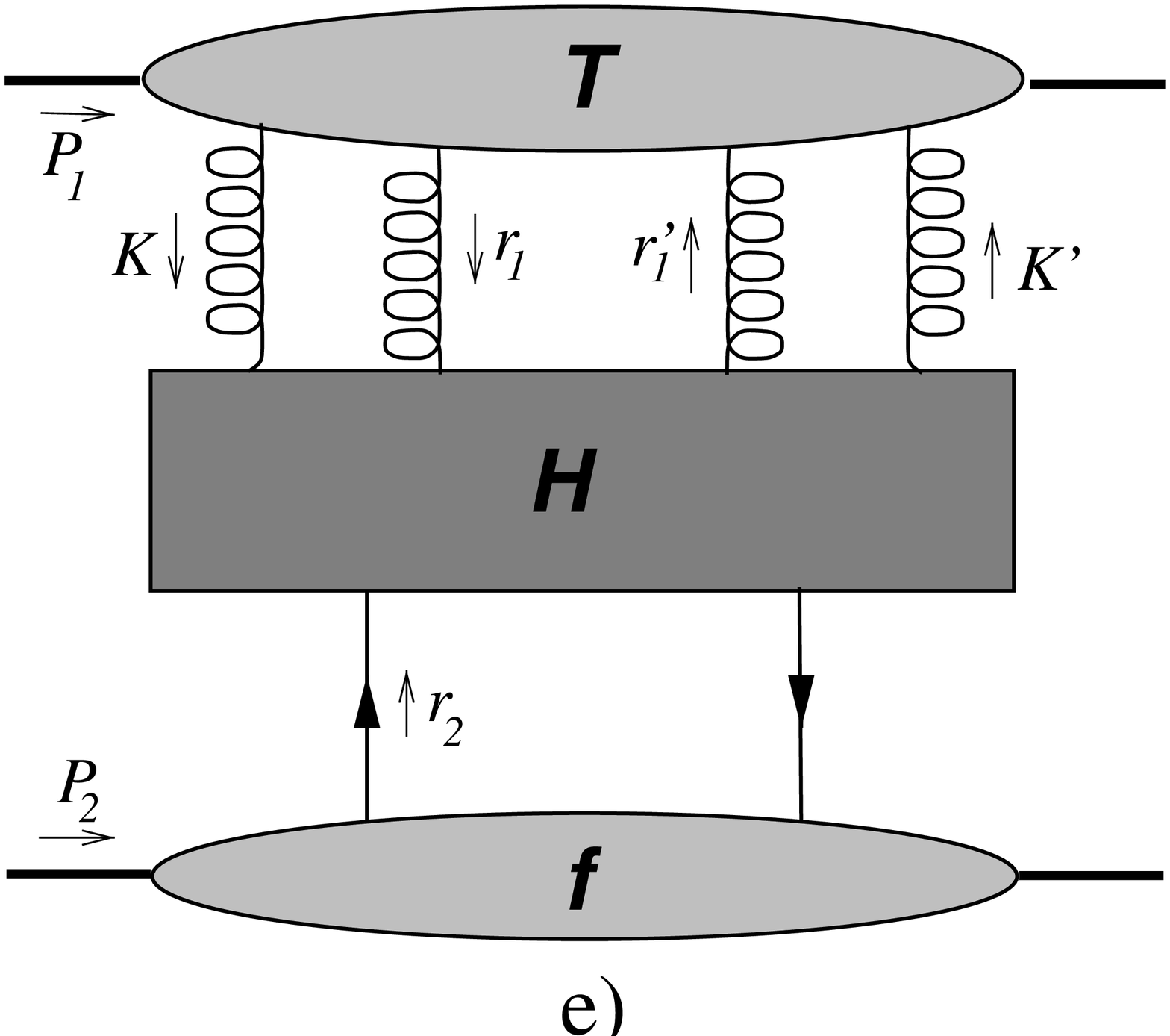,width=5cm}
  \caption{Collection of the schematic forward processes we want to consider 
    for double scattering. Factorization allows separate treatment of twist-4 
    and twist-2 matrix elements ($T$ and $f$) and hard part ($H$).}
  \label{overview}
\end{figure}

\begin{figure}[b]
  \epsfig{file=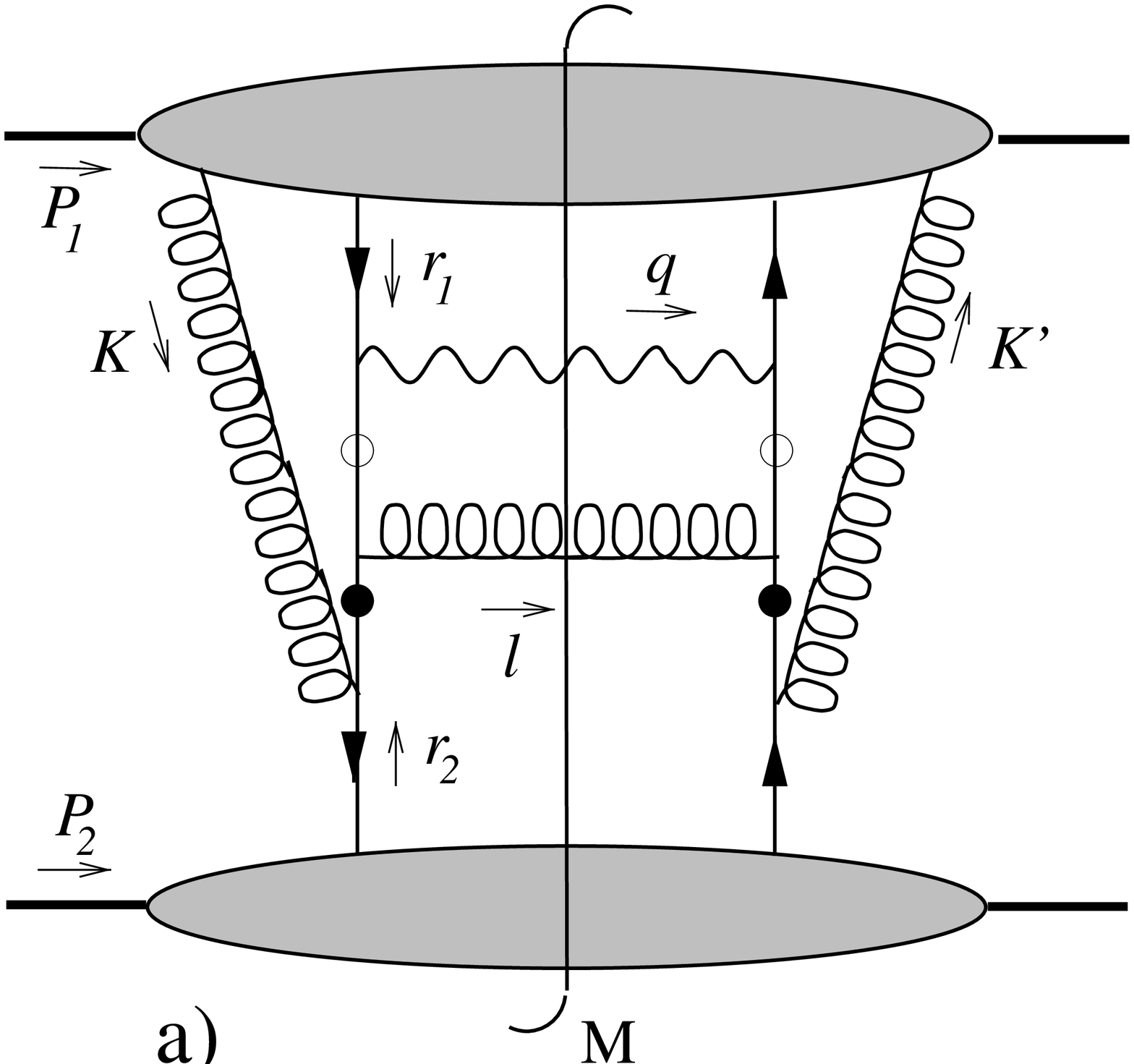,width=5cm}
  \epsfig{file=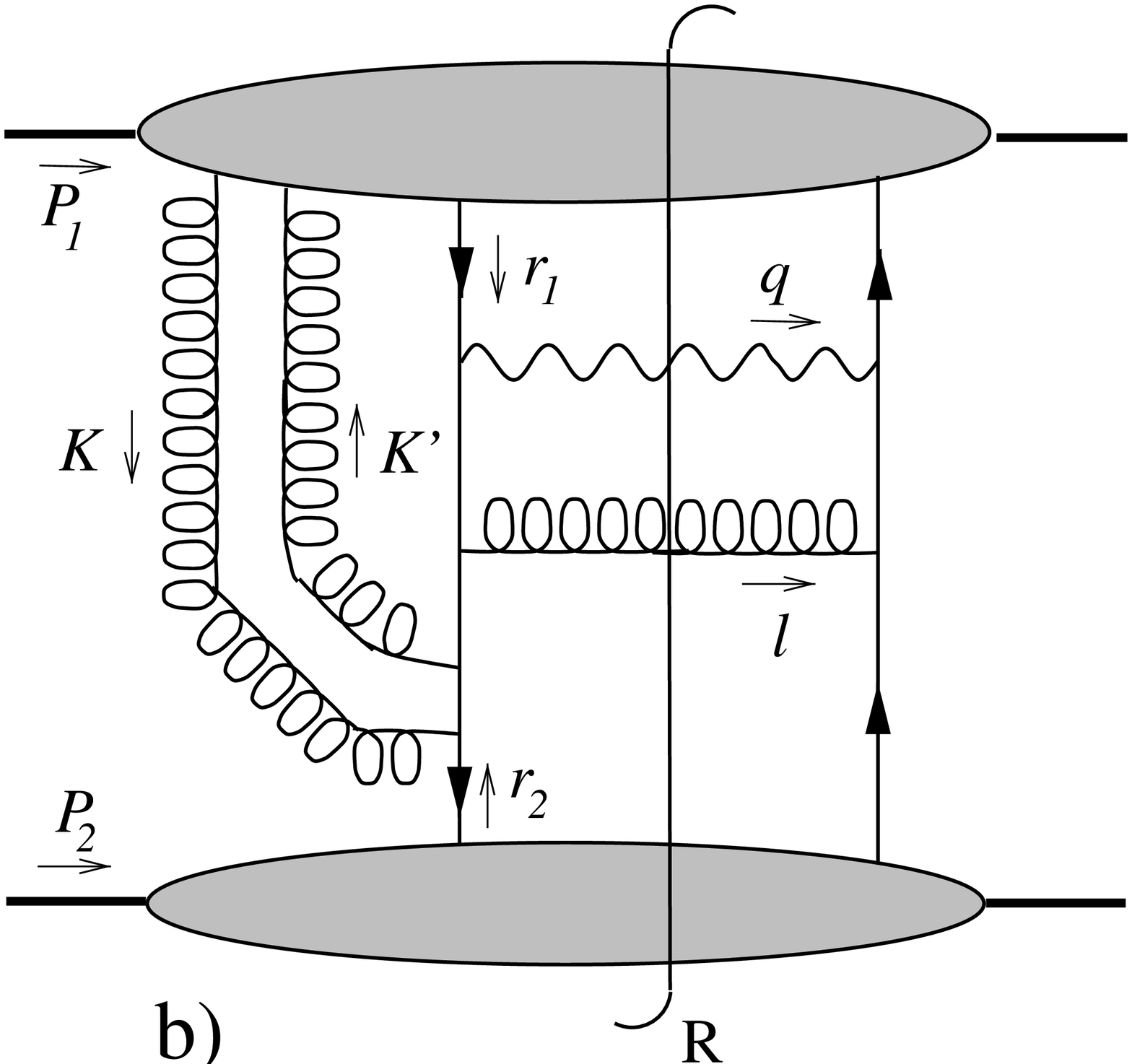,width=5cm}  
  \epsfig{file=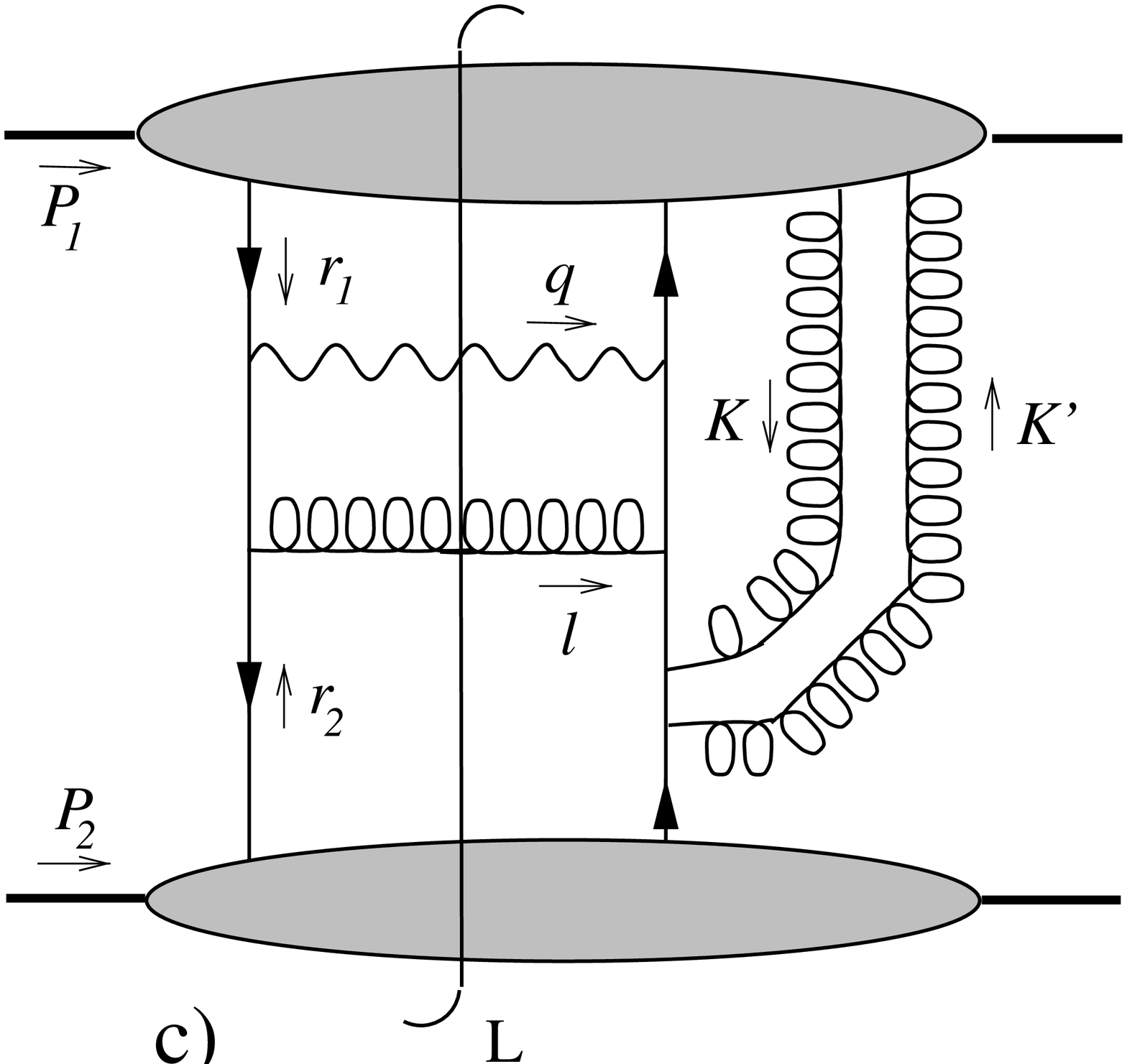,width=5cm}
  \caption{Examples for double-scattering diagrams with the partonic 
    process $qg+\bar q$ cut in the middle (a), right (b) 
    and left (c). Circles in Fig. (a) indicate poles from propagators 
    which lead to finite gluon momentum (open) or soft gluon momentum 
    (filled).}
  \label{exmpl}
\end{figure}

For the calculation of the double scattering contributions we use the 
approach of Luo, Qiu and Sterman which is presented e.\,g.\ in 
\cite{Luo94,Guo98}. We consider it worthwile to repeat some aspects here.
In Fig.~\ref{overview} we illustrate the forward diagrams that need to be 
calculated for double scattering. The box denotes all possible partonic 
subprocesses. We have to introduce a cut to obtain the corresponding hadronic 
tensors. This is not only possible in the middle but we also have to include 
interference graphs which appear if the forward diagram is cut in an 
asymmetrical way. In Fig.~\ref{exmpl} we give concrete examples for graphs 
which involve a quark-gluon matrix element in the nucleus and an antiquark 
coming from the single hadron. The first diagram is cut in the middle (M), 
the others are cut left (L) and right (R).
\begin{figure}[t]
  \epsfig{file=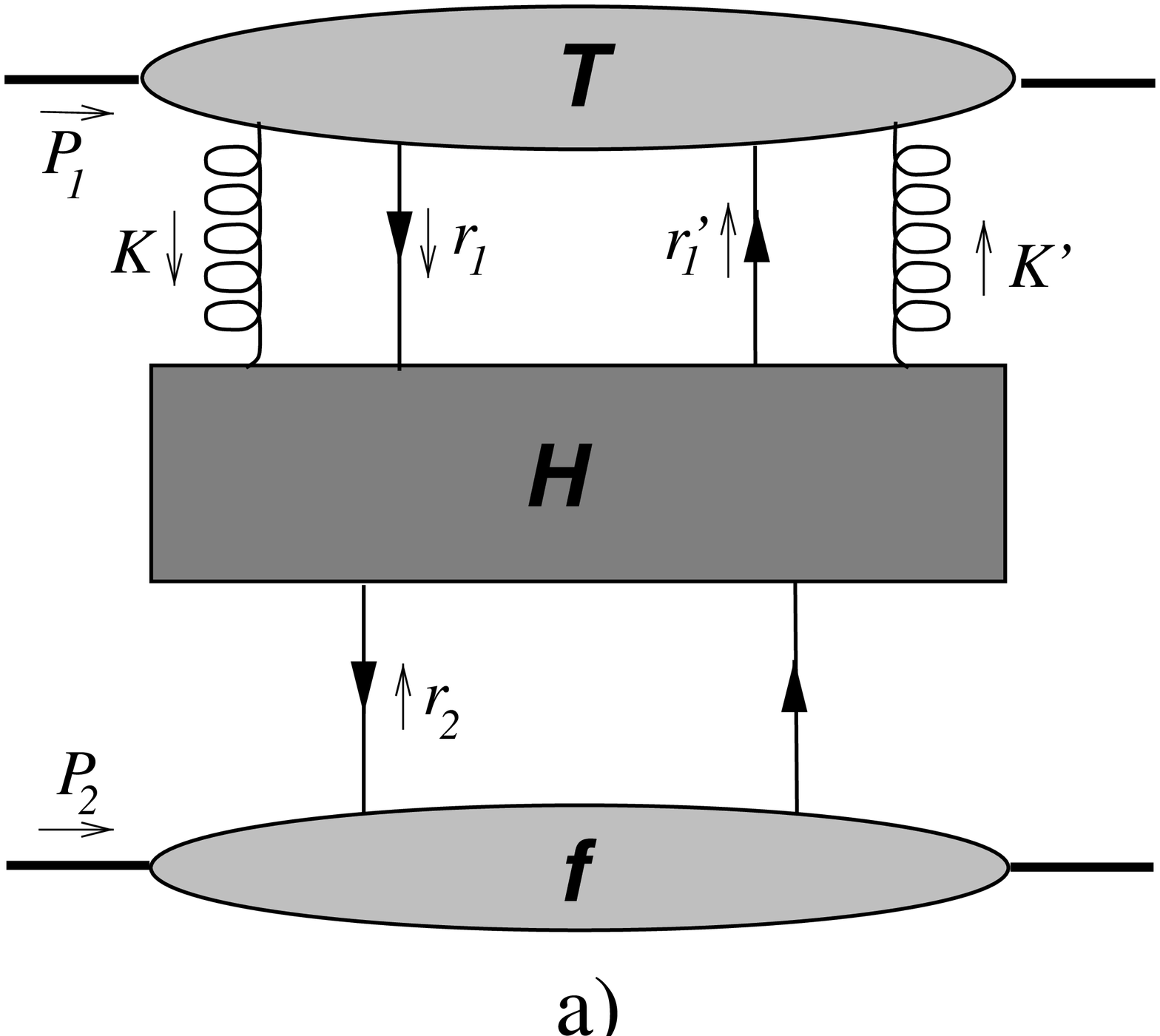,width=5cm}
  \epsfig{file=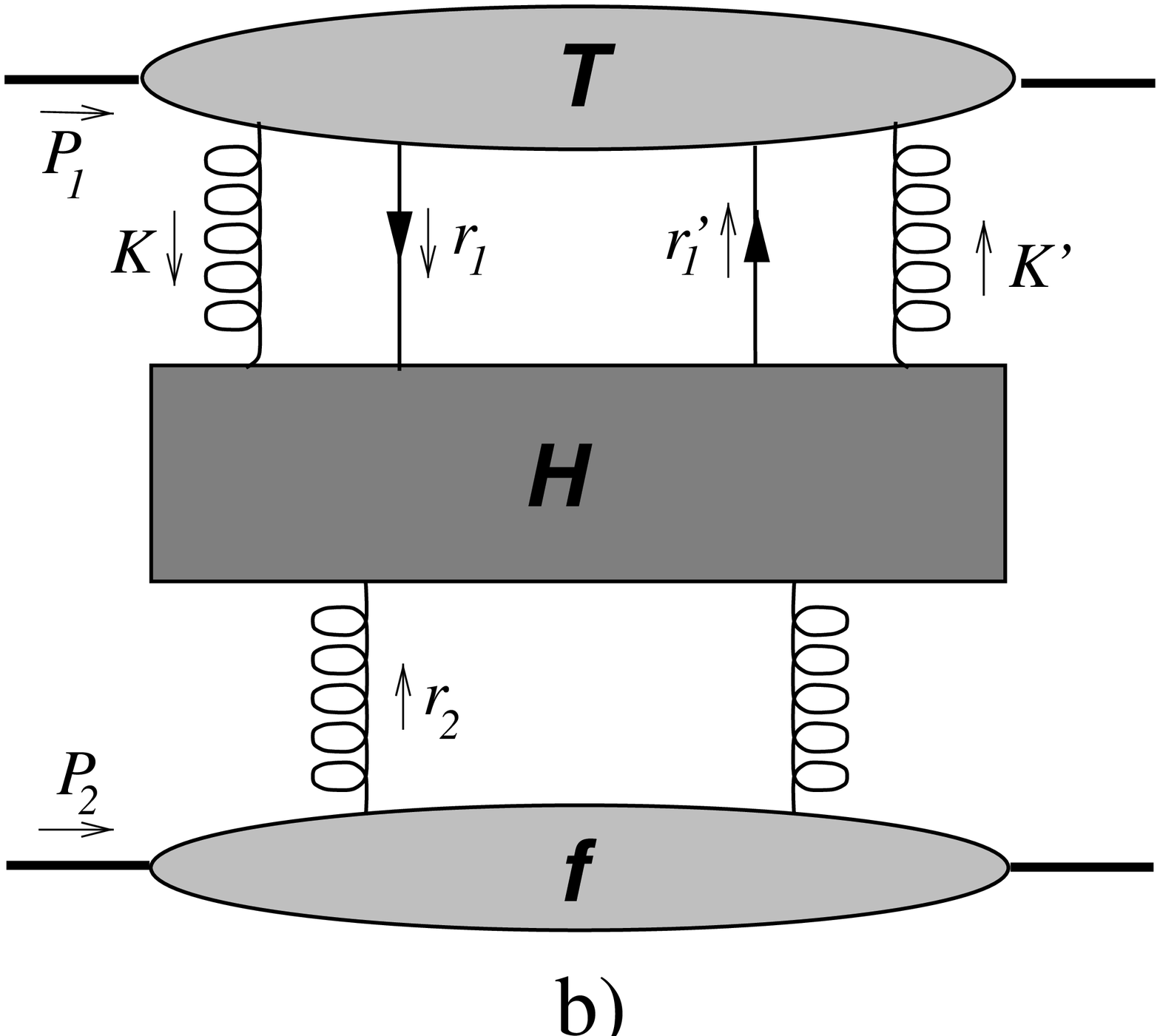,width=5cm}  
  \epsfig{file=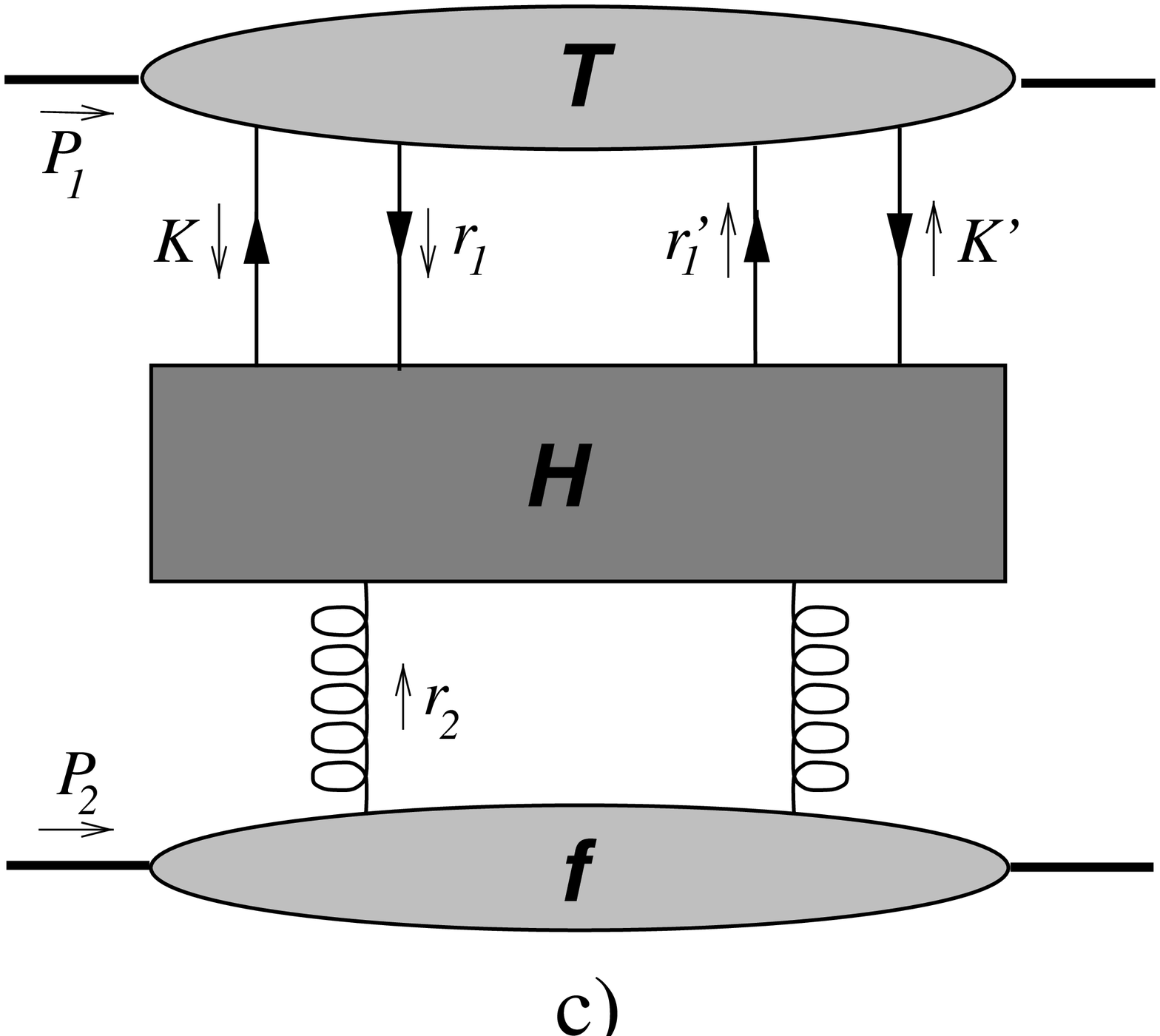,width=5cm}
\end{figure}

The starting point for the calculation of the hadronic tensor is the forward
scattering amplitude in fourth order of the strong coupling 
constant $g$:
\begin{equation}
  \label{forwtensor}
  \begin{split}
  T_{\mu\nu} &= \frac{g^4 e_q^2}{4!}\int \rd^4 x \rd^4 x_1 \rd^4 x_2 \rd^4 x_3 
  \rd x_4 e^{iqx} 
  \bra{P_1 P_2} T\left[ J_\rho (x_3) J_\alpha (x_1) j_\mu(0) j_\nu(x) 
  \right. \\  
  & \left. J_\beta (x_2) J_\sigma(x_4) 
  A^\rho(x_3) A^\alpha(x_1) A^\beta(x_2) A^\sigma(x_4) + \ldots \right]
  \ket{P_1 P_2}
  \end{split}
\end{equation}
where $j$ ($J$) are electromagnetic (color) quark
currents and $A$ denotes gluon fields. $T$ is the time-ordering operator. 
Dots indicate further contributions involving more gluon fields with gluon 
self interaction and less quark currents such that the power of $g$ remains
four. By application of Wick's theorem we obtain all diagrams which 
potentially contribute to the forward process. 

We classify diagrams by their partonic subprocess $ab+c$ where $a$ and $b$ are
partons from the nucleus and $c$ comes from the single hadron. Our notation 
for the parton momenta is shown in Fig.~\ref{overview}. 
We concentrate first on $qg+\bar q$ which can be realized e.g.\ by the 
diagrams of Fig.~\ref{exmpl}.
As a first step towards factorization we perform Fierz transformations 
in color and Dirac space for the fermion lines entering and leaving the blobs: 
\begin{equation}
\label{fierz}
  q^i_\alpha(0) \bar{q}^j_\beta (z) = -\frac{1}{4} \left(\gamma_\mu
  \right)_{\alpha\beta} \bar{q}^{j'}(z) \gamma^\mu q^{i'}(0) \left( 
  \frac{1}{N_c} \delta_{j'i'} \delta_{i j} +2 t^A_{j'i'} t^A_{ij} \right) + 
  \ldots.
\end{equation}
Here the $t^A$ are the $SU(N_c)$ Gell-Mann matrices, and $N_c$ is the number 
of colors. Latin characters are color indices, greek one are Dirac indices.
We keep only Dirac structures which give leading contributions to 
the unpolarized Drell-Yan process. After Fourier transformations of the fields
we end up with
\begin{multline}
  \label{forw2}
  T_{\mu\nu}^{qg+\bar q}  =  \int \frac{\rd^4 l}{(2\pi)^4} 
  \frac{\rd^4 r_1}{(2\pi)^4} \frac{\rd^4 r_2}{(2\pi)^4} 
  \frac{\rd^4 K} {(2\pi)^4} \frac{\rd^4 K'}{(2\pi)^4}
  \int \rd^4 y \rd^4 z_1 \rd^4 z_3 \rd^4 z_4 \\    
   (2\pi)^4 \delta^{(4)}(r_1+r_2+K-l-q)
  e^{ir_1 z_1} e^{-ir_2 y} e^{iK z_3}  e^{-i{K'} z_4} 
  \frac{1}{2}\bra{P_1} \bar{q}(0) \gamma^\eta q(z_1) A^\rho(z_3) 
  A^\sigma(z_4) \ket{P_1} \\
   \big(-\frac{1}{2}\big) \bra{P_2} \bar{q}(0) \gamma^\tau q(y) \ket{P_2} 
  e_q^2 g^4 \frac{1}{N_c^2}\frac{\delta^{RS}}{N_c^2-1} \sum_d
  S(d)^{RS}_{\nu\sigma\tau\rho\mu\eta}.  \qquad
\end{multline} 
We used the fact that forward matrix elements only depend on the relative 
coordinates. So, after shifting positions in each matrix element separately, 
we introduced a new set of space-time coordinates ($y$,$z_1$,$z_3$,$z_4$). 
$l$ is the momentum of the unobserved radiated parton.
The index $d$ runs over all diagrams of the class $qg+\bar q$ and 
$S(d)^{RS}_{\rho\mu\epsilon\nu\sigma\tau}$ contains all perturbative 
propagators and vertices of the hard part of diagram $d$, with $\gamma_\eta$ 
($\gamma_\tau$) inserted for the upper (lower) blob,  and traces over color
 and Dirac indices.
For the diagram of Fig.~\ref{exmpl}(a) we have e.\,g.\
\begin{equation}
  \label{hardex}
  \begin{split} 
  S_{\nu\sigma\tau\rho\mu\eta}^{RS} = \frac{1}{4}\tr \left[ (-i\gamma_\nu) 
  \frac{i(\not r_1 + \not K -\not K' -\not q)}{(r_1+K-K'-q)^2 +i\epsilon}
  (-i\gamma_\beta t^B) \frac{i(-\not r_2 -\not K')}{(r_2+K')^2+i\epsilon}
  (-i\gamma_\sigma t^S) \right. \\ \left.
  \gamma_\tau (-i\gamma_\rho t^R) 
  \frac{i(-\not r_2 -\not K)}{(r_2+K)^2+i\epsilon} (-i\gamma_\alpha t^A)
  \frac{i(\not r_1 -\not q)}{(r_1-q)^2 +i \epsilon} (-i \gamma_\mu)
  \gamma_\eta  \right] 
  \frac{i (-g^{\alpha\beta})\delta^{AB}}{l^2+ i\epsilon}.
  \end{split}
\end{equation}
$R$ and $S$ are color indices of gluon fields.
Note that we have coupled quark and gluon field operators in the nuclear 
matrix elements separately to color singlets. We have omitted a term 
proportional to 
$(d^{RES}+if^{RES}) \bra{P_1} \bar{q} \gamma^\eta t^E q A^{R\rho} A^{S\sigma} 
\ket{P_1}$ 
which would induce color forces between quark and gluon in the
nucleus. This would not allow to factorize quarks and gluons in color space to 
test the size of the nucleus. Such contributions are not expected to show 
nuclear enhancement, i.e.\ scaling by $A^{4/3}$. 

Next we perform the light cone expansion, i.e. we use that the partons move 
almost collinear to the hadron in the infinite momentum frame.
We introduce the parton momentum fractions by setting
$r_1 = \xi_1 P_1$, $r_2 = \xi_2 P_2$, $K = x P_1 + K_\perp$ and 
$K' = x' P_1 + K_\perp'$. $r'_1 = r_1+ K -K'$ is fixed by momentum 
conservation in the upper forward matrix element.
Note that we have to allow one parton (called $b$ here) from the nucleus to 
be soft with essentially zero longitudinal momentum. Therefore we must take 
into account transverse momenta $K_\perp$ and $K'_\perp$ to get a contribution
in higher orders of the collinear expansion as we will see later.
Now one can integrate the trivial degrees of freedom in 
eq.~(\ref{forw2}).
To complete the factorization we perform a Sudakov decomposition of the 
$\gamma$-matrices in the forward matrix elements on the light cone and retain 
only the leading terms $\gamma^\eta = \frac{P_1^\eta}{P_1^+}
\gamma^+$, $\gamma^\tau = \frac{P_2^\tau}{P_2^-}\gamma^-$.
Then we can introduce twist-2 distribution functions $f_{c/h}(\xi_2)$, given 
in eq.~(\ref{pardis}), for the lower blob. 
Finally we cut the forward diagram. This sets the radiated unobserved parton
with momentum $l$ on the mass shell and in the hard part all expressions right
to the cut receive a complex conjugation. 

Hence after these first steps we have factorized the hadronic tensor in a 
straightforward way. We get
\begin{equation}
\label{hadr1}
  \begin{split}
  W^{ab+c}_{\mu\nu} = (4\pi\alpha_s)^2 e_q^2 \int \frac{\rd\xi_1}{\xi_1} 
  \frac{\rd\xi_2}{\xi_2} \rd x P_1^+ \rd^2K_\perp \rd x' P_1^+ 
  \rd^2 K'_\perp f_{c/h} (\xi_2) (2\pi) \delta(l^2)  \\
  \bar{T}_{ab}(\xi_1,x,x',K_\perp,K'_\perp)
  \bar{H}_{\mu\nu}^{ab+c}(\xi_1,\xi_2,x,x',K_\perp,K'_\perp).
  \end{split}
\end{equation}
We have derived eq.~\ref{hadr1} for our examples of Fig.~\ref{exmpl}, but this equation holds for all processes shown in Fig.\ref{overview}.
We have separated the soft part involving the twist-2 distribution function 
$f_{c/h}$ and the nuclear matrix element $\bar T_{ab}$ from the hard part
$\bar H^{ab+c}_{\mu\nu}$.

For our example $qg+\bar q$ we are left with a quark-gluon matrix element in 
the nucleus
\begin{equation}
  \label{rawqgsoft}
  \begin{split}
  \bar{T}_{qg}(\xi_1,x,x',K_\perp,K'_\perp) = \int \frac{\rd 
  z_1^-}{2\pi} \frac{\rd z_3^- \rd^2 z_3^\perp}{(2\pi)^3} 
  \frac{\rd z_4^- \rd^2 z_4^\perp}{(2\pi)^3} e^{ir_1^+ z_1^-} e^{iK^+ z_3^-}
  e^{-iK_\perp z_3^\perp} e^{-i{K'}^+ z_4^-} e^{iK'_\perp z_4^\perp} \\
  \frac{1}{2} \bra{P_1} \bar{q}(0) \gamma^+  q(z_1^-) A^\rho(z_3^-,z_3^\perp) 
  A^\sigma(z_4^-,z_4^\perp) \ket{P_1}.
  \end{split}
\end{equation}
For other classes of processes we also need the quark-quark and gluon-gluon 
correlators $\bar T_{gg}$ and $\bar T_{qq}$ which are given by 
eq.~(\ref{rawqgsoft}) with the second line replaced by
\begin{equation}
\label{rawsoft}
  \begin{split}
    \frac{1}{\xi_1 P_1^+} \bra{P_1} F^{\omega +} (0) F^+_{\>\omega}(z_1^-) 
    A^\rho(z_1^-,z_1^\perp) A^\sigma(z_4^-,z_4^\perp) \ket{P_1} 
    \quad\text{for~}\bar T_{gg} \\
    \frac{1}{2} \bra{P_1} \bar{q}(0) \gamma^+  q(z_1^-) 
    \bar{q}(z_3^-,z_3^\perp) \gamma^\kappa  q(z_4^-,z_4^\perp)  
    \ket{P_1} \quad\text{for~}\bar T_{qq}.
  \end{split}
\end{equation}
Again here the field operators of each single parton form color singlets 
separately. We have arranged the first two field operators referring to the
nuclear parton $a$ in the way they appear in the corresponding twist-2 
structure function while the remaining two field operators of parton $b$ are 
still in their original form.

The hard part for the $qg+\bar{q}$-process is 
\begin{equation}
  \label{rawqghard}
  \bar{H}^{qg+\bar q}_{\mu\nu}(\xi_1,\xi_2,x,x',K_\perp,K'_\perp) = 
  \frac{1}{N_c^2}\frac{1}{N_c^2-1}
  (-g^{\alpha \beta}) r_1^\eta r_2^\tau \frac{1}{4} \sum_d S(d)^{RR}_{
  \nu\sigma\tau\rho\mu\eta}\Big\vert_{\rm cut}.
\end{equation}
Hard parts for other processes are obtained analogously.
The delta function in eq.~(\ref{hadr1}) can be rewritten for the cases in which
the diagrams are cut in the middle (M), right (R) or left (L) as
\begin{eqnarray}
  \label{deltas}
  {\rm (M):}\qquad \delta(l^2) &=& (\xi_2 S+T-Q^2)^{-1} 
     \delta(\xi_1+x- \tilde \xi_b) \nn \\
  {\rm (R):}\qquad \delta(l^2) &=& (\xi_2 S+T-Q^2)^{-1} 
     \delta(\xi_1+x-x'- \xi_c) \\
  {\rm (L):}\qquad \delta(l^2) &=& (\xi_2 S+T-Q^2)^{-1} 
     \delta(\xi_1- \xi_c) \nn
\end{eqnarray}
where
\begin{equation}
  \label{xis}
  \tilde\xi_b = -\frac{\xi_2(U-Q^2)+Q^2-2q K_\perp +K_\perp^2}{\xi_2 S+T-Q^2}
  \qquad\text{ and }\qquad \tilde\xi_c= \tilde\xi_b
  \big\vert_{K_\perp=0}.
\end{equation}

As pointed out in various papers \cite{Luo94b,Guo98} there are in principle 
two contributions to double scattering: 
There can be two hard scattering reactions of the parton $c$ from the single 
hadron or the outgoing unobserved parton on one side and partons $a$ and $b$ 
from the nucleus on the other side. This is called double-hard scattering. 
Furthermore there can occur processes where the parton $c$ from the single 
hadron or the outgoing parton picks up one soft nuclear parton $b$ in addition
to one hard scattering off the nuclear parton $a$. This is called soft-hard 
scattering.
These are not only qualitative notions but they can be rigorously derived from
the equations given above.  
One can understand them in terms of poles in the hard part of the scattering 
tensor. In eq.~\eqref{hadr1} we have four momentum integrations 
$\rd\xi_1\rd\xi_2\rd x\rd x'$ and four propagators. We will perform the
$\xi_1$-integration by making use of the delta function $\delta(l^2)$. The 
propagators provide poles
with respect to $x$ and $x'$. In general we expect two poles in $x$ and two 
poles in $x'$. Application of the residue theorem will fix the momenta 
and it is easy to check that two poles give large momenta $x=x_{\rm hard}=x'$,
respectively and the other ones give small momenta $x=x_{\rm soft}=x'$ with 
$x_{\rm hard}\gg x_{\rm soft}\ll 1$ as long as $q_\perp \sim Q$.

To be more precise the pole structure we expect for our example in 
Fig.~\ref{exmpl}(a) is
\begin{equation}
  \label{poles}
  W_{\mu\nu} \sim \int \rd x \frac{L(x)}{(x-x_{\rm soft}+i\epsilon)
  (x-x_{\rm hard}+i\epsilon)}
  \int \rd x' \frac{R(x')}{(x'-x_{\rm soft}-i\epsilon)(x'-x_{\rm hard}
  -i\epsilon)} ,
\end{equation}
where $L$ and $R$ stand for the numerators of the amplitudes left and right
of the cut depending on $x$ and $x'$. The residue theorem implies
\begin{equation}
  \label{poles2}
  W_{\mu\nu} \sim \left[ \frac{L(x=x_{\rm soft})}{x_{\rm soft}-x_{\rm hard}}
  -\frac{L(x=x_{\rm hard})}{x_{\rm soft}-x_{\rm hard}} \right]
  \left[ \frac{R(x'=x_{\rm soft})}{x_{\rm soft}-x_{\rm hard}}
  -\frac{R(x'=x_{\rm hard})}{x_{\rm soft}-x_{\rm hard}} \right] 
\end{equation}
Soft-hard and double-hard scattering correspond to the quadratic terms 
proportional to $L(x=x_{\rm soft})R(x'=x_{\rm soft})$ and $L(x=x_{\rm hard})
R(x'=x_{\rm hard})$ respectively.
We follow the arguments of reference \cite{Luo94b,Guo98} and neglect the 
remaining mixed terms of eq.~(\ref{poles2}) which are interference terms 
between soft and hard rescattering. 
This is permitted if $q_\perp$ is not too small. In the case 
$q_\perp^2 \ll Q^2$ however we obtain $x_{\rm soft} \approx x_{\rm hard}$.
 The interference will then be important and eventually spoil nuclear 
enhancement due to the negative signs in eq.~\eqref{poles2} \cite{Guo98}. 
Since the integrated Drell-Yan cross section does not exhibit nuclear 
enhancement, we even expect a suppression effect at low values of transverse 
momenta \cite{Guo98b}.

Most diagrams we naively get from Wick's theorem for double scattering show a 
different behaviour. For them the propagators do not provide four but 
only three or two poles ---  but at least one in each variable $x$ and $x'$. 
This implies that only soft-hard, only double-hard or neither of these 
contributions can occur. For example it is easy to understand from this 
analysis that only symmetrically (M) cut diagrams can contribute to 
double-hard 
scattering.
Consider the diagram of Fig.~\ref{exmpl}(b). The propagator on the right-hand
side has denominator $[(r_1'-q)^2-i\epsilon]$, from which $x$ and $x'$ 
dependences cancel out. On the left-hand side, the propagator at the bottom of
the diagram contributes a denominator $(r_2+K)^2+i\epsilon=\xi_2 S(x+
\frac{K_\perp^2}{\xi_2 S} +i\epsilon)$. This is a soft pole which fixes $x$ to 
$x_{\rm soft}=-\frac{K_\perp^2}{\xi_2 S}\ll 1$. The remaining two propagators 
on the left-hand side fix $x'$ to either a soft or a hard value. 
Therefore we have two contributions: (i) $x=x'$ is small. This
means that parton $b$ is soft --- in Fig.~\ref{exmpl}(b) this is the gluon 
from the nucleus. Together with the production of the virtual photon which is 
always a hard process, this constitutes a soft-hard double scattering process. 
(ii) $x$ is soft and $x'$ is hard. This is what we called a soft-hard 
interference because parton $b$ has different momenta on the left and the
right side of the diagram, i.e.\ momentum is transferred between partons $a$ 
and $b$. We omit this contribution as explained above.
We conclude that there is no double-hard scattering contribution from the 
diagram in Fig.~\ref{exmpl}(b) since $x$ cannot be hard. On the other hand the
diagram in Fig.~\ref{exmpl}(a) is an example in which all four poles occur, 
indicated by small circles.

\begin{figure}[h]
   \epsfig{file=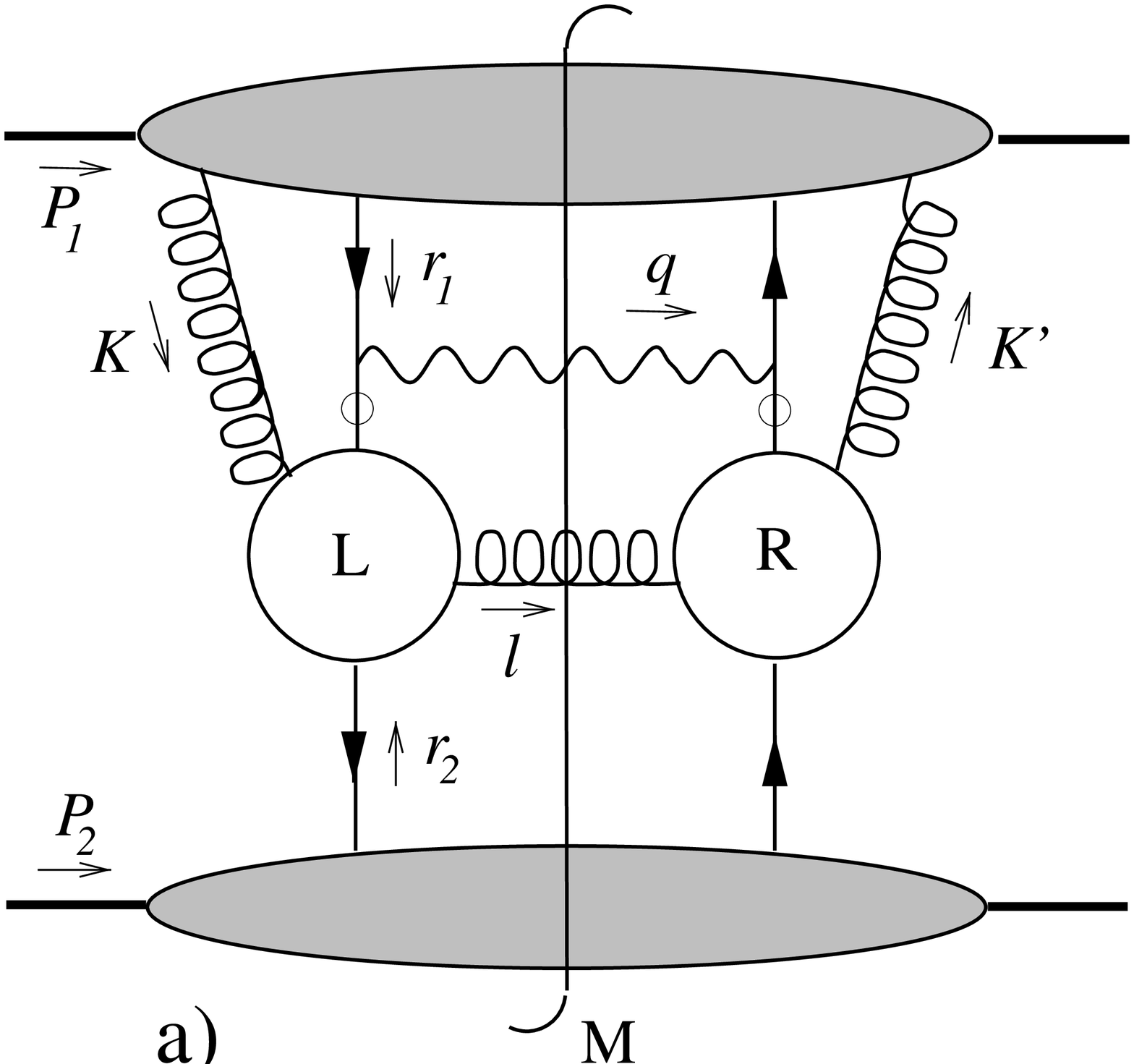,width=4cm}
   \epsfig{file=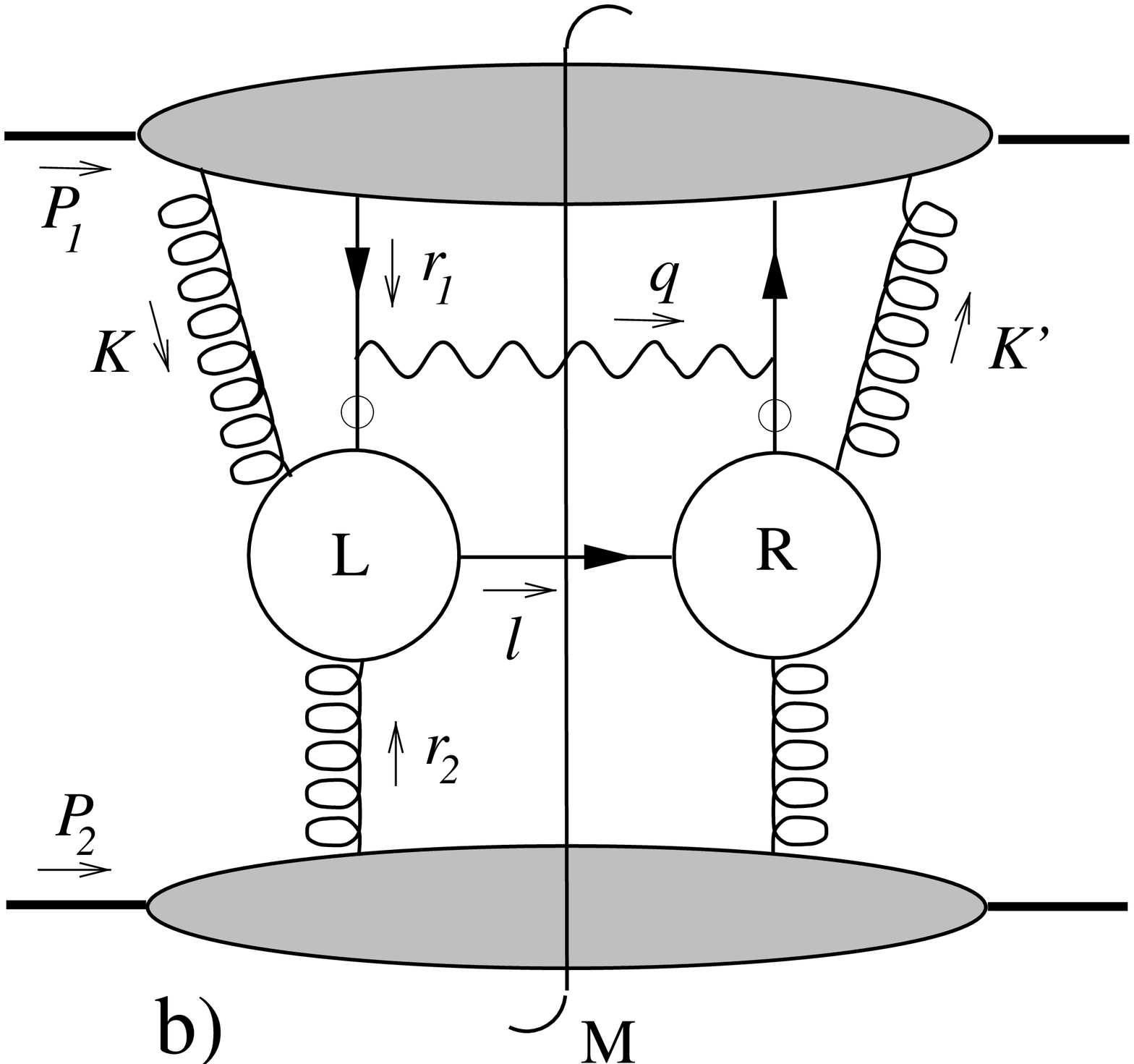,width=4cm}
   \epsfig{file=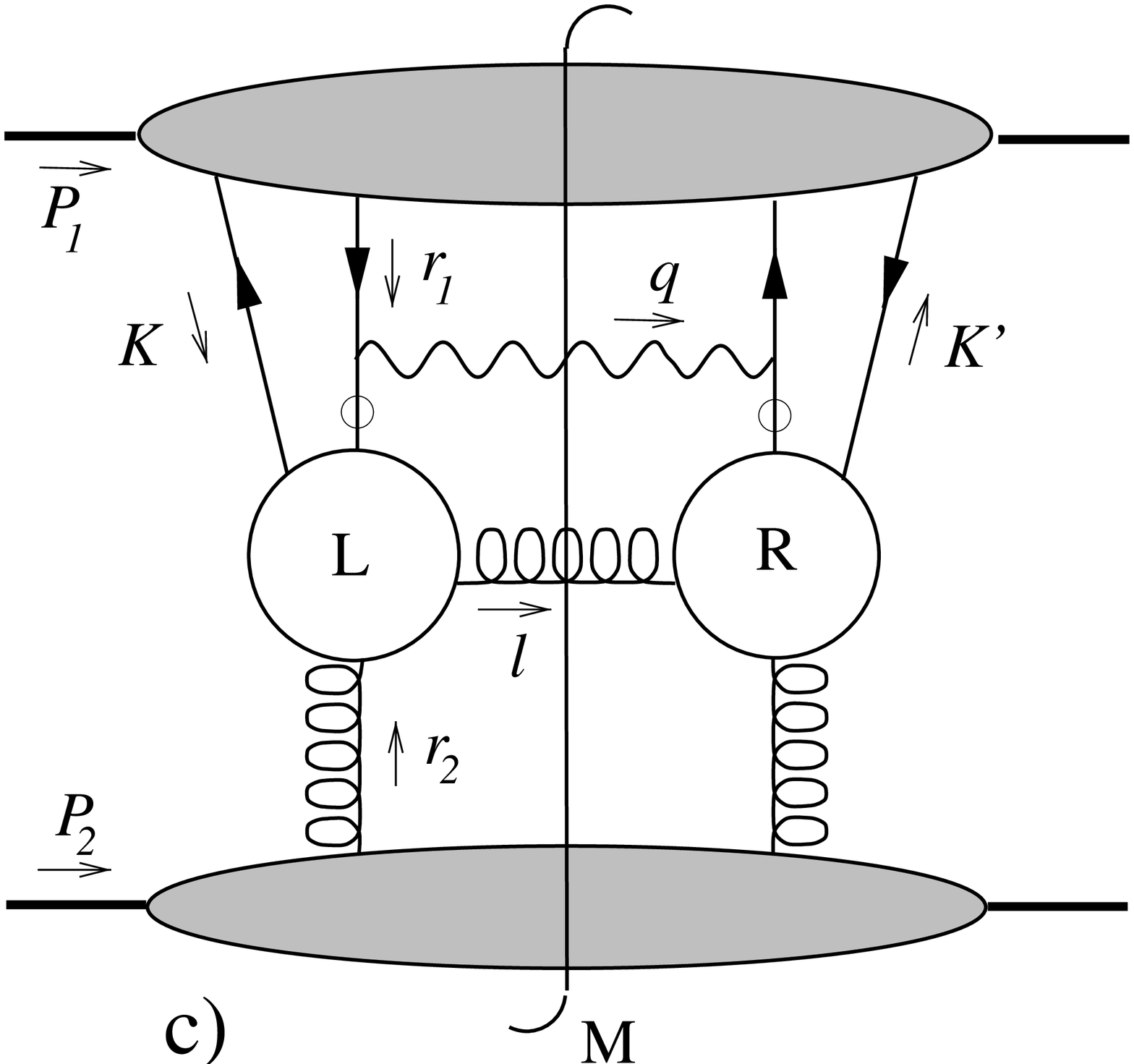,width=4cm}
   \epsfig{file=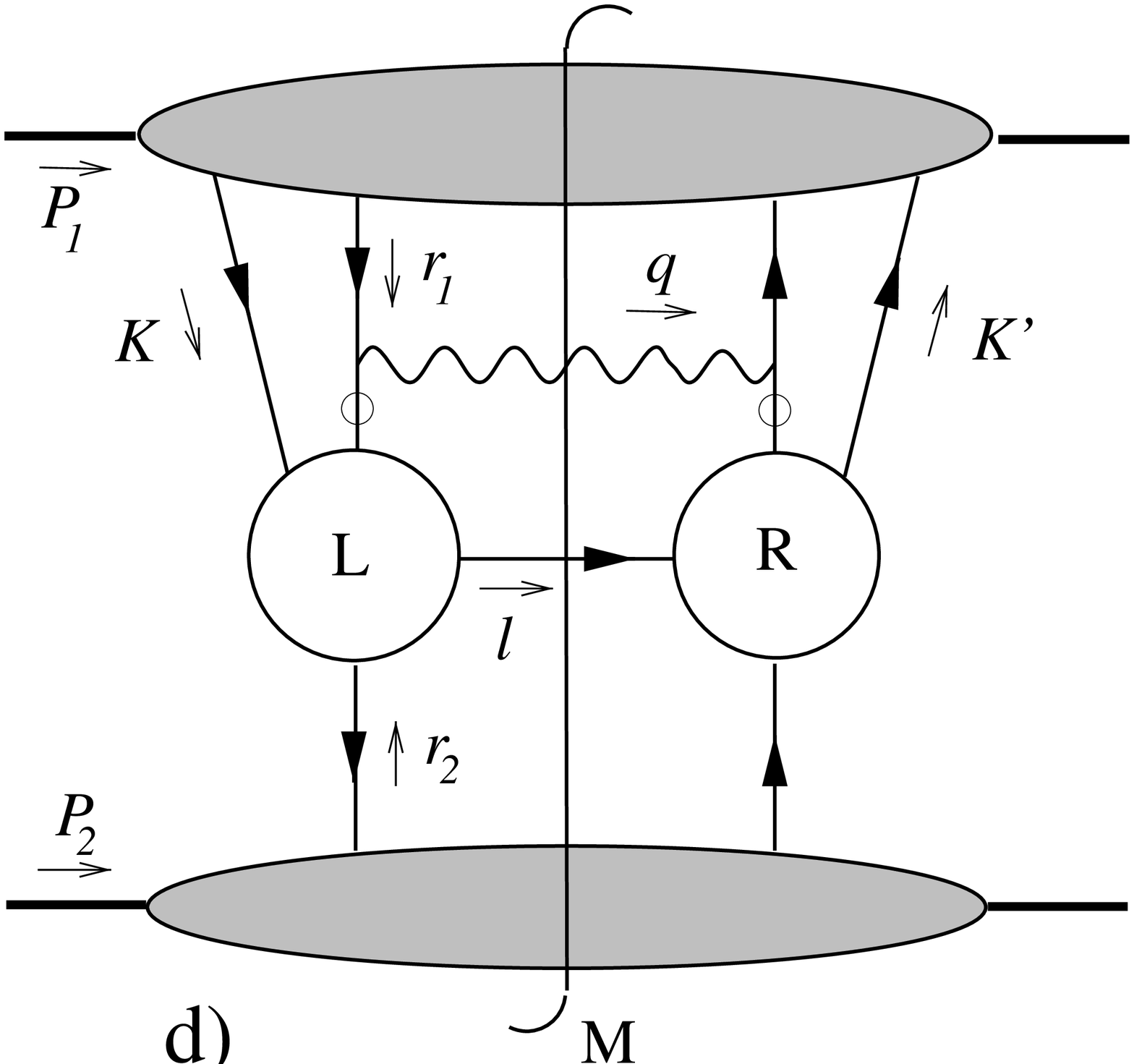,width=4cm}
   \caption{Graphs contributing to double-hard scattering: $qg+\bar q$, 
     $qg+g$, $q\bar q+g$ and $qq+\bar q$. There are additional processes with 
     structures similar to the last one, e.g.\ $q\bar q +q$. This is discussed 
     in the next section. The blobs (L) and (R) stand for all possible 
     tree-level diagramms. Propagators providing poles which are used to fix 
     hard momenta are indicated by small open circles.}
   \label{dhgraphs}  
\end{figure}

We now discuss another technical point which has some relevance for nuclear 
enhancement. Consider again eq.~\eqref{poles} and the example of 
Fig.~\ref{dhgraphs}(a) and note that the numerators $L$ and $R$ contain 
exponential functions $e^{ix P_1^+ z_3^-}e^{i(\tilde\xi_b-x)P_1^+ z_1^-}$ and 
$e^{-ix'P_1^+z_4^-}$. 
To apply the residue theorem we have to close the integration contours. 
For $z_3^- -z_1^- >0$ we have to close the $x$-integration in the upper 
complex $x$-plane with no poles encircled and therefore vanishing integral.
On the other hand for $z_3^- -z_1^- <0$ we must close in the lower complex 
$x$-plane with soft and hard pole encircled.
Therefore these poles always introduce additional theta functions in 
coordinate space, in our example $\Theta(z_1^- -z_3^-)$ from the $x$ 
integration and $\Theta(-z_4^-)$ from the $x'$ integral. 
The presence of these $\Theta$ functions in the matrix elements is an 
essential ingredient of the proof of the following rule about the cancellation
of final state interactions for soft-hard scattering:
If at least one soft parton from the twist-4 matrix element is directly 
coupled to the outgoing unobserved parton line these graphs do not contribute 
to soft-hard scattering. 
These graphs exhibit more than one possibility to set the cut and the nuclear 
enhancement cancels between the contributions with different cuts due to the 
particular arrangement of $\Theta$-functions in that case \cite{Luo94b}. 

\begin{figure}[b]
  \epsfig{file=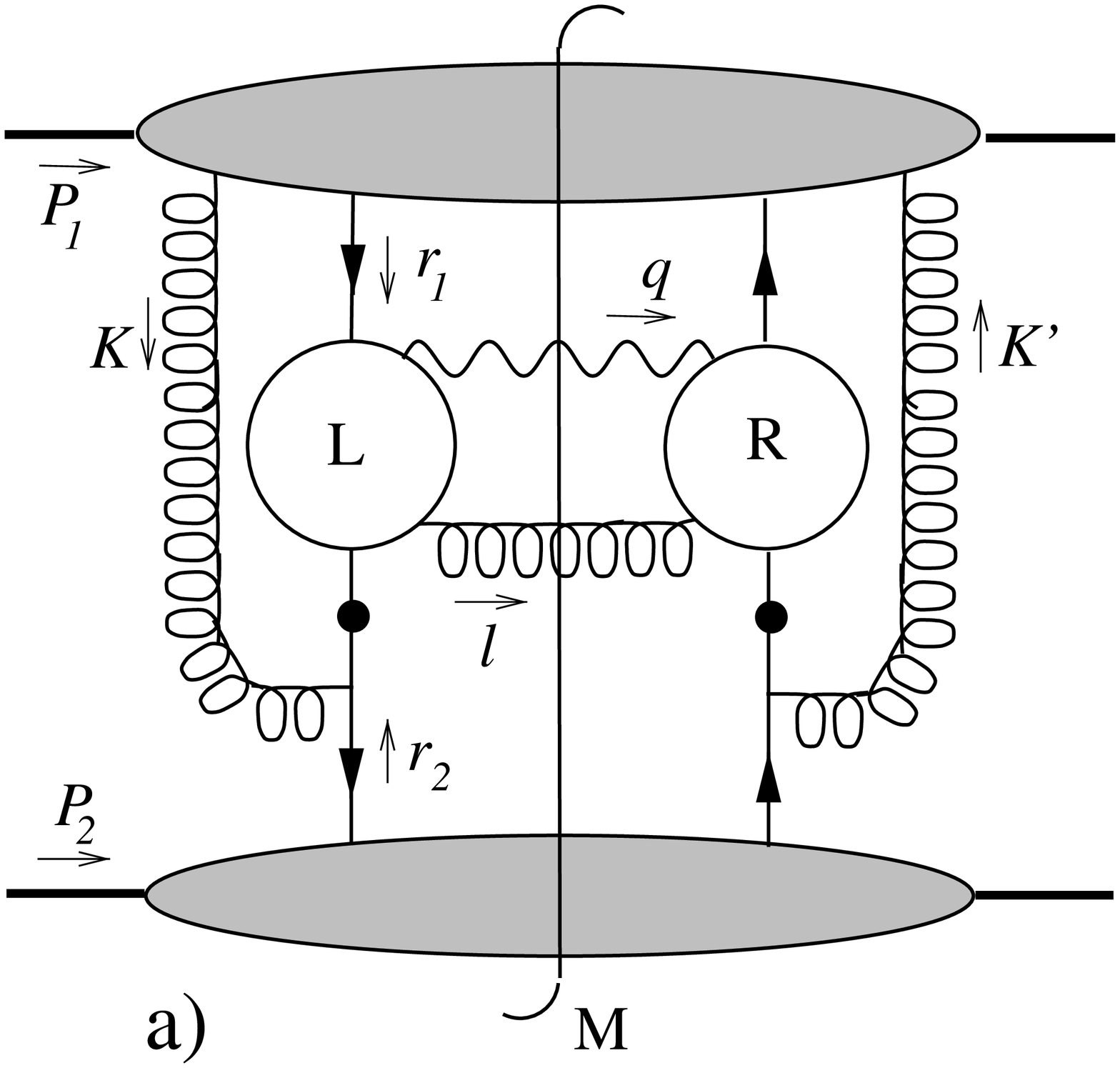,width=4cm}
  \epsfig{file=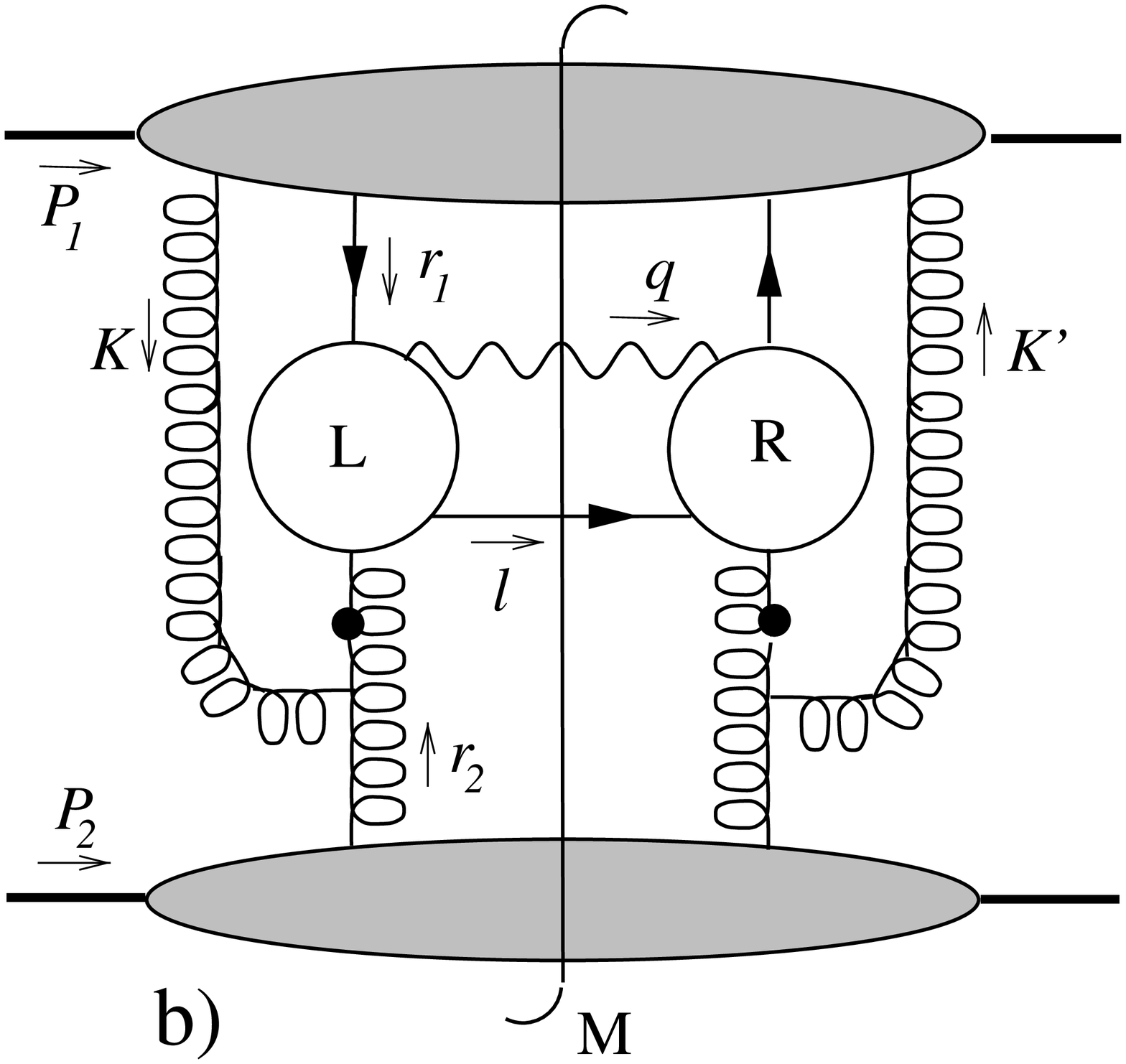,width=4cm}
  \epsfig{file=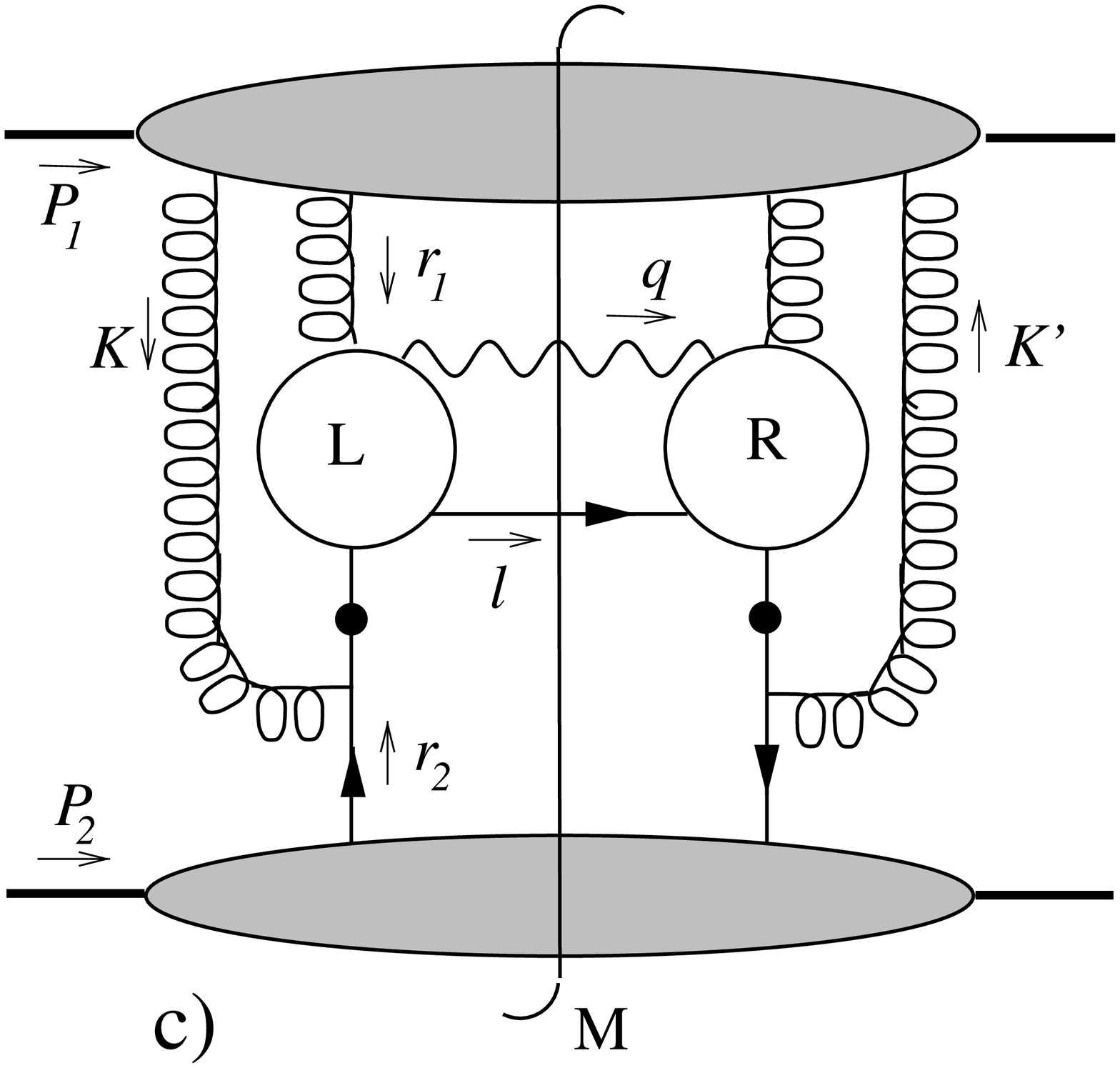,width=4cm}  
  \caption{Graphs contributing to soft-hard scattering: $qg+\bar q$, $qg+g$ 
  and $gg+q$. The blobs (L) and (R) stand for all possible tree-level 
  diagramms. Only symmetrically cut diagrams (M) are shown. Corresponding
  right (R) and left (L) cut diagrams have to be added. Propagators providing 
  poles which are used to fix soft momenta are indicated by small filled 
  circles.} 
  \label{shgraphs} 
\end{figure}

Hence, after a careful discussion of the pole structure of each diagram the 
number of diagrams we are finally forced to evaluate can be reduced. 
Figs.~\ref{dhgraphs} and \ref{shgraphs} show the diagrams that contribute to 
double-hard and soft-hard scattering, respectively. 
We marked the poles used to fix momenta. Also we omitted soft-hard contributions with soft quarks since they should be suppressed compared to their counterparts with soft gluons.


\subsection{Calculation of the double-hard processes}

First we investigate the case of double-hard scattering. Since all parton 
momenta are hard we can work in the collinear limit and set 
$K_\perp=0=K'_\perp$ immediately. We perform the integrations over $x$ 
and $x'$ only taking into account contributions from hard poles.
These poles are marked with circles in Fig.~\ref{dhgraphs}. They fix momenta 
to the values $x=x_h=x'$ and $\xi_1=x_a$ with
\begin{align}
  \label{dhfrac}
  x_a &= \frac{Q^2}{Q^2-T} \\
  x_h &= \tilde\xi_a -x_a = -\frac{\xi_2 (U-Q^2)+Q^2}{\xi_2 S+T-Q^2} 
       -\frac{Q^2}{Q^2-T}
\end{align}
Note that $x_a$ is determined by kinematical variables only whereas $x_h$ 
depends on the integration variable $\xi_2$. Two additional $\Theta$-functions
in the light cone space-time coordinates show up as explained above.

The field operators  of the second parton in the nuclear matrix element are 
selected in such a way that we get the usual twist-2 light cone operator 
$\bar q \gamma^+ q$ for quarks and field strength tensors $F^{\omega +} 
F^+_{\phantom{+}\omega}$ for gluons.
In the latter case we have to turn two factors of $P_1^+$ into light cone 
derivatives $(ix_h)^{-1} \partial_{z_3^-}$, $(-ix_h)^{-1} \partial_{z_4^-}$ 
and perform partial integrations to let them act on the gluon fields in the 
matrix element.
Hard gluons carry only physical polarizations, this allows us to set
$\partial^+ A^\rho(z_3^-) \partial^+ A^\sigma(z_4^-) = 
F^{\omega+}(z_3^-) F^+_{\phantom{+}\omega} (z_4^-) 
\frac{1}{2} (-g^{\rho\sigma} +\bar n^\rho n^\sigma
+n^\rho \bar n^\sigma )$ with the usual light cone basis vectors $\bar n^\rho=
P_1^\rho/P_1^+$, $\bar n^\rho=P_2^\rho/P_2^-$.

We are now able to give the final result for double-hard scattering. For a 
partonic subprocess of the type $ab+c$ we have
\begin{equation}
  \label{dhtensor}
  W^{ab+c}_{\alpha} = 
  \frac{(2\pi)^2(4\pi\alpha_s)^2 e_q^2}{Q^2-T} \int_B^1 \frac{\rd\xi_2}{\xi_2} 
  \frac{1}{\xi_2 S+T-Q^2} f_{c/H}(\xi_2) T^\rdh_{ab}(x_a,x_h) \frac{1}{x_h} 
  H^{\rdh/ab+c}_\alpha(x_a,x_h,\xi_2)
\end{equation}
with $\alpha \in \{\mu\nu,TL,L,\Delta,\Delta\Delta\}$.
Note that parton $a$ is always a quark or antiquark for double-hard scattering
and $e_q$ denotes its electric charge.
As in the helicity amplitudes which arise from contractions of the hadronic 
tensor $W_{\mu\nu}$ defined in eq.~\ref{helicity}, here we use the same 
contractions for the hard partonic part $H_{\mu\nu}$, defining amplitudes
$H_\rtl$, $H_\rl$, $H_\Delta$ and $H_{\Delta\Delta}$ depending on the frame.
The hard parts for the relevant diagrams can be written as
\begin{align}
  \label{dhhard1}
  H_{\mu\nu}^{\rdh,qg+\bar q} &= 
  \mathcal{C} \frac{1}{Q^2}\frac{1}{4} 
  \tr\left[ (\not r_1-\not q) R_{\sigma\beta} \not r_2
  L_{\rho\alpha} (\not r_1-\not q) \gamma_\mu \not r_1 \gamma_\nu 
  \right]_{\substack{\xi_1=x_a \\ x=x'=x_h}} \frac{1}{2}\,g^{\alpha \beta} 
  g^{\rho\sigma} \nn \\
  H_{\mu\nu}^{\rdh,qg+g} &= \mathcal{C}
  \frac{1}{Q^2}\frac{1}{4} 
  \tr\left[ (\not r_1-\not q) R_{\sigma\beta} \not l
  L_{\rho\alpha} (\not r_1-\not q) \gamma_\mu \not r_1 \gamma_\nu 
  \right]_{\substack{\xi_1=x_a \\ x=x'=x_h}} \frac{1}{2}\, g^{\alpha \beta} 
  g^{\rho\sigma} \\
  H_{\mu\nu}^{\rdh,q \bar q +g} &= \mathcal{C}
  \frac{1}{Q^2}\frac{1}{4} 
  \tr\left[ (\not r_1-\not q) R_{\sigma\beta} \not K
  L_{\rho\alpha} (\not r_1-\not q) \gamma_\mu \not r_1 \gamma_\nu 
  \right]_{\substack{\xi_1=x_a \\ x=x'=x_h}} \frac{1}{2}\, g^{\alpha \beta} 
  g^{\rho\sigma}. \nn
\end{align}
$L_{\rho\alpha}$ and $R_{\sigma\beta}$ indicate the sum over all possible tree
level diagrams for the blobs at the left and the right side of the 
corresponding diagram in Fig.~\ref{dhgraphs}. 
Each diagram provides an individual color factor which is indicated 
symbolically by the factor $\mathcal{C}$. 
For partonic processes with three participating fermions we have to 
distinguish between different possible combinations of quark and antiquarks 
and different flavours. In principle there are contributions written in our 
short notation as $q\bar q +q$, $q\bar q +\bar q$ and $qq + \bar q$. 
Each of these contributes four diagrams contained in 
Fig.~\ref{dhgraphs}(d).
In addition, there are processes of the type $q\bar q +q'$, $qq'+\bar q$ and 
$qq' + \bar q'$, which are each represented by a single diagram. In our 
notation $q$ is a quark or antiquark, $\bar q$ the corresponding antiparticle 
and $q'$ denotes a quark or antiquark of a different flavour than $q$.
As an example we give the hard part for the $q \bar q +q$ subprocess:
\begin{multline}
  \label{dhhard2}
  H_{\mu\nu}^{\rdh,q \bar q +q} = 
  \mathcal{C} \frac{1}{Q^2}\frac{1}{8} \left\{
  \tr\left[ (\not r_1-\not q) \gamma_\sigma \not K \gamma_\beta
  (\not r_1-\not q) \gamma_\mu \not r_1 \gamma_\nu \right]
  \tr\left[ \not r_2 \gamma_\rho \not l \gamma_\alpha \right] (r_1+K-q)^{-4} 
  - \right.  \\
  - \tr\left[ (\not r_1-\not q) \gamma_\sigma \not l \gamma_\beta \not r_2
  \gamma_\rho \not K \gamma_\alpha (\not r_1-\not q) \gamma_\mu \not r_1 
  \gamma_\nu \right] (r_1+K-q)^{-2}(r_2+K)^{-2}  +  \\ 
  + \left. (K \leftrightarrow l) \right\}_{\substack{\xi_1=x_a \\ 
  x=x'=x_h}} g^{\alpha \beta} g^{\rho\sigma}.
\end{multline}
Note the relative sign that is due to the different number of fermion loops in
 the corresponding diagrams.

The $T^\rdh_{ab}(x_a,x_h)$ are the universal nuclear matrix 
elements for double-hard scattering introduced in 
\cite{Luo94,mq86}. They depend on the momentum fractions $x_a$, $x_h$ of both 
hard partons from the nucleus and are given by
\begin{equation}
  \label{dhkorr}
  \begin{split}
  T^\rdh_{qg}(x_a,x_h) &= \frac{1}{x_h}\frac{1}{2}  \int \rd z_4^-
  \frac{\rd z_3^-}{2\pi} \frac{\rd z_1^-}{2\pi} \Theta(z_1^- -z_3^-)
  \Theta(-z_4^-) e^{ix_a P_1^+ z_1^-} e^{ix_h P_1^+(z_3^--z_4^-)}  \\ &
  \bra{P_1} F^{\omega +}(z_4^-) F^{+}_{\quad\omega}(z_3^-) 
   \bar{q}(0) \gamma^+  q(z_1^-)  \ket{P_1} \\
  T^\rdh_{q\bar q}(x_a,x_h) &= - \frac{1}{4} \int \rd z_4^- P_1^+
  \frac{\rd z_3^-}{2\pi} \frac{\rd z_1^-}{2\pi} \Theta(z_1^- -z_3^-)
  \Theta(-z_4^-) e^{ix_a P_1^+ z_1^-} e^{ix_h P_1^+(z_3^--z_4^-)}  \\ &
  \bra{P_1} \bar{q}(0) \gamma^+  q(z_1^-) 
  \bar q(z_3^-) \gamma^+ q(z_4^-) \ket{P_1} \\
  T^\rdh_{qq}(x_a,x_h) &= \frac{1}{4} \int \rd z_4^- P_1^+
  \frac{\rd z_3^-}{2\pi} \frac{\rd z_1^-}{2\pi} \Theta(z_1^- -z_3^-)
  \Theta(-z_4^-) e^{ix_a P_1^+ z_1^-} e^{ix_h P_1^+(z_3^--z_4^-)}  \\ &
  \bra{P_1} \bar{q}(0) \gamma^+  q(z_1^-) 
  \bar q(z_4^-) \gamma^+ q(z_3^-) \ket{P_1}.
  \end{split}
\end{equation}
We are now ready to explain the origin of nuclear enhancement in these matrix 
elements. As mentioned above the fields of both partons are coupled separately
to colour singlets.
This allows to probe distances up to the diameter of a large nucleus. 
Each spatial integration of the matrix element
arises from a Fourier transformation and we 
expect it to be accompanied by a rapidly oscillating exponential function 
(since $P_1^+ \longrightarrow \infty$). At first glance,
this seems to rule out coherence over the whole nucleus.
However only $z_1^-$ and $z_3^--z_4^-$ in eq.~(\ref{dhkorr}) are restricted by 
phase factors while $z_3^-+z_4^-$ is not. Hence, we can set 
$\int \rd(z_3^-+z_4^-)$ to be proportional to the nuclear radius 
$R_A \approx r_0 A^{1/3}$ and obtain an additional factor of $A^{1/3}$. Here
$r_0 \approx 1.1~\mathrm{fm}$ is the radius of a nucleon. 

To find a reasonable model for the double-hard matrix elements
imagine the insertion of a complete set of states 
$\sum_P \ket{P}\bra{P}/(2 P^+ V) = 1$ where $V$ is the volume of the
nucleus. Neglecting for a moment the $\Theta$-function and the integration 
over $z_3^- + z_4^-$ we can compare these definitions with the
common definitions of twist-2 structure functions in eq.~(\ref{pardis}).
Assuming that the sum over states is saturated by the lowest
lying nucleon state double hard matrix elements reduce to a 
product of two twist-2 structure functions times a dimensional factor 
which scales like $A^{4/3}$
\begin{equation}
  \label{dhmodel}
  T^\rdh_{ab}(x_a,x_h) = C A^{4/3} f_{a/A}(x_a) f_{b/A}(x_h).
\end{equation}
$f_{a/A}$ is the nuclear parton distribution per nucleon in a nucleus with 
mass number $A$ and proton number $Z$, $N=A-Z$, given by
\begin{equation}
  f_{a/A}(x) = \frac{Z}{A} f_{a/p}(x) + \frac{N}{A} f_{a/n}(x).
\end{equation}
In a primitive model one can simply set the remaining free integration 
with respect to $z_4^-+z_3^-$ equal to $R_A$ and obtains $C=3/(8 \pi r_0^2)
\approx 0.005~\gev^2$. Although the factorization of the double-hard matrix
element into a product of two twist-2 distribution functions and the scaling 
by $A^{4/3}$ seems to provide a reasonable model, the choice of the
normalization factor $C$ is a hot matter of discussion, see section 5.
The $\Theta$-functions provide an ordering of both single scattering events
on the light cone. They simply express the fact that scattering off parton $b$
takes place before the scattering off parton $a$. This corresponds to 
$z_4^-<0$ on the left side and to $z_3^- < z_1^-$ on the right side of each 
diagram.

Now we give the final results for the hard parts. For the evaluation it is 
important also to include ghosts, whenever one can close gluon lines through 
the non-perturbative blobs.
For $H^\rdh_\rtl=-\frac{1}{2} H^{\rdh\phantom{\mu}\mu}_{\phantom{\rdh}\mu}$, 
which enters the angular integrated cross section, we reproduce the results of
\cite{Guo98}. We have listed them in Appendix 7.3 for all subprocesses.
The helicity amplitude $W_\rtl$ is independent of the frame. 
Due to the special form of all diagrams contributing to double-hard scattering
we easily see that the contractions $P_\rli^{\mu\nu} H^\rdh_{\mu\nu}$ and 
$P_\rliii^{\mu\nu} H^\rdh_{\mu\nu}$ in eq.~(\ref{dhhard1}) always vanish, and for 
the last contraction we get
\begin{equation}
  P_\rlii^{\mu\nu} H_{\mu\nu}^\rdh = \frac{Q^2S\left[ (Q^2-T)(Q^2-U)-Q^2S 
  \right]}{{(Q^2-T)}^2{(Q^2-U)}^2} H_\rtl^\rdh = 
  \frac{Q^2q_\perp^2}{{(Q^2+q_\perp^2)}^2} H_\rtl^\rdh
\end{equation}
for all processes.
For the Collins-Soper frame this leads to the simple relations
\begin{eqnarray}
  W^\rdh_\rl =& \sin^2 \gamma_\rcs W^\rdh_\rtl &=
    \frac{q_\perp^2}{Q^2+q_\perp^2} W_\rtl^\rdh \\
  W_\Delta^\rdh =& \sin \gamma_\rcs \cos \gamma_\rcs W_\rtl^\rdh &=
    -\frac{Qq_\perp}{Q^2+q_\perp^2} W_\rtl^\rdh \\
  W_{\Delta\Delta}^\rdh =& \frac{1}{2} W^\rdh_L.
\end{eqnarray}
Immediately one observes that the Lam-Tung relation holds in the CS frame and 
all other frames considered here. For the 
Gottfried-Jackson frame the results get particularly simple. The double-hard 
contributions for deviations 
from the simple $1+\cos^2\theta$ behaviour vanish completely:
\begin{equation}
  W^\rdh_{\rm L} = 0\; ,\quad
  W^\rdh_\Delta = 0\; ,\quad
  W^\rdh_{\Delta\Delta} = 0\; .
\end{equation}
At first glance this seems to be rather surprising but it is a direct
consequence of the special form of all double-hard contributions
and therefore has a simple explanation.

The double hard process resembles the classical double
scattering picture. To understand this we consider again the process depicted 
in Fig.~\ref{scattering}. The antiquark from the hadron undergoes a
first scattering with a hard gluon from the nucleus. 
By subsequently radiating a gluon with large transverse momentum
the antiquark returns to the mass shell, but now with transverse momentum 
$q_\perp$. The second scattering then corresponds to the lowest-order 
annihilation process of two massless quarks. Therefore the double-hard 
process can be factored as the product of two single scattering cross sections,
supported by our model that the double-hard matrix element can
in good approximation be written as a product of two twist 2 distribution 
functions. This process naturally fulfills the Lam-Tung relation and depends
only on the structure function $W_\rtl \sim W_\mu^\mu$. 
Other helicity amplitudes differ only by kinematical factors, i.e.\ can 
be obtained from $W_\rtl$ in the Gottfried-Jackson frame by simple rotations.
In this probabilistic picture scaling by $A^{4/3}$ appears to be very natural.
The cross section of the first scattering obviously scales by the volume 
$\sim A$ of the nucleus. Afterwards the projectile parton travels further 
through the nucleus and has to penetrate the nuclear matter. 
The probability whether a second interaction takes place on the way to the 
boundary of the nucleus scales then by the nuclear radius $\sim A^{1/3}$.

\subsection{Calculation of the soft-hard processes}

For soft-hard scattering we have to keep $K_\perp$ in the expansion of the 
gluon momenta. The limit $K_\perp=0$ would lead to $K=0$ which gives no 
contribution to physical double scattering but just a contribution of the 
eikonal phase. Note that we are always using covariant gauge where these 
factors are present.
Starting from equation~(\ref{hadr1}) we do a collinear expansion 
\begin{equation}
  \label{expand}
  \tilde H = \tilde H\vert_{\substack{K_\perp=0\\K_\perp'=0}} + 
  K^\lambda_\perp \frac{\rd}{\rd K^\lambda_\perp} 
  \tilde H\vert_{K_\perp=0} + \frac{1}{2!} K_\perp^\lambda
  K_\perp^\kappa \frac{\rd^2}{\rd K_\perp^\lambda \rd K_\perp^\kappa} 
  \tilde H\vert_{K_\perp=0} + \dots.
\end{equation}
of the terms depending on intrinsic transverse momentum 
\begin{equation}
  \label{expand2}
  \tilde H = \int \frac{\rd x \rd x'}{\xi_1} e^{ir_1^+ z_1^-} e^{iK^+ z_3^-}
  e^{-i{K'}^+ z_4^-} \bar H(\xi_1,\xi_2,x,x',K_\perp,K_\perp').
\end{equation}
We have used $K_\perp=K_\perp'$ which allows us to write
\begin{equation}
  \label{expand3}
  K_\perp^\lambda \frac{\rd}{\rd K^\lambda_\perp} = K^\lambda_\perp 
  \frac{\partial}{\partial K^\lambda_\perp} + 
  {K'_\perp}^\lambda \frac{\partial}{\partial {K'_\perp}^\lambda}.
\end{equation}
The mixed terms of second order proportional to $K_\perp^\lambda 
{K'_\perp}^\kappa$ give the leading contribution to soft hard scattering. 
The factor $K_\perp^\lambda {K'_\perp}^\kappa$ can be converted into partial 
derivatives with respect to $z_3^-$, $z_4^-$. A partial integration and 
an expansion $A^\rho = A^+ P_1^\rho/P_1^+ + \ldots$ for soft gluon fields
in the nuclear matrix elements turns them into field strength tensors. 
The $K_\perp$ and $K_\perp'$ integrals can then be carried out.
Note that having soft quarks with transverse momentum and converting the 
corresponding $K_\perp$ into covariant derivatives acting on the quark fields 
would lead to operators contributing only at the level of twist-6 
\cite{Luo94}. 

We perform the pole integrations for the symmetrically cut graphs (M) in 
Figs.\ref{shgraphs} and the corresponding asymmetrically cut ones (R) and (L).
The residues fix the momentum fractions of the soft gluons to $x=x_s=x'$ 
while the hard partons carry $\xi_1=x_b$ for the symmetric diagrams and 
$\xi_1=x_c$ for the asymmetric diagrams and
\begin{align} 
  \label{shfrac}
  x_s &= \frac{k_\perp^2}{\xi_2 S},  \\
  x_b &= \tilde \xi_b - x_s = 
    -\frac{\xi_2(U-Q^2)+Q^2-2q K_\perp -k_\perp^2}{\xi_2 S+T-Q^2} 
    -\frac{k_\perp^2}{\xi_2 S},   \\
  x_c &= \tilde \xi_c = - \frac{\xi_2(U-Q^2)+Q^2}{\xi_2 S+T-Q^2}.
\end{align}
We have introduced the notation $k_\perp^2=-K_\perp^2 \ge 0$.
When we take a forward diagram 
for soft-hard scattering then the sum of differently cut diagrams (M), (R) and
(L) can be divided into two terms. This procedure is described in detail in 
\cite{Luo94b,Guo98}. In the first term, phase factors and 
$\Theta$-functions arrange in such a way that this term is bounded in 
space-time by geometrical arguments and cannot pick up an additional factor 
$A^{1/3}$. This contribution shows no nuclear enhancement. 
The second term is not bounded and only depends on the hard part of the 
symmetric diagram (M). We therefore neglect the first term and arrive at
\begin{multline}
  \label{shtensor}
  W_{\alpha}^{\rsh/ab+c} = (2\pi)^2(4\pi\alpha_s)^2 e_q^2 \int_B^1 
  \frac{\rd \xi_2}{\xi_2} 
  \frac{1}{\xi_2 S +T-Q^2} f_{c/H} (\xi_2)  \\
  \big( -\frac{g^{\lambda\kappa}}{4} \big) 
  \frac{\rd^2}{\rd K_\perp^\lambda \rd K_\perp^\kappa}\Big\vert_{K_\perp=0} 
  T^\rsh_{ab}(x_b) \frac{1}{x_b} H^{\rsh,ab+c}_{\alpha}(x_b,x_s,\xi_2)
\end{multline}
where $\alpha\in \{\mu\nu,\rtl,\rl,\Delta,\Delta\Delta\}$. 
The hard part for the $qg+\bar q$ process shown in Fig.~\ref{shgraphs}(a) is
\begin{equation}
  \label{shhard1}
  H_{\mu\nu}^{\rsh/qg+\bar q} = \mathcal{C} 
  \frac{1}{(\xi_2 S)^2} \frac{1}{4} \tr \left[
  R_{\nu\beta} (-\not r_2 -\not K) \not P_1 \not r_2 \not P_1 (-\not r_2 -
  \not K) L_{\alpha\mu} \gamma_\mu \not r_1 \gamma_\nu \right]_{\substack{
  \xi_1=x_b \\ x=x'=x_s}} (-g^{\alpha\beta} )
\end{equation}
$L_{\alpha\mu}$ and $R_{\nu\beta}$ stand for the tree-level diagrams on the 
left- and the right-hand side of the diagram in Fig.~\ref{shgraphs}(a) 
indicated by the blobs. $\mathcal{C}$ symbolizes the color factor of each 
single diagram. Hard parts for $qg+g$ and $gg+q$ subprocesses are obtained in 
an analogous fashion.

The universal matrix elements appearing for soft-hard scattering depend only 
on the momentum fraction $x_b$ of the hard parton and are given by
\begin{equation}
  \label{shkorr}
  \begin{split}
  T^\rsh_{qg}(x_b) &= \frac{1}{2} \int \rd z_4^- \frac{\rd z_1^-}{2\pi}
  \frac{\rd z_3^-}{2\pi} \Theta(z_1^- -z_3^-) \Theta(-z_4^-) 
  e^{i x_b P_1^+ z_1^-} \\
  & \bra{P_1} F^{\omega +}(z_4^-)\bar{q}(0) \gamma^+  q(z_1^-) 
  F^{+}_{\quad\omega}(z_3^-) \ket{P_1} \\
  T^\rsh_{gg}(x_b) &= \frac{1}{x_b P_1^+} \int \rd z_4^- \frac{\rd z_1^-}{2\pi}
  \frac{\rd z_3^-}{2\pi} \Theta(z_1^- -z_3^-) \Theta(-z_4^-) 
  e^{i x_b P_1^+ z_1^-} \\ 
  & \bra{P_1} F^{\omega +}(z_4^-) F^{+}_{\quad\omega}(z_3^-)
  F^{\lambda +}(0) F^{+}_{\quad\lambda}(z_1^-) \ket{P_1}
  \end{split}
\end{equation}
for scattering off a quark-gluon and a gluon-gluon pair in the nucleus,
respectively. We can repeat the arguments given below eq.~(\ref{dhkorr}) for 
the nuclear enhancement. 
Since the longitudinal momentum fraction of the gluons is essentially zero,
we even encounter only one phase factor $e^{i x_b P_1^+ z_1^-}$
contrary to the case of double hard scattering discussed before.
One therefore might expect that soft-hard matrix elements lead to an 
enhancement factor proportional to the square of the nuclear radius $A^{2/3}$ 
since only one out of three integrations is limited by phase factors.
However, the fields at positions $z_3^-$ and $z_4^-$ are correlated by color 
forces. The anticipated absence of long-range color correlations in a nucleus 
implies that integration over $z_4^--z_3^-$ is still restricted to a region 
of the size of a nucleon.

To build a model for the soft hard matrix elements one again can compare 
eq.~(\ref{shkorr}) with the definition of the usual twist-2 matrix elements 
eq.~(\ref{pardis}). Neglecting for a moment the $\Theta$-functions we see that
the twist-4 pieces differ only by an additional $x_b$ independent 
renormalization. Assuming that this can be factored into a term proportional to
the nuclear radius and a (possibly scale dependent) constant $\lambda^2$ 	
\begin{align}
  \label{lambda}
  A^{1/3} \lambda^2 &\sim  \int \rd z_4^- 
  \frac{\rd z_3^-}{2\pi} 
  F^{\omega +}(z_4^-) F^{+}_{\quad\omega}(z_3^-) 
\end{align}
the $x_b$ dependence of the twist-4 matrix element can be modeled by the usual
twist-2 structure function.
\begin{equation}
  \label{shmodel}
  T^\rsh_{ab}(x) = \lambda^2 A^\frac{4}{3} f_{a/A}(x)
\end{equation}
Eq. (\ref{lambda}) can be interpreted as the 
collective color-magnetic Lorentz force experienced by the projectile parton 
while traveling through nuclear matter \cite{Luo94}.

The only $K_\perp$-dependence of the hard part will be implicit in $x_b$ and 
$x_s$. Therefore we can rewrite the derivatives in the form
\begin{equation}
  \begin{split}
  W^{\rsh,ab+c}_{\alpha} =& (2\pi)^2(4\pi\alpha_s)^2 e_q^2 \int_B^1 
  \frac{\rd \xi_2}{\xi_2} 
  \frac{1}{\xi_2 S +T-Q^2} f_{c/H} (\xi_2)  \\
  & \frac{1}{2} \left[ \frac{\partial^2}{\partial x_c^2} \left( T^\rsh_{ab}
  (x_c) \frac{1}{x_c} H_{\alpha}^{\rsh,ab+c} (x_c,\xi_2) \right) 
  \frac{2q_\perp^2}{(\xi_2 S+T-Q^2)^2} + \right. \\
  & + \frac{\partial}{\partial x_c} \left( T^\rsh_{ab}(x_c) 
  \frac{1}{x_c} H_{\alpha}^{\rsh,ab+c} (x_c,\xi_2) \right) 
  \frac{2(Q^2-T)}{\xi_2 S(\xi_2 S+T-Q^2)} + \\
  & + \left. T^\rsh_{ab}(x_c) \frac{1}{x_c} {(H_{\alpha}^{\rsh,ab+c})}'
  (x_c,\xi_2) \frac{2}{\xi_2 S} \right]
  \end{split}
\end{equation}
where $H(x_c,\xi_2) =  H(x_c,x_s=0,\xi_2)$ and $H'(x_c,\xi_2) = \frac{
\partial}{\partial x_s} H(x_c,x_s=0,\xi_2)$. Note that in \cite{Guo98} the 
last term containing a derivative with respect to $x_s$ is absent since 
it turns out that $H^\rsh_\rtl$ does not depend explicitly on $x_s$. We give 
the results for the hard parts in the Collins-Soper frame in Appendix 7.4. 
There are some subtleties in removing the unphysical degrees of freedom
for processes of the class $qg+g$. There we have soft quarks from the nucleus
with dominating component $A^+$ and for the hard gluons from the single hadron
we have to take only the transverse polarizations.
For $W^\rsh_\rtl$ we reproduce the results of \cite{Guo98}.
Unlike for double-hard scattering the amplitudes $W^\rsh_L$, $W^\rsh_\Delta$ 
and $W^\rsh_{\Delta\Delta}$ cannot be obtained in a trivial way from $W_\rtl$. 
Related to that is an obvious violation of the Lam-Tung relation.

\section{Numerical Results}
 
In order to cover different energies $S$ we calculated helicity amplitudes and
angular coefficients for two different settings: for $252\gev$ $\pi^-$ on 
tungsten as in the Fermilab E615 experiment \cite{e615} 
and for future $p+A$ collider experiments at RHIC energies with $250\gev$ 
protons and $A \times 100\gev$ large nuclei.
The main task before doing the numerics is to insert a suitable model for the 
twist-4 matrix elements in the nucleus. The models introduced in the last 
sections for the double-hard matrix elements 
\begin{equation}
  \label{dhmodel1}
  T^\rdh_{ab}(x_a,x_h) =  C A^\frac{4}{3} f_{a/A}(x_a) f_{b/A}(x_h).
\end{equation}
and for the soft-hard matrix elements 
\begin{equation}
  \label{shmodel1}
  T^\rsh_{ab}(x) = \lambda^2 A^\frac{4}{3} f_{a/A}(x)
\end{equation}
are already given by LQS \cite{Luo94b,Guo98}. We tried to argue above that 
their dependence on the parton momentum fractions and their scaling behaviour 
with respect to $A$ seem to be quite natural. 

Nevertheless the correct normalization of the models is not yet clear.
The hope is that $C$ and $\lambda$ are universal constants.
In the early days of the model a value of $C=0.072\gev^2$ 
was taken from theory \cite{Guo98,mq86} while $\lambda^2$ was taken from 
comparison with experiment. This was done by analyzing 
dijet photoproduction on nuclei \cite{Luo94,Naples:1994uz}
leading to a value $\lambda^2=0.05-0.1 \gev^2$. 
Later an analysis of $q_\perp$-broadening in hadron-nucleon Drell-Yan 
processes \cite{Guo98b} implied a value of $\lambda^2=0.01 \gev^2$. 
Obviously there is a discrepancy, which may either point to a strong scale 
dependence or questions the universality of the soft-hard nuclear matrix 
elements. Note that the soft interaction besides the hard process 
in soft-hard double scattering is a final state interaction for dijet 
photoproduction but an initial state interaction for the Drell-Yan process.

In their latest work Guo, Qiu and Zhang \cite{gqz99} used the fact that there
is a formal connection between soft-hard and double-hard matrix elements via 
the limit
\begin{equation}
  \lim_{x_g\to 0} x_g T^\rdh_{qg}(x_q,x_g) = T^\rsh_{qg}(x_q)
\end{equation}
which leads with the models of eqs.~(\ref{dhmodel1}),(\ref{shmodel1}) to 
\begin{equation}
  C \approx \frac{\lambda^2}{x_g g(x_g)|_{x_g\approx 0}}.
\end{equation}
They take $x_g g(x_g)|_{x_g\approx 0} \approx 3$ as the gluon momentum 
density at $x_g\longrightarrow 0$.
Taking into account the later value of $\lambda^2$ this leads to a 
normalization factor $C$ which is an order of magnitude below the 
former theoretical value. But it is more consistent with the value of
$C\approx 0.005\gev^2$ which we obtained in 
section 4 from a naive normalization argument.

In another, more geometrical, model one can estimate the two-parton area 
density that can be seen by the projectile parton in the nucleus by the 
product of two single-parton area densities. Integration of the two-parton 
area density over the effective transverse area of the nucleus yields exactly
the right-hand side of eq.~(\ref{dhmodel1}) with a constant $C= 9/(8\pi r_0^2)
\approx 0.01\gev^2$.
Obviously $C$ is presently basically unknown and a true improvement can only 
be expected if comparisons of a larger number of observables with more precise
experimental data are available.

For soft-hard scattering we use the smaller value of $\lambda^2=0.01
\gev^2$ obtained from Drell-Yan data to maintain consistency.
For double-hard scattering we have chosen the maximal value $C=0.072\gev^2$
to give a feeling how big the effects of double scattering might be.
We take the twist-2 distribution functions from the CTEQ3M 
parameterization for the nucleons \cite{cteq3} and from the recent GRS fits 
for the pion \cite{grs}. Where nuclear effects on twist-2 distribution 
functions are taken into account the EKS98 parametrization \cite{eks98} is 
used.

\begin{figure}[t]
  \epsfig{file=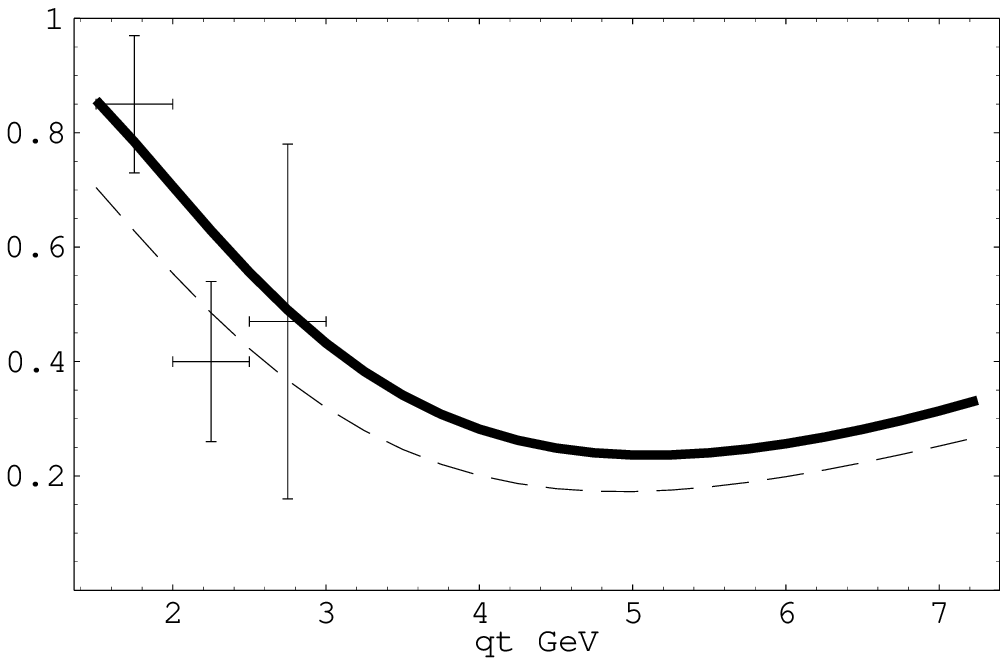,width=5.5cm}
  \epsfig{file=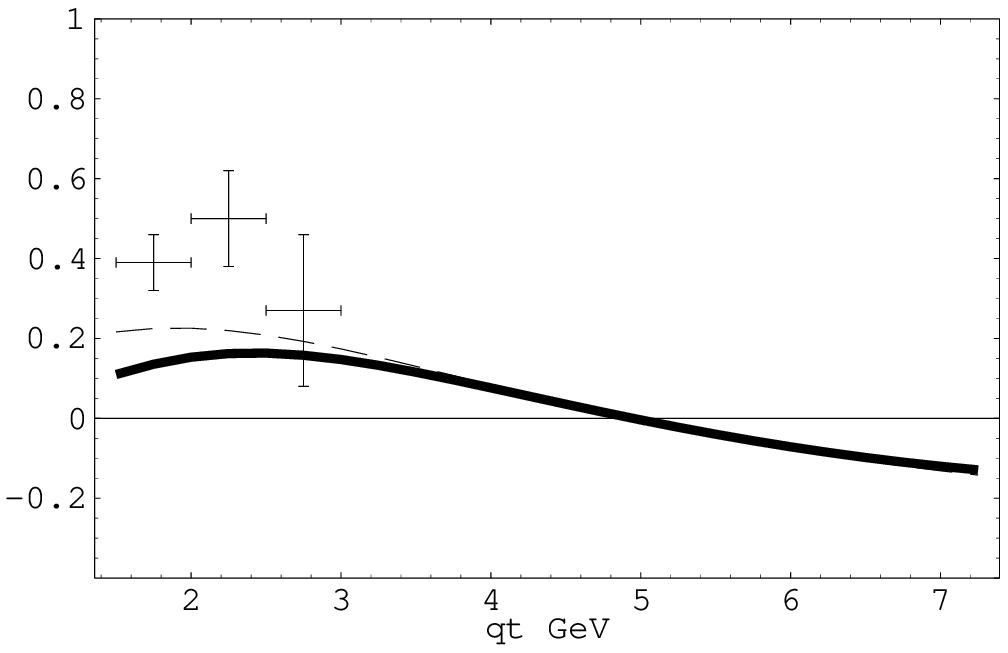,width=5.5cm}
  \epsfig{file=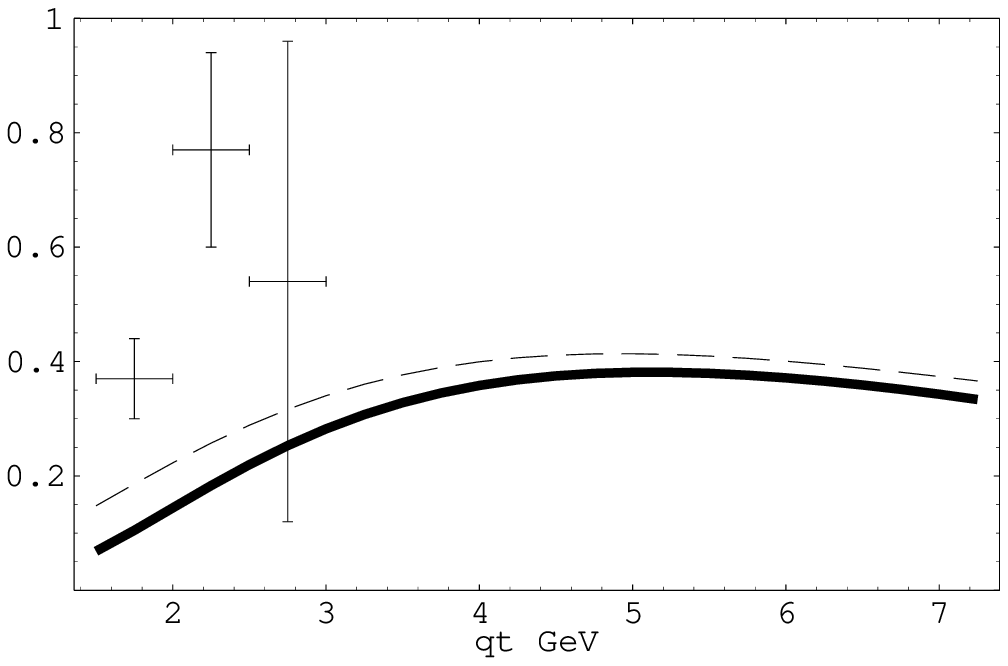,width=5.5cm}
  \caption{The angular coefficients $\lambda$, $\mu$ and $\nu$ 
    (from left to right) as functions
    of $q_\perp$ in the Gottfried-Jackson frame for $252\gev$ $\pi^-$ on 
    tungsten at $Q= 5\gev$, $y=0$. The dashed line is the result for 
    single-scattering and the thick line is given by the full calculation 
    with single- and double-scattering. We also show results of the 
    E615 experiment at low transverse momentum.}
  \label{picoeff}
\end{figure}

\begin{figure}[b]
  \epsfig{file=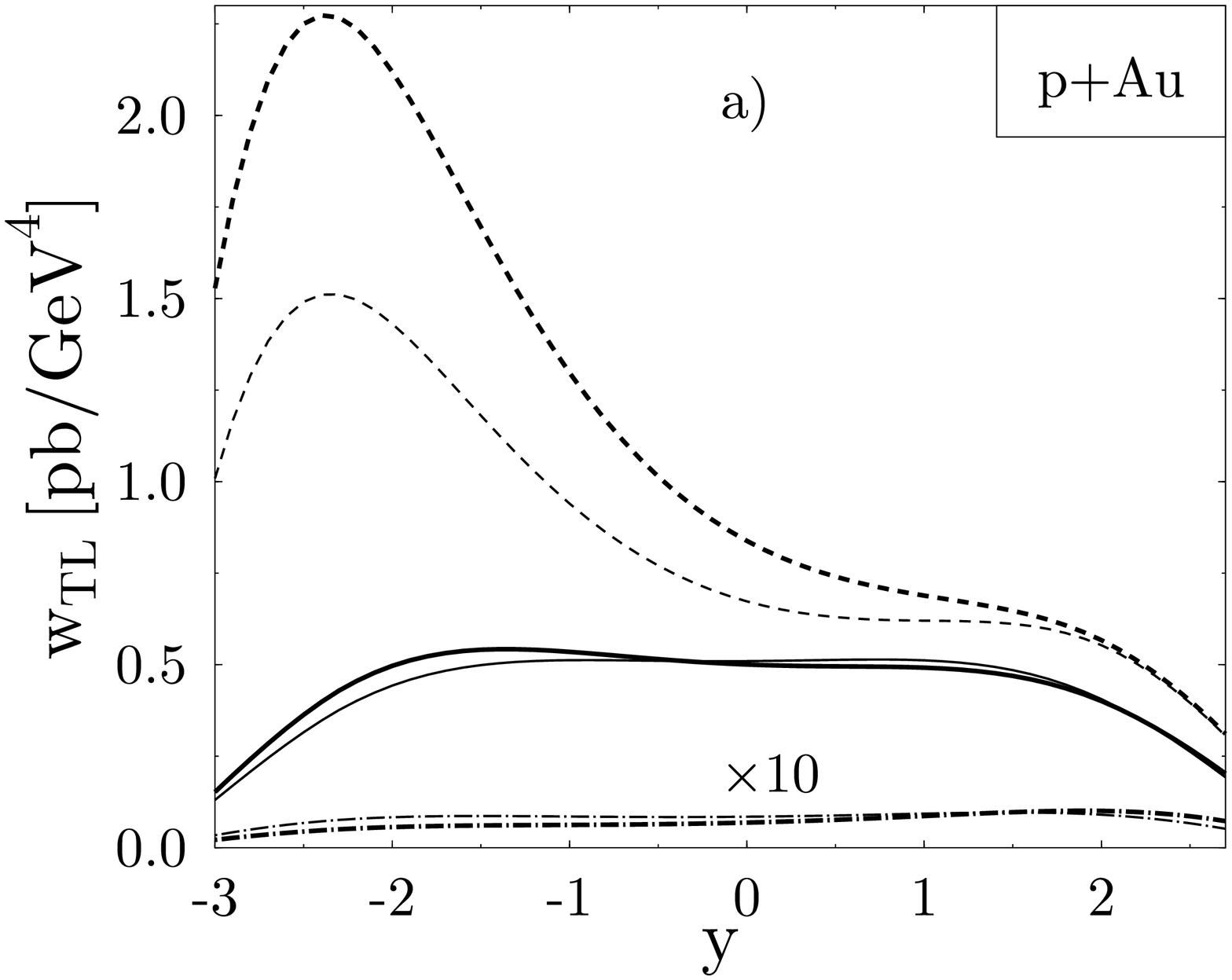,width=5.5cm}
  \epsfig{file=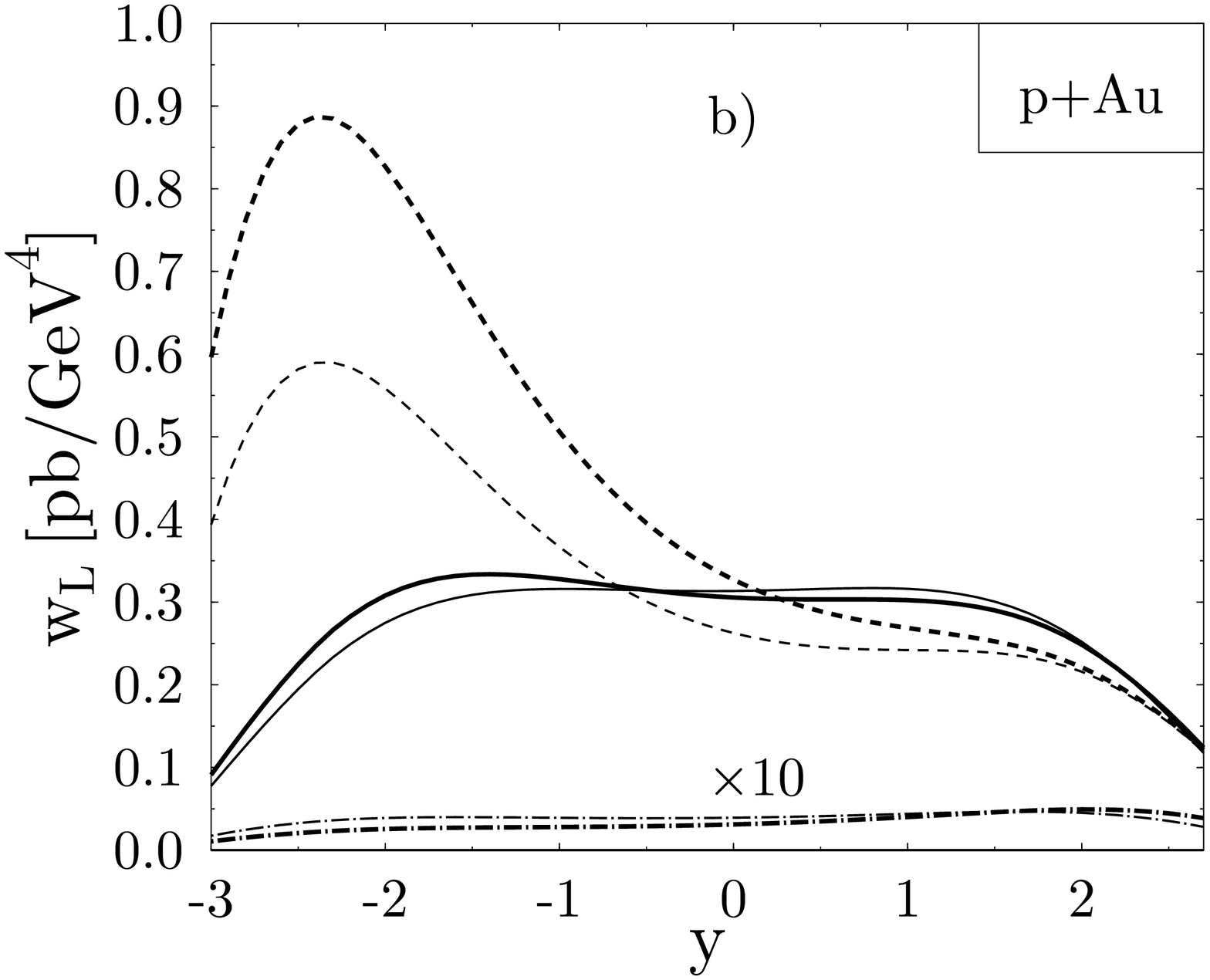,width=5.5cm} 
  \epsfig{file=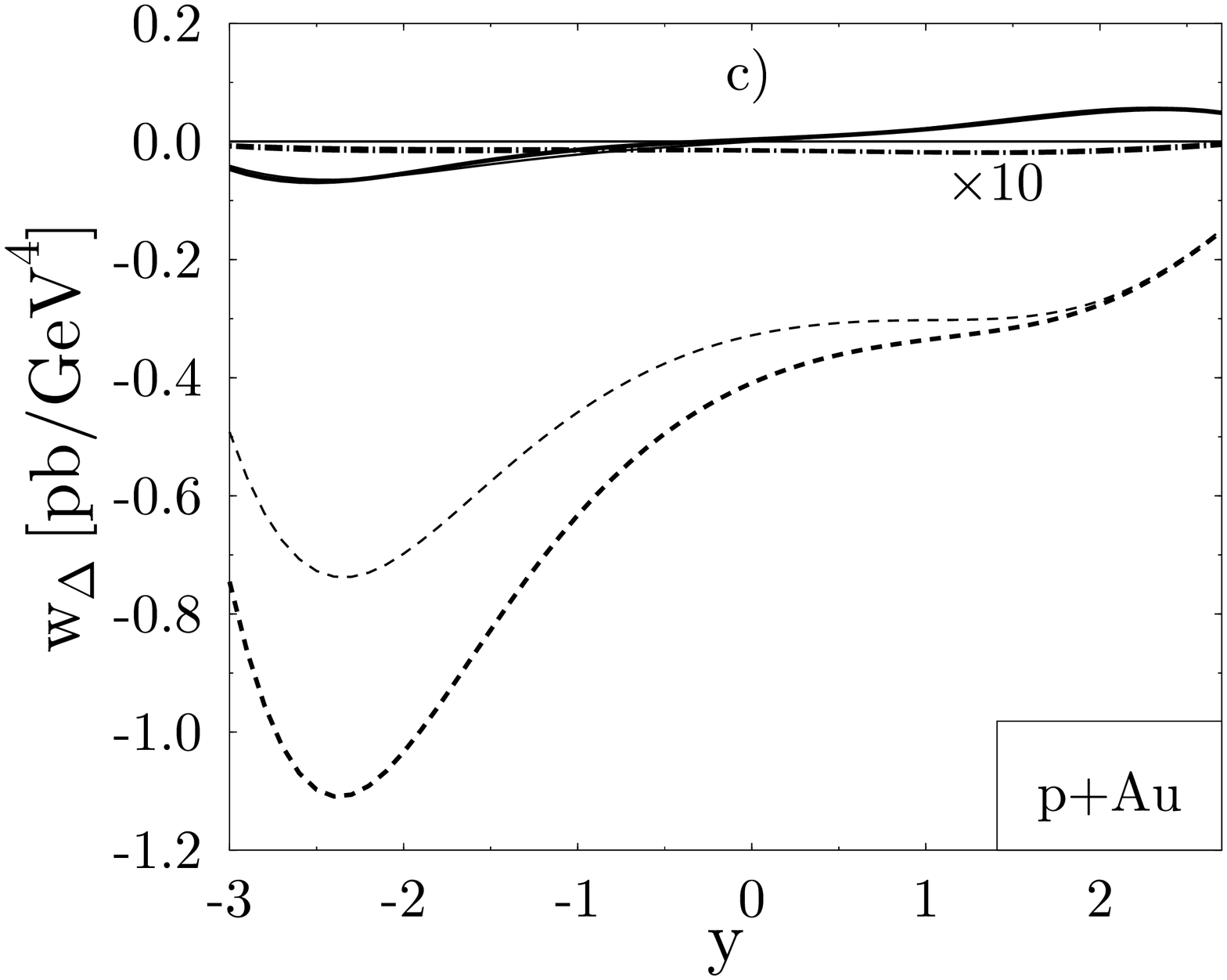,width=5.5cm} 
  \caption{The helicity amplitudes $w_\rtl$, $w_\rl$ and $w_\Delta$ 
    for $p+Au$ collisions at $S=10^5\gev$, $Q= 5\gev$ and 
    $q_\perp=4\gev$. We show
    the twist-2 contribution (solid lines), double-hard (dashed lines), and 
    soft-hard (dashed-dotted lines). Soft-hard is scaled up by a factor of 10.
    We give results with (thin lines) and without (thick lines) EKS98 
    modifications for nuclear parton distributions.} 
  \label{yfig}  
\end{figure}

Let us first address the question how this approach works with the pion data 
from E615. In Fig.~\ref{picoeff} we show our results for the angular 
coefficients $\lambda$, $\mu$ and $\nu$ in the Gottfried-Jackson frame for the
process $\pi^- + W$ at E615 energy. We also give experimental data from the 
E615 collaboration. We can fairly well describe the data for $\lambda$ at low 
transverse momentum but fail for $\mu$ and $\nu$. 
It seems to be an essential feature that corrections due to nuclear enhanced 
twist-4 matrix elements can be quite large for the helicity amplitudes but the 
effects largely cancel for the angular coefficients which are ratios
thereof. Therefore changes of the
angular coefficients in the range $q_\perp\approx Q$ are small compared with
the pure twist-2 calculation. 
Experimental values have large error bars above $q_\perp = 2 \gev$.
We already pointed out that data at $q_\perp/Q \approx 1$ would provide a much
better test for the twist-4 calculation because of the absence of soft-hard 
interference terms and the absence of double logarithmic corrections of the 
type $\log^2(Q^2/q_\perp^2)$ which have to be resummed. 
We have to conclude that nuclear enhanced double scattering contributions are
not able to yield substantial improvement for the old pion data.
Deviations of the $\pi-W$ data from the naive perturbative QCD-predictions 
might be explained by other approaches like the Berger-Brodsky mechanism 
which takes into account higher twist contributions of the pion wave function
\cite{bbkm94}, in terms of non-trivial spin correlations in the QCD-vacuum 
\cite{Brandenburg:1993cj} or via contributions of the chiral-odd $T$-odd
distribution function $h_1^\perp$ \cite{db:99}.

\begin{figure}[t]
  \epsfig{file=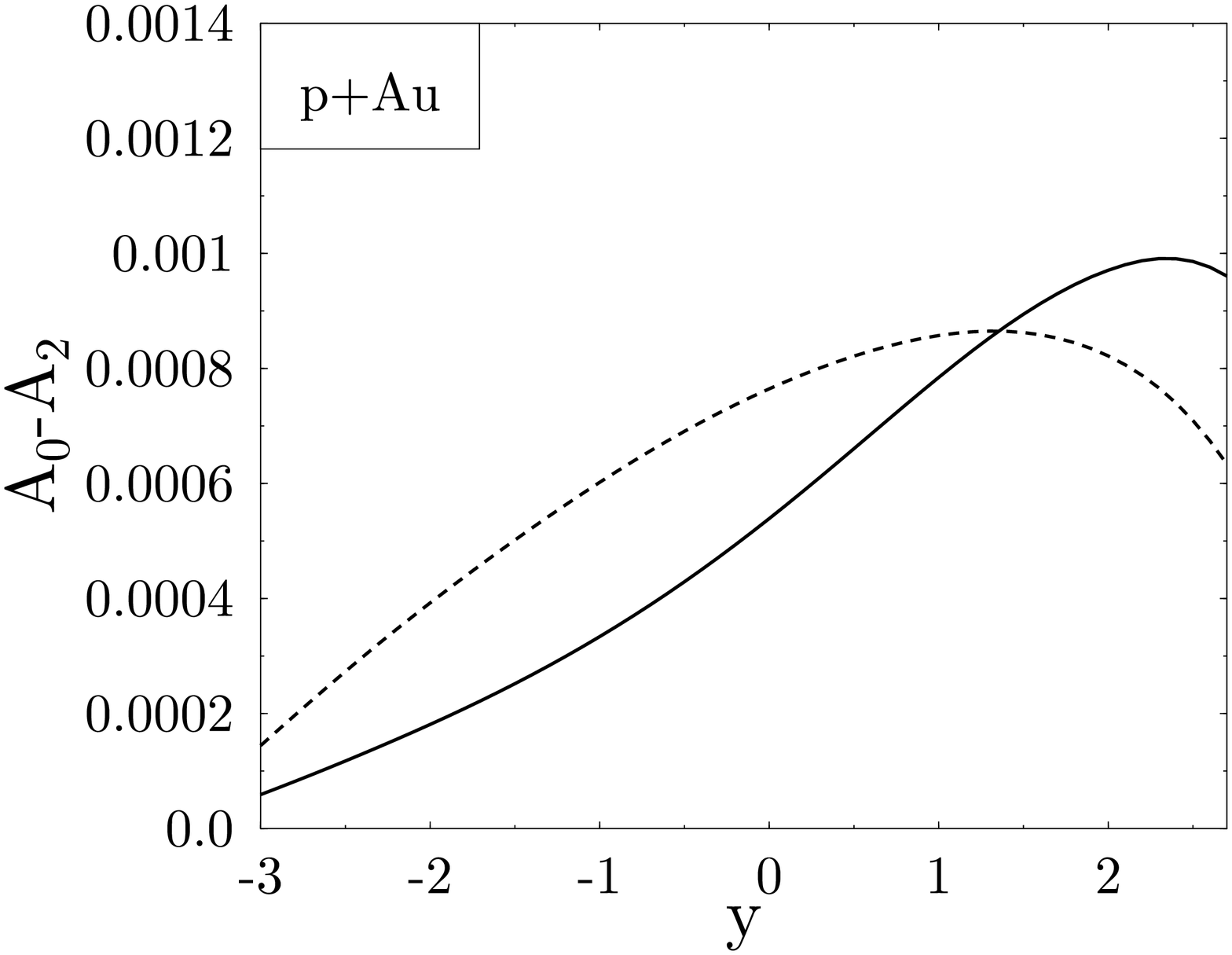,width=5.5cm}
  \caption{Violation of the Lam-Tung relation. Difference of angular 
  coefficients $A_0$ and $A_2$ is shown. Calculation are with (solid line) and
  without (dashed line) EKS98 corrections to nuclear parton distributions.}
  \label{lamtfig} 
\end{figure}

\begin{figure}[b]
  \epsfig{file=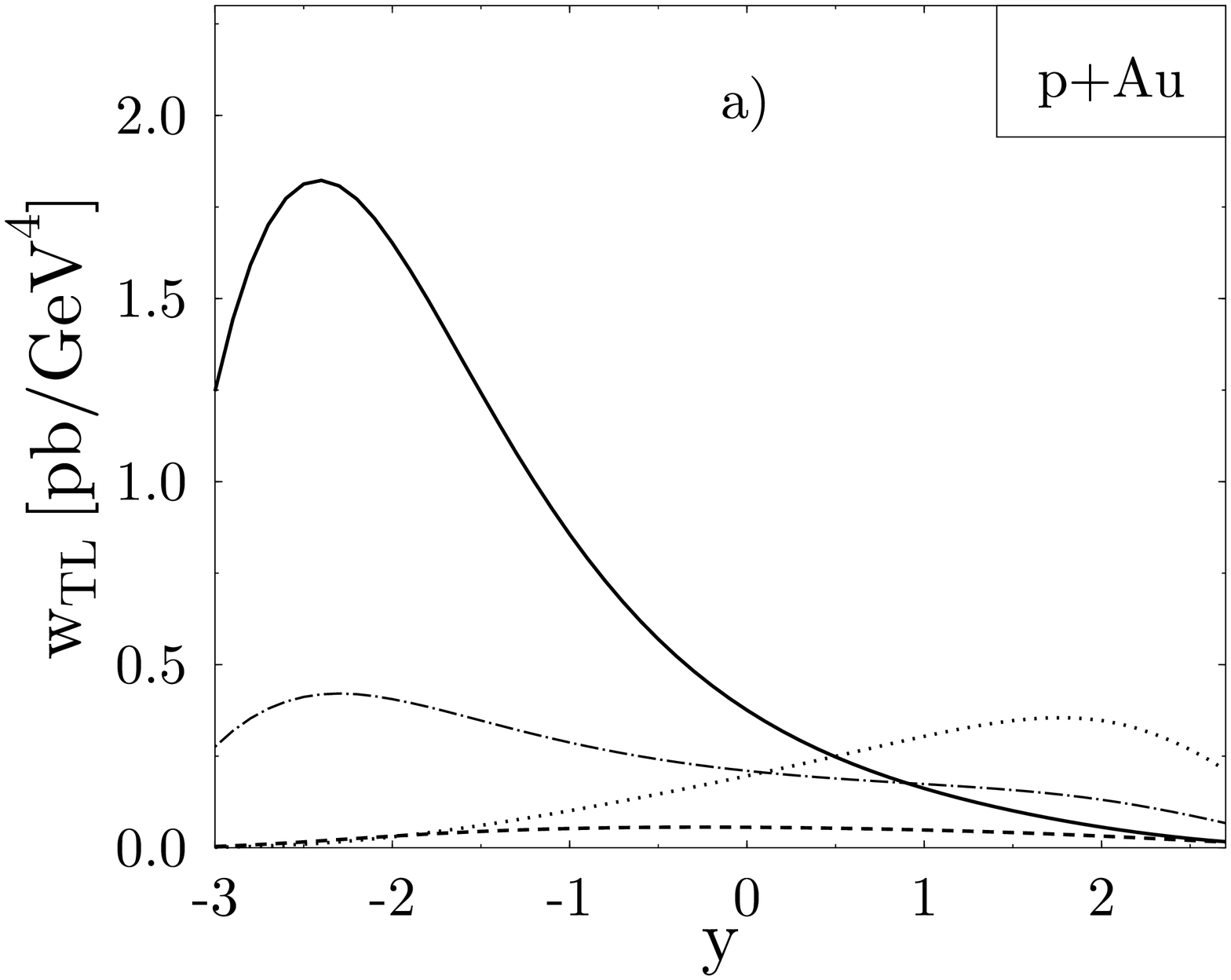,width=5.5cm}
  \epsfig{file=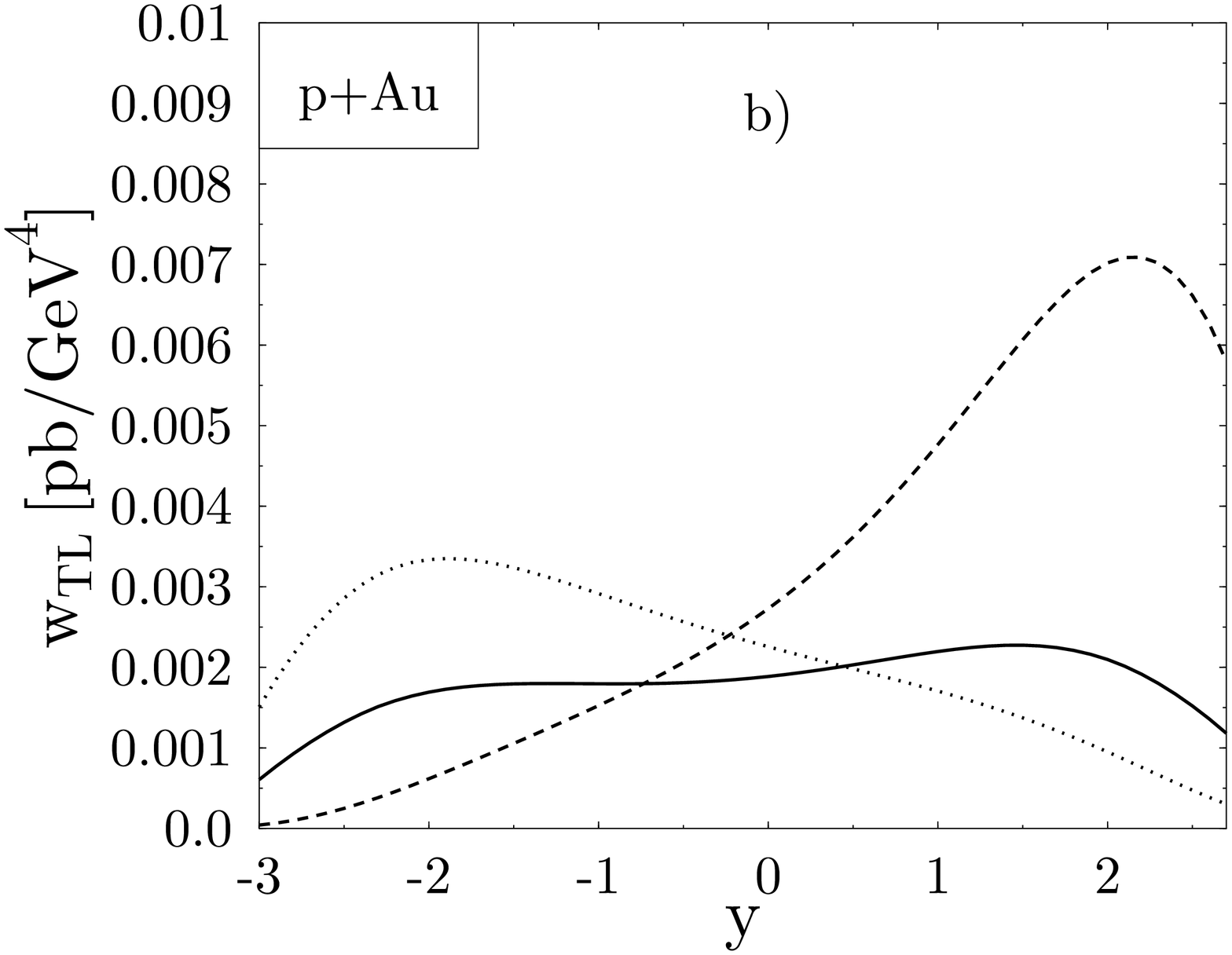,width=5.5cm}
  \caption{Contributions of different subprocesses to $w_\rtl$ for
  double-hard (a) and soft-hard (b) scattering. Double-hard subprocesses 
  are $qg+\bar q$
  (solid line), $qg+g$ (dashed line), $q\bar q+g$ (dotted line) and 
  $q\bar q+q$ plus all pure fermionic processes
  (dash-dotted line). Soft-hard processes are $qg+\bar q$
  (solid line), $qg+g$ (dashed line) and $gg+q$ (dotted line).}   
  \label{tlcontr}
\end{figure}

\begin{figure}[t]
  \epsfig{file=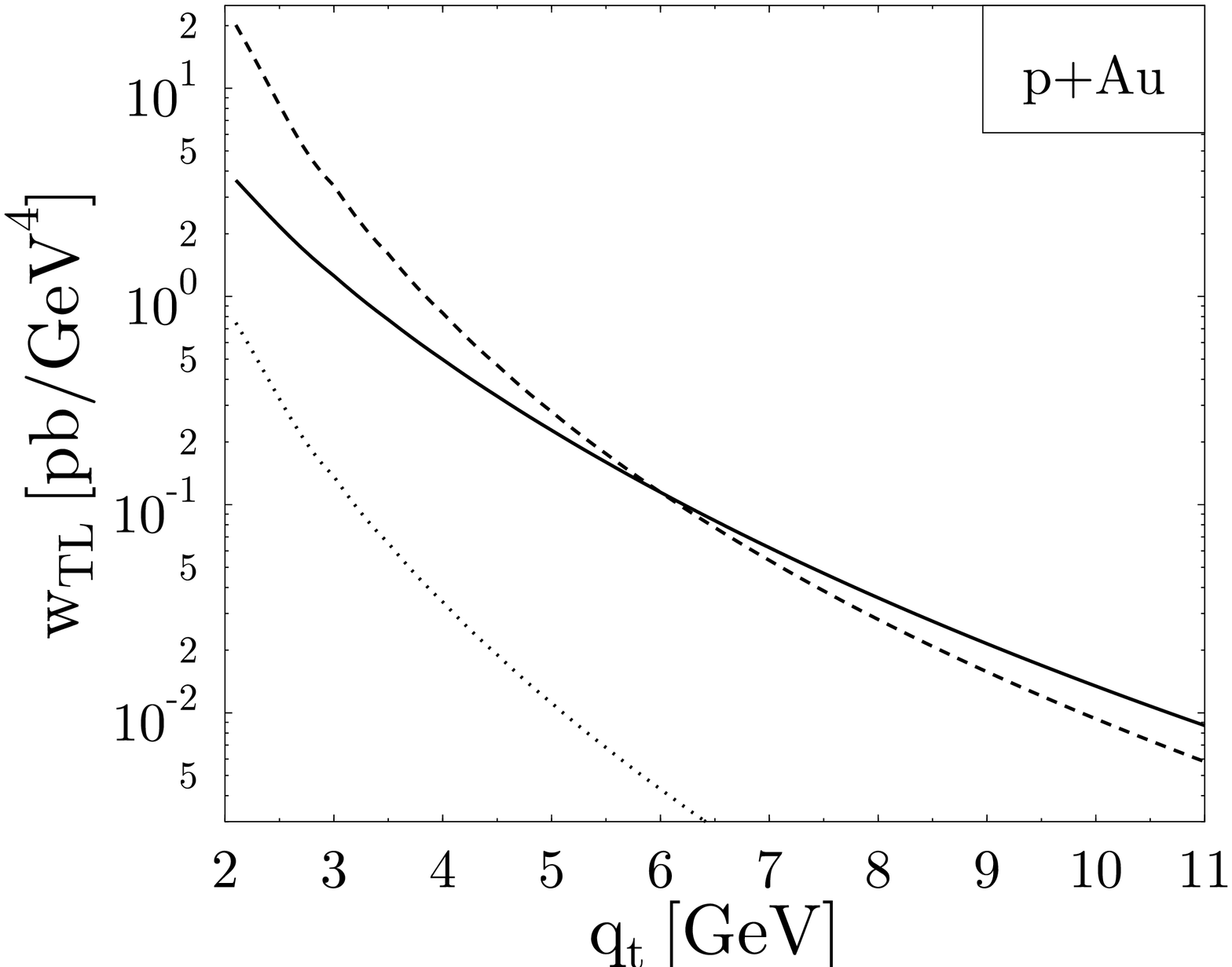,width=5.5cm}
  \epsfig{file=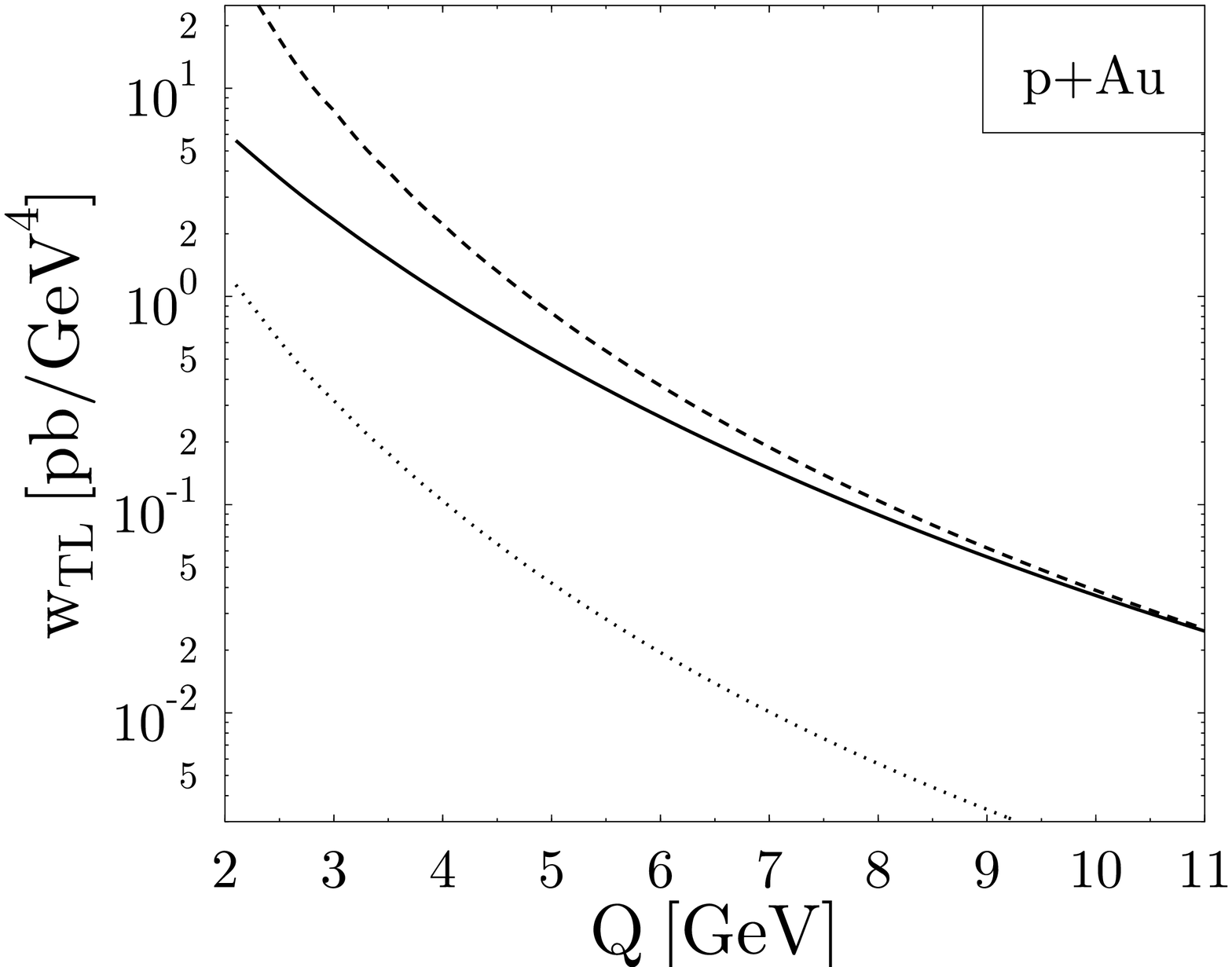,width=5.5cm}
  \caption{$w_\rtl$ as a function of transverse momentum $q_\perp$ at 
  fixed $Q=5 \gev$, $y=0$ 
  (left) and as a function of photon mass $Q$ at $q_\perp=4 \gev$ and $y=0$
  (right). 
  Curves are shown for single scattering (solid line), double-hard 
  (dashed line) and soft-hard scattering (dotted line).}
  \label{tlqt} 
\end{figure}

We turn to $p+Au$ collisions at RHIC with $S=10^5\gev^2$ per nucleon pair. We 
give helicity amplitudes instead of angular coefficients since these provide a
clearer measure of nuclear enhanced higher twist effects as explained 
above. To make numerical results of helicity amplitudes $W_\alpha$, $\alpha 
\in \{\rtl,\rl,\Delta,\Delta\Delta \}$ more meaningful, we introduce the 
prefactor given in eq.~\eqref{cross2},
\begin{equation}
  w_\alpha = \frac{\alpha^2}{64 \pi^3 SQ^2} W_\alpha,
\end{equation}
to convert the quantities $W_\alpha$ into differential cross sections.
In Fig.~\ref{yfig} we give single scattering, double-hard and soft-hard 
results for the helicity amplitudes $w_\rtl$, $w_\rtl$ and $w_\Delta$ as 
functions of rapidity $y$ in the Collins-Soper frame.
One can read off the main results already given in \cite{fmss99}. First the 
nuclear enhanced twist-4 contribution is absolutely important at RHIC 
energies. Even at quite large values of $Q=5\gev$ and $q_\perp=4\gev$ used
for the calculation its effect can exceed the leading-twist result for the 
parameters used. 
Second the double-hard contribution is by far larger than the soft-hard part 
in the kinematical range under consideration. This can be seen for all 
helicity amplitudes in Fig.~\ref{yfig}. Even if we use smaller values of $C$
proposed above double-hard scattering would reach the same order of magnitude
as single scattering.

We also give results with EKS98 modifications to nuclear parton distributions.
These para\-metrizations take into account shadowing, antishadowing and other 
nuclear effects on parton distributions, first described by the European Muon 
Collaboration. The numerical effect of these modifications on the single 
scattering results are small, but they turn out to be quite large for 
double-hard  scattering. Note that it is not clear whether these nuclear 
effects for twist-2 parton distributions should be carried over to the models 
for twist-4
matrix elements. We give results for modified and unmodified distributions to
indicate the further uncertainty of the twist-4 matrix elements. The 
sensitivity of double-hard scattering to these EKS98 modifications is simply 
related to the fact that they involve a product of two nuclear parton 
distributions. The general question about the correct small-$x$ behaviour of 
the twist-4 nuclear matrix elements is very interesting since it has a big
influence at RHIC energies. 

We return to the question of the Lam-Tung relation. We already saw analytically
that double-hard scattering respects this relation  while soft-hard does not. 
In Fig.~\ref{lamtfig} we show the difference $A_0-A_2$. It only picks up 
contributions from soft-hard scattering and therefore this quantity is very 
small. That means that violation of the Lam-Tung sum rule is a tiny effect for
the process under consideration.

The characteristic feature of the rapidity distribution is the peak of the 
double-hard distribution at negative rapidities (i.e.\ for the photon emitted 
in direction of the initial single hadron). The reason for this is that 
negative rapidity prefers small Bjorken-$x$ for partons from the nucleus. 
Parton distributions at small parton momentum fractions are large particularly
for gluons and they enter quadratically for double-hard scattering.
In Fig.~\ref{tlcontr} we separated the different partonic subprocesses 
contributing to double-hard and soft-hard scattering to enable a deeper 
analysis. One can see that double-hard processes with a quark and a gluon from
the nucleus dominate over those with two quarks at negative rapidities (i.e. 
small $x_h$). On the other hand $qg+\bar q$ is more important than $qg+g$ in 
this region since momentum fraction $\xi_2$ of the parton from the single 
hadron is large and quarks dominate over gluons in that case. 
All the rapidity spectra can be understood in this way.

Fig.~\ref{tlqt} shows dependence of $w_\rtl$ on photon mass $Q$ and transverse
momentum $q_\perp$. It can be seen that soft-hard and double-hard 
contributions as expected for higher-twist processes fall off more rapidly 
with $Q$ and $q_\perp$ than (twist-2) single-scattering.

\section{Summary}

We have calculated nuclear enhanced double-scattering contributions to 
Drell-Yan pair production in hadron-nucleus collisions using the framework of
 Luo, Qiu and Sterman. We found that the nuclear enhancement at RHIC energies 
and large transverse momentum makes double scattering a contribution that is 
equally important as single scattering. 
Measurements of angular coefficients and helicity amplitudes can distinguish 
double-hard and soft-hard 
scattering. Double-hard scattering has the interpretation of two sequential 
hard scatterings of onshell partons and is therefore related to probabilistic 
pictures of multiple scattering that are widely used in the phenomenology of 
heavy-ion reactions. In this spirit further comparision with alternative
approaches to multiple scattering \cite{bod89,kts99} would be very interesting.
 
Double-hard scattering respects the Lam-Tung relation and, in contrast to 
soft-hard scattering, does not show  non-trivial deviations from the 
$1+\cos^2\theta$ angular distribution. Numerically double-hard scattering 
dominates over the soft-hard contribution. Soft-hard scattering violates 
the Lam-Tung relation but this violation is an effect of order $10^{-3}$ 
for the model used. 

Double-hard scattering can be modeled as the product of two twist-2 structure
functions which give the probability for finding two partons in the nucleus.
The soft-hard matrix elements on the other hand encodes essential 
non-perturbative properties  of the nucleus, describing the influence of the 
collective color field on the transverse momentum of the produced lepton pair. 
Inserting reasonable models for the corresponding matrix elements we predict,
that collective color effects are negligible compared to the effects of 
subsequent independent hard scatterings in the kinematical range considered 
here. However the absolute normalization of the twist-4 matrix elements in 
these models is not yet clear.

The double-hard scattering shows an interesting asymmetry in the rapidity
distribution of the Drell-Yan pair. The pair is predominantly produced in the 
projectile, i.e. single hadron rapidity region.
Although the absolute counting rates for the DY pairs at large transverse 
momentum are not high \cite{fmss99} the predicted effect fits perfectly well 
in the PHENIX detector acceptance \cite{phenix}. This asymmetry therefore 
should be observable.

An analysis of $W_L$ and $W_\Delta$, particularly in the Gottfried-Jackson 
frame, as well as the measurement of $W_\rtl$ (i.e.~the angular integrated 
cross section) as a function of rapidity can check whether double-hard 
scattering is the dominant contribution. A possible violation of the 
Lam-Tung sum rule provides a further possibility to check the picture of 
soft-hard and double-hard scattering in general. If our predictions with
respect to the ratio of soft-hard and double-hard scattering are confirmed by 
experiment it would provide support for the description of heavy ion reactions 
in terms of incoherent multiple scattering of free partons.
Note that the prediction of a trivial frame dependence of double-hard 
scattering and the conservation of the Lam-Tung relation are independent of 
the models for the twist-4 distribution functions and provide good tests 
for the entire LQS approach.
\vspace*{5mm}

{\bf Acknowledgements.}
We are grateful to X.~Guo for helpful correspondence. 
In course of the work we benefitted from discussions with V.~Braun, 
L.~Syzmanowsky, and O.~Teryaev. 
E.~S.\ thanks G.~Sterman and X.N.~Wang for useful discussions.

\section{Appendix}

\subsection{Transformation of amplitudes to different frames}

Here we list the matrices $M_{(\beta)}$ defined in eq.~(\ref{mirkes2}) that 
map the invariant projections of the hadronic tensor to the set of helicity 
amplitudes ($W_\rtl$, $W_L$, $W_\Delta$, $W_{\Delta\Delta}$). They are
\begin{equation}
  \label{mapcs}
  \begin{split}
  M_\rcs &= \displaystyle \begin{pmatrix} \frac{1}{2} & 0 & 0 & 0 \\
        0 & \frac{1}{4 \cos^2\gamma_\rcs} & \frac{1}{4 \cos^2\gamma_\rcs} &
        -\frac{1}{4 \cos^2\gamma_\rcs} \\
        0 & -\frac{1}{4 \sin\gamma_\rcs\cos\gamma_\rcs} & 
        \frac{1}{4 \sin\gamma_\rcs\cos\gamma_\rcs} & 0 \\
        \frac{1}{2} & -\frac{1+\cos^2\gamma_\rcs}{8\cos^2\gamma_\rcs
        \sin^2\gamma_\rcs} & -\frac{1+\cos^2\gamma_\rcs}{8\cos^2\gamma_\rcs
        \sin^2\gamma_\rcs} & \frac{1-3\cos^2\gamma_\rcs}{8\cos^2\gamma_\rcs
        \sin^2\gamma_\rcs}
           \end{pmatrix}    \\
  \medskip
  M_\rgj &= \begin{pmatrix} \frac{1}{2} & 0 & 0 & 0 \\ 0 & 1 & 0 & 0 \\
        0 & -\cot \gamma_\rgj & 0 & -\frac{1}{2\sin\gamma_\rgj} \\
        \frac{1}{2} & -\frac{1+\cos^2\gamma_\rgj}{2\sin^2\gamma_\rgj} &
        -\frac{1}{\sin^2\gamma_\rgj} & 
        \frac{\cos\gamma_\rgj}{\sin^2\gamma_\rgj}
           \end{pmatrix} 
  \end{split}
\end{equation}
for the Collins-Soper and the Gottfried-Jackson frame, respectively. 
$\gamma_\rcs$ denotes the angle between ${\mathbf P}_1$ and the $\mathbf Z$ 
axis in the photon rest frame and $\gamma_\rgj$ is the angle between 
${\mathbf P}_1$ and ${\mathbf P}_2$. Obviously $2\gamma_\rcs + \gamma_\rgj = 
\pi$ and in terms of hadronic Mandelstam variables we have
\begin{eqnarray}
  \cos\gamma_\rcs = \sqrt{\frac{Q^2 S}{(Q^2-T)(Q^2-U)}}, &\quad &
  \sin\gamma_\rcs = -\sqrt{1-\frac{Q^2 S}{(Q^2-T)(Q^2-U)}} \\
  \cos\gamma_\rgj = 1-\frac{2Q^2 S}{(Q^2-T)(Q^2-U)},     &\quad &
  \sin\gamma_\rgj = \frac{2Q^2 S}{(Q^2-T)(Q^2-U)}\sqrt{\frac{(Q^2-T)(Q^2-U)}{
   Q^2S} -1}. \nn
\end{eqnarray}

\pagebreak

\subsection{Single-scattering results}

Here we list the results for the hard parts $H_\alpha$ of single-scattering 
helicity amplitudes in the Collins-Soper frame. They are defined in 
eq.~(\ref{t2result}). We distinguish annihilation and Compton processes. 
\begin{equation}
\label{loresults}
\begin{split}
H_\rtl^{q+\bar{q}} &= \mathcal{C}_1 \frac{1}{t u }
  \left((Q^2 -t)^2 + (Q^2-u)^2\right)
   \\
H_\rl^{q+\bar{q}} &= \mathcal{C}_1 \left(
  \frac{Q^2 -t}{Q^2-u} + \frac{Q^2 -u}{Q^2-t}  \right)
   \\
H_{\Delta}^{q+\bar{q}} &= \mathcal{C}_1 \sqrt{\frac{Q^2 s}{t u}}
  \left( \frac{Q^2 -u}{Q^2-t}  - \frac{Q^2 -t}{Q^2-u} \right)
   \\
H_{\Delta\Delta}^{q+\bar{q}} &= \frac{1}{2} H_L^{q+\bar{q}}
   \\
H_\rtl^{q+g} &= - \mathcal{C}_2 \frac{(Q^2-s)^2+(Q^2-t)^2}{st}
   \\
H_\rl^{q+g}  &= - \mathcal{C}_2 \frac{u}{s}
  \frac{(Q^2+s)^2+(Q^2-t)^2}{(Q^2-t)(Q^2-u)}
  \\  
H_{\Delta}^{q+g}  &= - \mathcal{C}_2 \sqrt{\frac{Q^2 u}{st}} 
  \frac{2(Q^2-t)^2-(Q^2-u)^2}{(Q^2-t)(Q^2-u)} 
   \\
H_{\Delta\Delta}^{q+g} &= \frac{1}{2} H_L^{q+g}
   \\
H_\rtl^{g+q} &= - \mathcal{C}_2 \frac{(Q^2-s)^2+(Q^2-u)^2}{su}
   \\
H_\rl^{g+q}  &= - \mathcal{C}_2 \frac{t}{s}
  \frac{(Q^2+s)^2+(Q^2-u)^2}{(Q^2-t)(Q^2-u)}
   \\
H_{\Delta}^{g+q}  &= \mathcal{C}_2 \sqrt{\frac{Q^2 t}{su}} 
  \frac{2(Q^2-u)^2-(Q^2-t)^2}{(Q^2-t)(Q^2-u)} 
   \\
H_{\Delta\Delta}^{g+q} &= \frac{1}{2} H_L^{g+q}.
\end{split}
\end{equation}
$\mathcal{C}_1 = C_F/N_c = 4/9$ and $\mathcal{C}_2 = 1/(2N_c)=1/6$ are 
essentially the color factors.

\subsection{Double-hard results}

In this section we give the hard parts $H_\rtl$ of the projections
$W^\rdh_\rtl$ for double-hard processes. They do not depend on the frame. 
\begin{equation}
  \begin{split}
  H_\rtl^{DH,qg+\bar q} &= 
  \frac{2}{27} \left( \frac{(Q^2-T)}{\xi_2 S} + \frac{\xi_2 S}{Q^2-T}
  \right) + \frac{1}{6} \frac{{(Q^2-T)}^2+(\xi_2 S)^2}{\xi_2 S+T-Q^2} \\
  H_\rtl^{DH,qg+g} &= \frac{1}{36} \left( \frac{\xi_2 S+T-Q^2}{Q^2-T}+
  \frac{Q^2-T}{\xi_2 S+T-Q^2} \right) - \frac{1}{16} \frac{{(Q^2-T)}^2+
  {(\xi_2 S+T-Q^2)}^2}{(\xi_2 S)^2} \\
  H_\rtl^{DH,q \bar q +g} &= \frac{2}{27} \left( 
  \frac{\xi_2 S+T-Q^2}{\xi_2 S} + \frac{\xi_2 S}{\xi_2 S+T-Q^2} \right)
  + \frac{1}{6} \frac{(\xi_2 S)^2 + {(\xi_2 S+T-Q^2)}^2}{{(Q^2-T)}^2} \\
  H_\rtl^{DH,q \bar q +q} &= \frac{2}{27} \left( \frac{(\xi_2 S)^2 + 
  {(\xi_2 S+T-Q^2)}^2}{{(Q^2-T)}^2} + \frac{{(Q^2-T)}^2+
  {(\xi_2 S+T-Q^2)}^2}{(\xi_2 S)^2} \right) + \\
  & + \frac{4}{81} \frac{{(\xi_2 S+T-Q^2)}^2}{\xi_2 S(Q^2-T)}   \\
  H_\rtl^{DH,q\bar q +\bar q} &= \frac{2}{27} \left( \frac{(\xi_2 S)^2 + 
  {(\xi_2 S+T-Q^2)}^2}{{(Q^2-T)}^2} + \frac{{(Q^2-T)}^2+
  {(\xi_2 S)}^2}{(\xi_2 S+T-Q^2)^2} \right) - \\
  & -\frac{4}{81} \frac{{(\xi_2 S)}^2}{(Q^2-T)(\xi_2 S+T-Q^2)}  \\
  H_\rtl^{DH,qq +\bar q} &= \frac{2}{27} \left( \frac{{(Q^2-T)}^2 + 
  {(\xi_2 S+T-Q^2)}^2}{{(Q^2-T)}^2} + \frac{{(Q^2-T)}^2+
  {(\xi_2 S)}^2}{(\xi_2 S+T-Q^2)^2} \right) + \\
  & +\frac{4}{81} 
  \frac{{(Q^2-T)}^2}{\xi_2 S(\xi_2 S+T-Q^2)}  \\
  H_\rtl^{DH,q\bar q +q'} &= \frac{2}{27} \frac{{(\xi_2 S)}^2 + 
  {(\xi_2 S+T-Q^2)}^2}{{(Q^2-T)}^2}  \\
  H_\rtl^{DH,qq' +\bar q'} &= \frac{2}{27} \frac{{(Q^2-T)}^2 + 
  {(\xi_2 S+T-Q^2)}^2}{{(\xi_2 S)}^2} \\
  H_\rtl^{DH,qq' +\bar q} &= \frac{2}{27} \frac{{(Q^2-T)}^2 + 
  {(\xi_2 S)}^2}{{(\xi_2 S+T-Q^2)}^2}  
  \end{split}
\end{equation}

\subsection{Soft-hard results}

Here we list the hard parts $H(T_\alpha)$ of the invariant projections 
$T^\rsh_\rtl$, $T^\rsh_\rli$, $T^\rsh_\rlii$ and $T^\rsh_\rliii$ of the 
hadronic tensor, defined in section 2.2, for soft-hard processes. The results 
can be projected into a specific frame via eq.~(\ref{mirkes2})
and the matrices given in Appendix 7.1. This is mathematically trivial but it 
complicates the expressions substantially. For this reason we give only the 
invariant projections. Note that especially $H_\rtl = H(T_\rtl)/2$ for every 
frame described in section 2.2.
\begin{equation}
  \begin{split}
  H(T_\rtl^{qg+\bar q}) &= \frac{4}{27} \left( \frac{(Q^2-T)(x_c-x_a)}{
    x_c (\xi_2 S+T-Q^2)} + \frac{2 x_a \xi_2 S}{(x_c-x_a) (\xi_2 S+T-Q^2)}
    + \frac{x_c(\xi_2 S+T-Q^2)}{(Q^2-T)(x_c-x_a)} \right)
  \\
  H(T_\rli^{qg+\bar q}) & = \frac{8}{27} \frac{x_a \xi_2 S}{x_c (Q^2-T)}
  \\ 
  H(T_\rlii^{qg+\bar q}) &= \frac{8}{27} \left( \frac{\xi_2 S+T-Q^2}{
    (x_c+x_s) S +U-Q^2} \frac{x_s Q^2 S}{(Q^2-U)^2}
    - \frac{x_a}{x_a-x_c} \frac{x_c S ((x_c+x_s) S +U-Q^2)}{
    (Q^2-U)^2} \right. \\
    & \left. - \frac{x_a}{x_a-x_c} \frac{S^2}{(Q^2-U)^2} 
    \frac{(x_c+x_s)((x_c-x_s)(Q^2-T) + \xi_2 (Q^2-U) -x_c\xi_2 S) - 
    (x_c-x_s) Q^2}{\xi_2 S+T-Q^2} \right)
  \\  
  H(T_\rliii^{qg+\bar q}) &= - \frac{8}{27} \left( 
    \frac{x_a}{x_a-x_c} \frac{\xi_2 S^2}{(Q^2-T)(Q^2-T)} \frac{(x_c-x_s)
    (Q^2-T) + \xi_2 (Q^2-U) -Q^2 -x_c\xi_2 S}{\xi_2 S+T-Q^2} + \right. \\
    & \left. + \frac{x_a s}{x_c (Q^2-U)} \frac{(x_c+x_s)(\xi_2 S+T-Q^2)+\xi_2
    ((x_c+x_s) S +U-Q^2)-Q^2+x_c\xi_2 S }{\xi_2 S+T-Q^2} \right)
  \\
  H(T_\rtl^{qg+g}) &= \frac{1}{8} \left( \frac{x_c \xi_2 S}{(x_c-x_a)(Q^2-T)} 
    - \frac{2x_a(\xi_2 S+T-Q^2)}{\xi_2 S(x_c-x_a)} +
    \frac{(x_c-x_a)(Q^2-T)}{x_c \xi_2 S}  \right)
  \\
  H(T_\rli^{qg+g}) &= \frac{1}{4} \frac{x_a(\xi_2 S+T-Q^2)}{x_c(Q^2-T)}
  \\ 
  H(T_\rlii^{qg+g}) &= \frac{1}{4} \left( \frac{x_s Q^2}{x_c(Q^2-U)}
    \frac{(x_c+x_s) S +U-Q^2}{\xi_2 (Q^2-U)} - \right. \\
    & - \frac{x_a}{x_a-x_c} \frac{(x_c+x_s) S +U-Q^2}{\xi_2(Q^2-U)}
    \frac{(x_c-x_s)(Q^2-T) + \xi_2 (Q^2-U)-Q^2 - x_c \xi_2 S}{Q^2-U} - \\
    & \left. - \frac{x_a x_c}{x_a-x_c} \frac{((x_c+x_s) S +U-Q^2)
    (\xi_2 S+T-Q^2)+(Q^2-T)(Q^2-U)-Q^2 S}{\xi_2(Q^2-U)^2} \right)
  \\  
  H(T_\rliii^{qg+g}) &= \frac{1}{4} \left( \frac{x_a}{x_a-x_c} 
    \frac{\xi_2 S+T-Q^2}{Q^2-T)} \frac{(x_c+x_s)(Q^2-T)-\xi_2(Q^2-U)-Q^2
    +x_c \xi_2 S}{\xi_2 (Q^2-U)} + \right. \\
    & \left. + \frac{x_a}{x_c} \frac{(x_c+x_s)(\xi_2 S+T-Q^2)+\xi_2
    ((x_c+x_s) S +U-Q^2)+Q^2-x_c\xi_2 S}{\xi_2(Q^2-U)} \right)
  \\
  H(T_\rtl^{qg+g}) &=  \frac{1}{18} \left( \frac{\xi_2 S}{\xi_2 S+T-Q^2} -
    \frac{2Q^2(Q^2-T)(x_c-x_a)}{x_c^2 \xi_2 S(\xi_2 S+T-Q^2)} 
    \frac{\xi_2 S+T-Q^2}{\xi_2 S} \right)
  \\
  H(T_\rl^{qg+g}) &= -\frac{2}{9} \frac{x_a*(x_a-x_c)}{x_c^2} 
  \\ 
  H(T_{\Delta}^{qg+g}) &= \frac{1}{9} \frac{Q^2}{Q^2-U} \left( 
    \frac{(x_c+x_s) S +U-Q^2}{\xi_2 (Q^2-U)} + \frac{x_s S}{(Q^2-U)}
    \frac{S}{\xi_2 S+T-Q^2} - \right. \\
    & -\frac{(x_c S+U-Q^2) ((x_c+x_s)(\xi_2 
    S+T-Q^2)+\xi_2 ((x_c+x_s) S +U-Q^2)+Q^2-x_c\xi_2 S)}{x_c \xi_2 (Q^2-U) 
    (\xi_2 S+T-Q^2)} - \\
    & \left. \frac{2x_s Q^2((x_c+x_s) S +U-Q^2)}{x_c^2 \xi_2 (Q^2-U) 
    (\xi_2 S+T-Q^2)} \right)
  \\  
  H(T_{\Delta\Delta}^{qg+g}) &= -\frac{1}{9} \frac{x_a}{\xi_2 x_c} \left(
    \frac{2x_c \xi_2 (Q^2-U) +x_c(x_c+x_s)(Q^2-T) + x_c^2 \xi_2 S 
    +2x_2 Q^2}{Q^2-U} + \right. \\
    & \left. + \frac{\xi_2((x_c+x_s) S +U-Q^2)(2Q^2+x_c (Q^2-T)) 
    +3x_c Q^2 (Q^2-T) +2Q^4 +x_c^2\xi_2^2 S^2}{(Q^2-U) (\xi_2 S+T-Q^2)} 
    \right)
  \end{split} \nn
\end{equation}

\vfill
\eject

\end{document}